\documentclass[twocolumn,tighten]{aastex6}

\newcommand{\lint}{$L_{\rm int}$}
\newcommand{\lbol}{$L_{\rm bol}$}
\newcommand{\lacc}{$L_{\rm acc}$}
\newcommand{\lsun}{$L_{\sun}$}
\newcommand{\msun}{$M_{\sun}$}
\newcommand{\rsun}{$R_{\sun}$}

\newcommand{\vsys}{$v_{\rm sys}$}
\newcommand{\yro}{yr$^{-1}$}

\usepackage{natbib,graphicx,morefloats,multirow,pbox,booktabs}
\usepackage{hyperref}
\usepackage{breakurl}
\usepackage{soul}
\usepackage{color}

\begin{document}
  
\title{CO Outflow survey of 68 Very low luminosity objects: A Search for Proto-Brown Dwarf Candidates}

\author{Gwanjeong Kim\altaffilmark{1,2,3,7}, Chang Won Lee\altaffilmark{2,3,10}, Gopinathan Maheswar\altaffilmark{4}, Mi-Ryang Kim\altaffilmark{2}, Archana Soam\altaffilmark{2,9}, Masao Saito\altaffilmark{5}, Kazuhiro Kiyokane \altaffilmark{5,6}, and Sungeun Kim\altaffilmark{8}} 

\altaffiltext{1}{Nobeyama Radio Observatory, National Astronomical Observatory of Japan, National Institutes of Natural Sciences, Nobeyama, Minamimaki, Minamisaku, Nagano 384-1305, Japan}
\altaffiltext{2}{Korea Astronomy and Space Science Institute, 776 Daedeokdae-ro, Yuseong-gu, Daejeon, 34055, Republic of Korea}
\altaffiltext{3}{Department of Astronomy and Space Science, University of Science \& Technology, 217, Gajeong-ro, Yuseong-gu, Daejeon, 34113, Republic of Korea}
\altaffiltext{4}{Indian Institute of Astrophysics, II Block, Koramangala, Bengaluru 560 034, INDIA}
\altaffiltext{5}{National Astronomical Observatory of Japan, National Institutes of Natural Sciences, 2-21-1 Osawa, Mitaka, Tokyo 181-8588, Japan}
\altaffiltext{6}{Department of Astronomy, The University of Tokyo, 7 Chome-3-1 Hongo, Bunkyō, Tokyo 113-8654, Japan}
\altaffiltext{7}{Department of Physics and Astronomy, Sejong University, 209, Neungdong-ro, Gwangjin-gu, Seoul, 05006, Republic of Korea}
\altaffiltext{8}{Department of Astronomy and Space Science, Sejong University, 209, Neungdong-ro, Gwangjin-gu, Seoul, 05006, Republic of Korea}
\altaffiltext{9}{SOFIA Science Centre, USRA, NASA Ames Research Centre, MS N232-12 Moffett Field, CA 94035, USA}
\altaffiltext{10}{corresponding author}

\begin{abstract}
We present the results of a systematic search for molecular outflows in 68 Very Low Luminosity Objects (VeLLOs) from single-dish observations in CO isotopologues which find 16 VeLLOs showing clear outflow signatures in the CO maps. With additional three VeLLOs from the literature, we analyzed the outflow properties for 19 VeLLOs, identifying 15 VeLLOs as proto-Brown Dwarf (BD) candidates and four VeLLOs as likely faint protostar candidates. The proto-BD candidates are found to have a mass accretion rate ($\sim 10^{-8} - 10^{-7}$ {\msun} yr$^{-1}$) lower than that of the protostar candidates ($\gtrsim 10^{-6}$ {\msun} yr$^{-1}$). Their accretion luminosities are similar to or smaller than their internal luminosities, implying that many proto-BD candidates might have had either small accretion activity in a quiescent manner throughout their lifetime, or be currently exhibiting a relatively higher (or episodic) mass accretion than the past. There are strong trends that outflows of many proto-BDs are less active if they are fainter or have less massive envelopes. The outflow forces and internal luminosities for more than half of the proto-BD candidates seem to follow an evolutionary track of a protostar with its initial envelope mass of $\sim$0.08 {\msun}, indicating that some BDs may form in less massive dense cores in a way similar to normal stars. But, because there also exists a significant fraction (about 40\%) of proto-BDs with much weaker outflow force than expected by the relations for protostars, we should not rule out the possibility of other formation mechanism for the BDs.
\end{abstract}

 
\section{Introduction \label{sec:intro}}
The \textit{Spitzer} C2D legacy programme \citep{Evans:2003bo} has discovered a new class of objects referred to as Very Low Luminosity Objects (VeLLOs). The VeLLO is defined as an object with an internal luminosity of $\leq$0.1 {\lsun} deeply embedded in a dense molecular cloud \citep[e.g.,][]{Young:2004bc, DiFrancesco:2007vg}. 

Presently, there is no clear consensus on the nature of the VeLLOs mainly because of the lack of detailed studies conducted on them. In literature there exist only a few number of sources ($\sim$9) that have been studied in detail so far, namely, L1014-IRS \citep{Young:2004bc,Bourke:2005kg}, IRAM 04191+1522 \citep{Andre:1999tr,Dunham:2006jl}, L1521F-IRS \citep{Bourke:2006kn,Takahashi:2013fm}, L328-IRS \citep{Lee:2009bq,Lee:2013gx,Lee:2018iq}, L673-7-IRS \citep{Dunham:2010fk,Schwarz:2012hm}, L1251-IRS4 \citep{Lee:2010fc}, L1148-IRS \citep{Kauffmann:2011kz}, IC348-SMM2E \citep{Palau:2014kp}, and IRAS 16253-2429 \citep{Hsieh:2016if}. These studies have revealed that their bolometric temperatures and spectral energy distributions are very similar to those of protostars, while the luminosities of VeLLOs are much fainter than the expected luminosity ($\sim$1.6 {\lsun}) of the least mass protostar ($\sim$0.08 {\msun}) with a radius of $\sim$3 {\rsun} and a constant mass accretion rate of $\sim2\times10^{-6}$ {\msun} {\yro} in the standard star-formation theory \citep[e.g.,][]{Shu:1987dp,Dunham:2006jl}. Therefore the VeLLOs can be an extreme case exhibiting the ``luminosity problem'' identified in Young Stellar Objects (YSOs) by \citet{Kenyon:1990ki}. One viable solution suggested to resolve the luminosity problem seen in the VeLLOs is an episodic accretion scenario. According to this the accretion rate is usually low, but sparsely interrupted by bursts of relatively large accretion rates, making the average internal luminosity of a VeLLO much smaller than that expected from a uniform accretion rate \citep{Dunham:2010bx,Dunham:2012ic,Vorobyov:2017dt}. Therefore, a majority of the VeLLOs may be at present in a low (quiescent) accretion stage while there may have been occasional massive accretions sometime in the past. To investigate whether the episodic accretion could be a viable solution to the luminosity problem, comprehensive surveys of mass accretion rates in the vicinity of the VeLLOs are required. 

The direct measurement of the mass accretion rate is difficult because the accretion processes occur over very small spatial scales that are highly obscured inside the dense cores. However, the molecular outflows originating from the embedded sources could be used as an indirect way to infer the mass accretion rate. Molecular outflows are known to be generated when a portion of the matter accreted by the central object is ejected along the poles \citep[e.g.,][]{Cabrit:1992uw,Bontemps:1996vb}. These outflows may provide information to the past accretion history of the central object \citep{Dunham:2006jl,Dunham:2010fk,Lee:2010fc}. In addition to this, the knowledge of the accretion rate and the masses of both the central object and its envelope could provide a potential diagnostic to determine the present status and the final fate of the central object.

So far outflows have been found in nine VeLLOs in CO emission. Their properties inferred from the detected outflows are found to be somewhat diverse. For example, IRAM 04191+1522 and L673-7-IRS have the accretion rates ($\sim 10^{-6}$ {\msun} yr$^{-1}$) of normal stars with their envelope masses of a few solar masses, and thus are proposed to be faint low mass protostars \citep{Andre:1999tr,Dunham:2010fk,Schwarz:2012hm}. However, L328-IRS has a small mass accretion rate of $\sim 3.6\times10^{-7}$ {\msun} yr$^{-1}$ with an envelope mass of $\sim 0.09$ {\msun} and thus is suggested to be a proto-brown-dwarf candidate \citep{Lee:2013gx}. The results from these studies indicated that VeLLOs are possibly the progenitors of either young low mass protostars or proto-brown-dwarfs. Therefore the study of the VeLLOs may provide vital information not only on the observed low luminosity of the VeLLOs, but also on our understanding of low-mass stars or brown dwarf formation \citep{Andre:1999tr,Bourke:2005kg,Dunham:2010fk,Schwarz:2012hm,Kauffmann:2011kz,Lee:2010fc,Lee:2013gx,Takahashi:2013fm,Palau:2014kp,Hsieh:2016if}.

Although the VeLLOs are thought to be essential targets to study the formation of low-mass stars or substellar objects, a census of outflows in the VeLLOs, an important parameter to understand them, has not been carried out so far. For example, \cite{Schwarz:2012hm} carried out a search for CO outflows in low luminosity objects which include the VeLLOs taken from the catalog compiled by \citet{Dunham:2008ks}, and obtained full CO maps for the two VeLLOs, namely, L673-7-IRS and L1251-IRS4. Therefore, to characterize the nature of the VeLLOs, a further systematic survey of outflows in a larger number of VeLLOs is earnestly required. Using photometric data from the \textit{Spitzer} and  the \textit{Herschel} archives and spectroscopic radio observations using the \textit{KVN} and the \textit{Mopra} telescopes, \citet{Kim:2016fh} carried out a search for VeLLOs in all the nearby molecular clouds in the Gould Belt region and made a comprehensive list of VeLLOs. This paper presents the results of a systematic search that we carried out for molecular outflows toward the VeLLOs chosen from the list made by  \citet{Kim:2016fh} in $^{12}$CO, $^{13}$CO, and/or C$^{18}$O 2-1 and/or 3-2 lines using the single dish telescopes. Our primary goals of this work are to find outflow activities in the dense envelopes of the VeLLOs and to estimate the mass accretion rates and other related physical quantities of the outflows. Characterizing the properties of outflows in the VeLLOs would help us to understand the processes involved in the early stages of formation of low-mass stars and substellar objects. 

The paper is organized into five sections. Section \ref{sec:obs} describes the methodology used in the sample selection and in observations. Section \ref{sec:res} explains how we identify the outflows and their physical properties. In Section \ref{sec:dis}, we discuss the implications of the observed features on the VeLLOs in the formation of the stars and substellar objects. Finally, we summarize the results of this study in Section \ref{sec:con}.
 
\section{Observations}\label{sec:obs}
The survey was carried out on a large number of VeLLOs using various telescopes. This section explains how we selected the targets and how the survey was executed on different telescopes. 
 
\subsection{Targets, Observing Telescopes and Lines, {\&} Survey Strategy}\label{sec:intro:target}

We selected our targets from the catalog of VeLLOs compiled by \citet{Kim:2016fh}. This catalog contains the largest number of VeLLOs (95 sources) identified so far. Of the 95, we observed a total of 68 VeLLOs. We could not observe all the targets from the catalogue due to the limited observing time available on various telescopes during the period of our study. In this regard our survey is not complete. Of the 68 VeLLOs, 59 are new and the remaining nine are already well studied ones (e.g., IRAM 04191$+$1522 and L328-IRS). The selected VeLLOs are faint ({\lint}$\leq$0.2 {\lsun}), embedded in low-mass envelopes ($M_{\rm env}\leq$0.7 {\msun}) and are distributed over the Gould belt clouds (d$\leq$450 pc). Accounting for a probable uncertainty of a factor of 2 in the estimation of {\lint} \citep[e.g., ][]{Dunham:2008ks}, we relaxed the criterion of {\lint}$\leq0.1$ {\lsun} and included the sources having an internal luminosity up to 0.2 {\lsun}. The basic information of the targets, adopted from \citet{Kim:2016fh}, is listed in Table \ref{tbl:obj}. We used $^{12}$CO lines for the detection of molecular outflows as $^{12}$CO molecules are believed to be one of the best tracers to identify the existence of extended outflows in molecular clouds in single dish surveys, mostly due to their strong brightness \citep[e.g., ][]{Bontemps:1996vb, Narayanan:2012ej}. The signature of the CO outflow is usually seen in a form of wide wings in $^{12}$CO lines. However, it is not always simple to disentangle such outflow wing components in the CO profiles unless we know the exact envelope components that are usually seen in a Gaussian form. In this regard we used the rare isotopologue $^{13}$CO and C$^{18}$O lines in our survey. Because these lines are optically thinner and thus usually seen in a Gaussian form, they would provide assistance to identify the outflow wings in $^{12}$CO lines.

We executed our survey observations in two phases. In the first, we made single pointing observations of all our target sources in the CO isotopologue lines and looked for an outflow signature in the line wings. In the second phase, the sources that showed an outflow signature in the line wings were mapped fully in the $^{12}$CO lines. The survey was performed with four single dish telescopes, the Seoul Radio Astronomy Observatory (SRAO), the Caltech Submillimeter Observatory (CSO), the James Clerk Maxwell Telescope (JCMT), and the Atacama Submillimeter Telescope Experiment (ASTE). Table \ref{tbl:gen} summarizes all the necessary information for the survey, the observing lines used for the survey, observing bandwidth, and spectral (velocity) resolution, HPBWs and main beam efficiencies of the observing telescopes, and the observing periods. The procedures followed for the observations at different observatories were somewhat different and are described in detail in the subsections below.
 
\subsection{\textit{SRAO} Observation}\label{sec:intro:srao}
The \textit{SRAO} is a six-meter single dish telescope located at the Seoul National University in South Korea. A SIS dual polarization receiver and a Fast Fourier Transform spectrometer were used \citep{Lee:2013gs}. A standard chopper wheel calibration was used to estimate the antenna temperature of a source corrected for atmospheric extinction. Pointing accuracy was checked every two hours using SiO sources close to our targets and found to be better than $\sim10${\arcsec}. The DR21 and ORION KL were used as standard sources to check the status of observing systems and source brightness every three hours.

To subtract sky background emissions, we used a frequency-switching mode. However, a telluric line was sometimes detected close to the systemic velocity of the source. Because the observed lines could be confused with the telluric line, in those cases, we used a position-switching mode which enabled us to subtract the contribution from the telluric line completely. We used the CO survey map of the Galactic plane \citep{Dame:2001bg} to identify a CO-emission-free position within an angular distance of five degrees from the target to execute the position-switching mode. 

We started our survey with single pointing observations of 14 VeLLOs in $^{12}$CO 2-1, $^{13}$CO 2-1, and C$^{18}$O 2-1 molecular lines. We found broad wing feature in the $^{12}$CO line in 12 out of 14 sources. Consequently, we made full mapping observations for them in the same line. CLASS program\footnote{See \burl{https://www.iram.fr/IRAMFR/GILDAS/}} was used for the basic data reduction such as the subtraction of baselines, averaging of the spectra, and resampling to achieve a velocity resolution of $\sim$ 0.1 km s$^{-1}$.
 
\subsection{\textit{CSO} Observation}\label{sec:cso}
The \textit{CSO} is a 10.4 meter single dish telescope near the summit of Mauna Kea, Hawaii. An heterodyne receiver SIDECAB and a Fast Fourier Transform Spectrometer FFT1 were used to enable simultaneous observations in $^{12}$CO 2-1, $^{13}$CO 2-1, and C$^{18}$O 2-1 lines \citep{Kooi:2004bz}. The telescope pointing was checked with five pointing observations around a compact CO source close to each target every two hours to achieve an accuracy of about 3{\arcsec}. 

We began our survey by making single pointing observations of 14 VeLLOs in $^{12}$CO 2-1, $^{13}$CO 2-1, and C$^{18}$O 2-1 lines. Four of the 14 VeLLOs displayed a broad wing feature in their $^{12}$CO 2-1 spectrum which were subsequently mapped in the $^{12}$CO 2-1 line. We subtracted the sky background emission using the same method which we adopted for the SRAO observations and used the CLASS for the basic data reduction.
 
\subsection{\textit{ASTE} Observation}\label{sec:aste}
The \textit{ASTE} is a 10 meter single dish telescope on the Atacama desert in Chile. Two heterodyne receivers CATS345 and DASH345, and a XF-type digital spectro-correlator MAC were used \citep{Inoue:2008ts}. The telescope pointing was checked using the same method which we employed for the CSO observation and then resulted in a pointing accuracy of about 2{\arcsec}. 

We started our survey with single pointing observations of 15 VeLLOs in $^{12}$CO 3-2 molecular line with a position-switching mode and additional simultaneous observations in $^{13}$CO 3-2 and C$^{18}$O 3-2 lines. We found seven out of 15 sources manifesting a broad wing feature. We made full mapping observations of these in the $^{12}$CO 3-2 line. Initially the data were in a NEWSTAR\footnote{See \burl{http://alma.mtk.nao.ac.jp/aste/research_e.html}} program format and were converted into CLASS data format for further reduction of the data.
 
\subsection{\textit{JCMT} Observation}\label{sec:jcmt}
The \textit{JCMT} is a 15-meter single-dish telescope near the summit of Mauna Kea, Hawaii. A heterodyne 4$\times$4 array receiver HARP with a spacing grid of 30{\arcsec} and an ACSIS digital autocorrelation spectrometer were used \citep{Buckle:2009cn}. The telescope pointing was checked every hour with the same procedure used for the CSO observations and found to be better than about 2{\arcsec}. 

We made single pointing observations of 31 VeLLOs with the HARP in $^{12}$CO 3-2 line with a position-switching mode and additional pointing observations in $^{13}$CO 3-2 and C$^{18}$O 3-2 lines. The observations covered a field of view of 90{\arcsec}$ \times $90{\arcsec} which allowed us to obtain partially mapped data for each object. Unfortunately, because of a breakdown of three pixels in the HARP, a single pointing observation produced the data for 13 pixels instead of 16 pixels. From these observations, we found two targets showing broad wing features in the spectra and we made full mapping observations for these sources in raster-scan mode with the CO 3-2 line. All data were reduced by the standard pipeline ``oracdr$\_$acsis'' in the version 2015B of the ``starlink'' package \citep{Jenness:2015kh}\footnote{See \burl{http://starlink.eao.hawaii.edu/starlink/2015B/}} and then converted into CLASS format for further data analysis.
 
\section{RESULTS}\label{sec:res}
We observed a total of 68 targets in the single pointing mode and all the sources were detected in $^{12}$CO 2-1 and/or 3-2. Of the 54 targets observed in 2-1 and/or 3-2 lines of $^{13}$CO and C$^{18}$O, again in the single pointing mode, all of them were detected in $^{13}$CO and 52 sources were detected in C$^{18}$O line. In 52 sources some hints of outflow wing features were seen. These sources were mapped to cover the full extent of them in $^{12}$CO 2-1 and/or 3-2 lines and their acquired spectra were used for further analysis to obtain outflow properties. More detailed information on the statistics of the observed targets is given in Table \ref{tbl:obs}.

\subsection{Characteristic Shapes of CO profiles {\&} Identification of Outflow}\label{sec:det}

In this section we present the CO isotopologue line profiles toward the VeLLOs obtained from the single pointing observations and the results of $^{12}$CO mapping observations, and describe how the presence of an outflow was identified from these.

Figure \ref{fig:Lineprofile1} shows spectra of $^{12}$CO, $^{13}$CO, and/or C$^{18}$O 3-2 and/or 2-1 toward all the 68 VeLLOs obtained from single-pointing observations. The $^{12}$CO spectra are found to show very complex features such as a wide line width and multiple components which may be due either to the self-absorption by the high optical depth of the $^{12}$CO line profile or to other velocity components in the line of sight. The real reason for the origin of the multiple components can be investigated by comparing the $^{12}$CO spectra with the optically thinner $^{13}$CO and C$^{18}$O lines and examining whether the peak of the thinner lines is located at the peak positions of $^{12}$CO or between them. The $^{13}$CO line widths of our target sources are found to be much narrower than the $^{12}$CO line. The $^{13}$CO line profiles often show a single Gaussian profile located between the double peaked $^{12}$CO profiles, although there are some cases where the $^{13}$CO line also shows a double-peaked profile similar to that of the $^{12}$CO line. In contrast, in all the cases except two, the C$^{18}$O line profiles show optically thin feature, i.e., a single Gaussian form in their spectra. Two exceptional sources, J032832 and J160115, are found to display double-peaked or saturated features in the spectra.

The majority of the $^{12}$CO spectra are found to display the self-absorbed dip profiles. Interestingly, 27 of the sources show blue asymmetry profiles in their $^{12}$CO spectra which are indicative of inward motions in their envelopes. The blue asymmetry is characterized by having a blue-shifted peak higher than a red-shifted one in an optically thick line and a single Gaussian-like profile in an optically thin line peaked around the self-absorbed dip of the optically thick line \citep[e.g., ][]{Lee:1999de, Lee:2001jm}. In contrast, eight VeLLOs display red asymmetry profiles in $^{12}$CO spectra, which are the same as the blue asymmetry but the red peak is higher than the blue one, indicative of outward motions. We marked `B' for blue asymmetry and `R' for the red asymmetry in the column 12 of the Table \ref{tbl:obs}. The seven sources which show multiple velocity components in CO isotopologue spectra, we marked them as `mc' in the table.

In addition to the self-absorbed feature of the $^{12}$CO spectra, finding wide wing feature in the spectra is also interesting as this may be indicative of the outflow motions in our targets. The broad wing components in $^{12}$CO lines were identified visually by comparing them with their rare isotopologue line profiles. For example, a profile is said to have `outflow wing' if the $^{12}$CO line shows wide wing feature in comparison with the spectral shape of the $^{13}$CO and/or C$^{18}$O that is more or less Gaussian. The target sources showing the wing feature in the $^{12}$CO spectra are identified with a `Y' in the column 13 of the Table \ref{tbl:obs}. Due to the limited observing time available to us, for 14 targets we could carry out only $^{12}$CO 3-2 line observations (as shown in Figure \ref{fig:Lineprofile1} which are listed in the Table \ref{tbl:obs}). As a result,  we could neither probe the presence of outflow or infall asymmetry nor derive physical quantities (as done in Section \ref{sec:pro} for other sources) for them except for one source (J162145) which was mapped in $^{12}$CO 3-2 line only.
For the reason that our identification of outflow features in sources with single pointing observations only could be somewhat subjective, we decided to make mapping observations for 52 sources. Although four sources do not show any signatures of outflow, their mapping data were also acquired by JCMT observations with the multi-beam receiver in single pointing mode. Table \ref{tbl:obs} lists the summary of our observations for the single pointing and mapping observations. The information such as the telescopes used for observations, transition of CO isotopologues, the peak intensity and observed rms noise of $^{12}$CO, $^{13}$CO, and C$^{18}$O in main beam temperature scale, the line width, the systemic velocity, the description of characteristic features found in CO profiles, and the mapped area towards each source are also given. 

The velocity channel maps of our sources are useful to identify the outflow sources in a more precise manner and also constrain their physical quantities better by looking for blue or red lobe patterns of the outflows. To illustrate, in Figure \ref{fig:chmap} we present the velocity channel map of J154216 which shows the blue and red lobes of its outflow over different velocity channels. From such bipolarity of the outflows in channel maps, the approximate velocity range of the outflows can be determined as indicated by vertical dashed lines in Figure \ref{fig:Lineprofile1}. These values are used for the estimation of the physical quantities of the outflows discussed in the next section. The maximum velocities of the blue and red-shifted components of the outflow can be easily determined by looking for the velocities where the wing emission reaches to its rms noise level. On the other hand, their minimum velocities are chosen to be the values where the outflow lobe pattern in the velocity channel map begins to disappear. 

Figure \ref{fig:Contourmap1} displays 16 sources with an outflow pattern in the blue or red component. Note that not all of the 16 sources show clear bipolar outflows. For example, one VeLLO, J182943, shows only a red component of outflow while three targets show the blue and red outflow components overlapping on each other. Column 17 in Table \ref{tbl:obs} describes the results of our mapping observations and gives a specific description on the identified outflow features. We have several other sources that do not show such clear outflow pattern in the mapping data but show a wide wing feature in a single dish spectrum. These are marked as `NY' in the column 17 to indicate that the presence of their outflow is not yet clearly identified. One possible reason could be that the size of the outflow is too small compared to the beam sizes of the telescopes we used \citep[e.g., L1148-IRS,][]{Kauffmann:2011kz}. We found that the line profiles of a significant number of the target sources are affected by strong outflows from the neighbouring YSOs, and thus the identification of the outflow components in those targets becomes highly uncertain \citep[e.g., IC348-SMM2E,][]{Palau:2014kp}. To check whether known YSOs are present in the vicinity of our target sources, we made a search around them in the SIMBAD astronomical database\footnote{\burl{http://simbad.u-strasbg.fr/simbad/}} with a search radius of 100 arcsec. We found that 29 of our targets are located close to the known YSOs having outflow activities which could contaminate the velocity components of our target sources. We identified them with a `C' in the column 17 of Table \ref{tbl:obs}. A total of 16 sources were found to show multiple velocity components or do not show any wing features. We identified them as `mc' or `N' in the column 17 of Table \ref{tbl:obs}, respectively. Note that a small-scale bipolar outflow was identified by interferometric observations in three cases (L1014-IRS, L1148-IRS, IC348-SMM2E) but the outflow is not apparent in our single dish observations due to the coarse spatial resolution of our observations. In these cases, we used the properties of outflows derived from the previous studies for our further discussion. Presence of any outflow activities in sources labeled as `C', `NY', or `mc' cannot be completely ruled out and hence needs to be confirmed using interferometers with higher sensitivity and angular resolution like ALMA or SMA. From our survey for outflow in 68 VeLLOs, we found that the outflow detection rate is approximately 24{\%}. This detection rate seems to be higher than those from the previous survey for outflows in VeLLOs \citep[7-13{\%}, ][]{Schwarz:2012hm} or unbiased survey for outflows in protostars \citep[$\sim$18{\%}, ][]{Hatchell:2007cz, Curtis:2010ia}, while lower than those of biased survey for outflow in protostars \citep[$\sim$73{\%}, ][]{Bontemps:1996vb}. The outflow properties of 19 sources (including L1014-IRS, L1148-IRS, and IC348-SMM2E) will be discussed further in the next section.

\subsection{Outflow Properties}\label{sec:pro}
In this section we discuss the procedures used to derive the physical quantities of the outflows detected toward the VeLLOs, especially the outflow mass, the outflow force, the mass accretion rate, and the accretion luminosity which are the essential parameters required to understand the nature of the VeLLOs.
 
\subsubsection{Column Density {\&} Mass of Outflow}\label{sec:col}
To estimate the outflow mass, we require the values of $H_2$ column density of the outflows. The $H_2$ column density can be derived for each velocity channel within a given velocity range at each spatial position. We assume that the outflow wing part in $^{12}$CO line profile is optically thin under the local thermal equilibrium. Then the column density can be calculated by the following equation;
\begin{equation}\label{eq:col}
N (H_2)_{\rm i,j} = {\rm f} (J_{\rm low},~X_{\rm co},~T_{\rm ex}) T_{\rm i,j} ~ \delta v~{\rm cm^{-2}},
\end{equation} 
where ${\rm f} (J_{\rm low},~X_{\rm co},~T_{\rm ex})$ is a function of the quantum number of the low rotational transition, $J_{\rm low}$, the $^{12}$CO abundance relative to $H_2$, $X_{\rm co}$, and the excitation temperature of the outflow materials, $T_{\rm ex}$ \citep[See Appendix C, ][]{Dunham:2014gh}. The $T_{\rm i,j}$ is the main beam temperature of each velocity channel i at each spatial position j, and $\delta v$ is the velocity resolution.

We adopted the value of $X_{\rm CO}$ as $10^{-4}$ \citep[e.g.,][]{Frerking:1982iq}. The excitation temperature $T_{\rm ex}$, in the case of two transitional observations with $^{12}$CO 2-1 and 3-2 lines, can be derived by the following equation \citep[e.g.,][]{Choi:1993hq};
\begin{equation}\label{eq:tex}
\frac{T^{32}_{\rm i}}{T^{21}_{\rm i}}= {\rm \phi} \frac{[J_{32}(T_{\rm ex,i})-J_{32}(T_{\rm bg})]}{[J_{21}(T_{\rm ex,i})-J_{21}(T_{\rm bg})]}\frac{3}{2}\frac{1 - {\rm exp}(-{\rm h}\nu_{32}/{\rm k}T_{\rm ex,i})}{{\rm exp}({\rm h}\nu_{21}/{\rm k} T_{\rm ex,i})-1},
\end{equation}
where $T^{32}_{\rm i}/T^{21}_{\rm i}$ is the ratio of main beam temperatures in $^{12}$CO 3-2 and $^{12}$CO 2-1 lines in each velocity channel, $\phi$ is the ratio of beam-filling factors in two CO lines, $J(T)$ is the Planck function of temperature, $T_{\rm ex,i}$ is the excitation temperature in each velocity channel i, $T_{\rm bg}$ is the temperature of the cosmic microwave background radiation, h is the Planck constant, k is the Boltzmann constant, $\nu_{32}$ and $\nu_{21}$ are the frequencies of two CO lines. As for $T^{32}_{\rm i}/T^{21}_{\rm i}$, we used both $^{12}$CO 2-1 and $^{12}$CO 3-2 spectra of three VeLLOs (J162648, IRAS16253-2429, and CB130-3-IRS) observed in both the rotational transitions of $^{12}$CO line. As listed in Table \ref{tbl:gen}, because both $^{12}$CO 2-1 and $^{12}$CO 3-2 spectra were obtained in different beam sizes and different velocity domain, it is essential to bring them both to same spatial and velocity domains. Thus, using the CLASS program, first we convolved 3-2 data (in $\sim$22{\arcsec} beam size) into the larger beam size ($\sim$48{\arcsec}) of 2-1 data and then re-sampled the two data to bring them to a same velocity domain. $\rm \phi$ is a numerical factor which is on the order of unity because the respective transitional data are convolved into a single beam size and thus the different beam filling factors of two telescopes are expected to be almost identical. Taking these into account, $T_{\rm ex}$ was numerically determined from the equation \ref{eq:tex} with $T^{32}_{\rm i}/T^{21}_{\rm i}$.

Figure \ref{fig:tex} shows the ratios of two main beam temperatures in the top panel and the excitation temperatures in the bottom panel as a function of velocity with respect to {\vsys}. Over the velocity range of $|v-${\vsys}$| \leq$ 4 km s$^{-1}$, the excitation temperatures are estimated to be 9.3-16.1 K with their average value of 12.2 K. Therefore we calculate the column densities with the median excitation temperature of 12.2 K by assuming that the excitation temperature in outflow material would be the same for all the VeLLOs. The equation (\ref{eq:col}) then can be rewritten as;
\begin{eqnarray} 
N (H_2)_{\rm i,j} = 5.5\times10^{18} ~ T^{21}_{\rm i,j}~ \delta v ~ {\rm cm^{-2}} ~{\rm for ~ J=2-1}\label{eq:col21},\\
N (H_2)_{\rm i,j} = 9.6\times10^{18} ~ T^{32}_{\rm i,j}~ \delta v ~ {\rm cm^{-2}} ~{\rm for ~ J=3-2}\label{eq:col32}.
\end{eqnarray}

It should be noted that the optically thin assumption considered for the $^{12}$CO lines in the above calculations may not be valid for some part of the outflow gas which has fairly high column density and result in an underestimation of the column density. In such a scenario it is important to correct for the underestimation of the column density. Below we explain the methodology we followed to infer the optical depths in outflow wings and how we corrected it in the estimation of the column density.

Assuming that the beam filling factors of $^{12}$CO and $^{13}$CO lines are the same, both the lines are in local thermal equilibrium with a single excitation temperature, and $^{13}$CO line is optically thin, the optical depth ($\tau_{12}$) of $^{12}$CO line can be estimated using the following equation \citep[e.g.,][]{Goldsmith:1984dq, Choi:1993hq, Dunham:2014gh};
\begin{equation}\label{eq:tau}
\frac{T^{12}_{\rm i}}{T^{13}_{\rm i}} = \frac{1-{\rm exp}(-\tau_{\rm 12,i})}{1-{\rm exp}(-\tau_{\rm 12,i}/{\rm X})},
\end{equation}
where $T^{12}_{\rm i}$ and $T^{13}_{\rm i}$ are main beam temperatures of $^{12}$CO and $^{13}$CO in each velocity channel i, respectively, X is the abundance ratio between $^{12}$CO and $^{13}$CO, $[\rm ^{12}CO]/[\rm ^{13}CO] = 62$ \citep{Langer:1993cv}. 

The optical depths at the velocity channels in $^{12}$CO 2-1 line were derived by the ratios of main beam temperatures from using average spectra of $^{12}$CO 2-1 and $^{13}$CO 2-1 at the central positions of 14 VeLLOs, as shown in the left panels of Figure \ref{fig:tau}. The optical depths at the velocity channels in $^{12}$CO 3-2 line were also derived with the same method for $^{12}$CO 2-1 using the average spectra of $^{12}$CO 3-2 and $^{13}$CO 3-2 at the central positions of 12 VeLLOs, as shown in the right panels of Figure \ref{fig:tau}. We found that these optical depths can be explicitly derived at the velocity channel where both $^{12}$CO and $^{13}$CO lines are detected. But it is difficult to derive the optical depths at high velocity wings where $^{13}$CO lines are not detected. Instead we can derive only the upper limits of the optical depths at such high velocity wings by giving 3$\sigma$ value of $^{13}$CO line emission shown in Figure \ref{fig:tau}. However, such an approach would give highly uncertain values of optical depths. Thus we used the least squares fit method in which the fit results obtained from the temperature ratios at the low velocity regime are applied  to derive mean optical depths of $^{12}$CO lines at the high velocity regime where $^{13}$CO emission is not detected \citep[e.g., ][]{Arce:2001eo, Dunham:2014gh}. We found second order polynomial functions which gave the least squares fit to the temperature ratios at velocity channels. These fits are shown using the green curves in the upper panels of the Figure \ref{fig:tau}. The fitted polynomial functions have the following forms; 
\begin{eqnarray} 
\frac{T^{12}_{\rm i}}{T^{13}_{\rm i}}= 4.7((v_{\rm i}-v_{\rm sys})-0.1)^2+1.7 ~{\rm for ~ J=2-1} \label{eq:tau21},\\
\frac{T^{12}_{\rm i}}{T^{13}_{\rm i}}= 3.9((v_{\rm i}-v_{\rm sys})+0.1)^2+1.2 ~{\rm for ~ J=3-2}. \label{eq:tau32}
\end{eqnarray} 
For the above fits, we considered the temperature ratios within $\pm$1 km s$^{-1}$ for J=2-1 and $\pm$0.8 km s$^{-1}$ for J=3-2 with respect to the {\vsys} as the $^{12}$CO and $^{13}$CO spectra are bright enough in these velocity ranges and thus the estimated ratios are reliable. The polynomial functions thus obtained are used to get reasonable values of the ratios and hence the optical depths at high velocity range (as shown with green lines in low panels of Figure \ref{fig:tau}). From this work the inferred optical depths of $^{12}$CO lines are found to be optically thin ($<0.1$) at high velocity ($\rm 3.5 - 3.9~km~s^{-1}$) wings while the optical depths are found to be high ($\geq$10) at low velocity ($\rm \leq1.0~km~s^{-1}$) wings. This shows that the correction for the high optical effects on the column density of outflow gas at the low-velocity regime is required.

To correct the opacity in calculation of the column density, we numerically calculated $\tau_{\rm 12,i}$ at each velocity channel in all the $^{12}$CO spectra using the equations (\ref{eq:tau21}) and (\ref{eq:tau32}) and then multiplied the velocity-dependent correction factor $\tau_{\rm 12,i}/(1-{\rm exp}(-\tau_{\rm 12,i}))$ to the equations (\ref{eq:col21}) and (\ref{eq:col32}), respectively. This correction has resulted in an increase in the column density by a factor of about 3.5. 

If the column density $N (H_2)_{\rm i,j}$ is calculated by the equations (\ref{eq:col21}) and (\ref{eq:col32}) as described above, then the outflow mass within each velocity channel i for each spatial position j can be estimated by the following equation;
\begin{equation}\label{eq:mass}
M_{\rm i,j} = \mu_{\rm H_2} m_{\rm H} N (H_2)_{\rm i,j} {\rm A},
\end{equation} 
where $\mu_{\rm H_2}$ is the mean molecular weight per hydrogen molecule \citep[2.8 for gas composed of 71\% hydrogen, 27\% helium, and 2\% metals, ][]{Cox:2000ua}, $m_{\rm H}$ is the mass of a hydrogen atom, and A is the area of a beam. The total outflow mass (${\rm M=\sum}M_{\rm i,j}$) is calculated by summing up the derived masses at each velocity channel i over given outflow velocity ranges at each spatial position j over the outflow region. The errors in the integrated intensity, abundance of CO relative to hydrogen molecule, excitation temperature, optical depth, lower and upper limits of the outflow velocities adopted and the pointing accuracy are the factors responsible for making the estimated values of the outflow mass uncertain. Of these, the major factor contributing to the uncertainty is the CO abundance which is uncertain by a factor of $\sim 3$ \citep[e.g.,][]{Frerking:1982iq} while the other parameters are uncertain by 10-20{\%}.  
 
\subsubsection{Outflow Force, Mass Accretion Rate, {\&} Accretion Luminosity}\label{sec:oth}

The key physical properties of the outflow such as the outflow force, mass accretion rate, and accretion luminosity are calculated using the values of the outflow mass obtained above. To begin with, we calculated the outflow force at each spatial position j using the following equation;
\begin{eqnarray} 
F_{\rm j} = \frac{M_{\rm i,j}|v_{\rm i}-v_{\rm sys}|}{t_{\rm dyn_{\rm j}}} \frac{{\rm sin}~i}{{\rm cos}^{2}~i} \label{eq:for}, 
\end{eqnarray} 
where $|v_{\rm i}-v_{\rm sys}|$ is an individual velocity channel with respect to {\vsys}, $t_{\rm dyn_{\rm j}}$ is a dynamical time at each spatial position j, and $i$ is the inclination of the outflow axis. The $t_{\rm dyn_{\rm j}}$ can be estimated by dividing the distance of each spatial position from the given position of a source ($R_{\rm j}$) by a characteristic outflow velocity $v_{\rm char_{\rm j}}$ which is given by an intensity-weighted velocity with $\int T_{\rm i,j} v_{\rm i} dv/ \int T_{\rm i,j} dv$. The total outflow force (${\rm F=\sum}F_{\rm j}$) is derived by summing up $F_{\rm j}$ over spatial positions encompassing the outflow. 

From the outflow force F, the mass accretion rate $\dot{M}_{\rm acc}$, and the accretion luminosity $L_{\rm acc}$ are estimated using the following equations \citep{Bontemps:1996vb,Dunham:2010fk, Lee:2013gx};
\begin{eqnarray} 
\dot{M}_{\rm acc} = \frac{1}{f_{\rm ent}} \frac{\dot{M}_{\rm acc}}{\dot{M}_{\rm w}}\frac{1}{v_{\rm w}}F \frac{{\rm sin}~i}{{\rm cos^{2}}~i} \label{eq:dmacc} ,\\ 
L_{\rm acc} = \alpha \frac{G M_{\rm acc} \dot M_{\rm acc}}{\rm R} \frac{{\rm sin}~i}{{\rm cos^{3}}~i} \label{eq:lacc},
\end{eqnarray} 
where $f_{\rm ent}$ is the entrainment efficiency which is a quantity indicating the fraction of the momentum flux of jet/wind from the central source transferred into the entrained ambient gas), $\dot{M}_{\rm acc}/\dot{M}_{\rm w}$ is the ratio between the protostellar mass accretion rate and the mass-loss rate in the jet/wind, $v_{\rm w}$ is the velocity of the jet/wind, G is the gravitational constant, R is the protostellar radius, $\alpha$ is the energy transfer rate of the accretion materials onto the central object from the envelope, and $i$ is the inclination of the outflow axis.

The $f_{\rm ent}$ is suggested to be typically 0.1-0.25 \citep{Andre:1999tr, Dunham:2010fk}. Here we adopted $f_{\rm ent}=0.25$ which would result in deriving the minimum value of the mass accretion rate. It is known that the $v_{\rm w}$ typically ranges between 100-300 km s$^{-1}$ \citep[e.g.,][]{Andre:1999tr, Schwarz:2012hm} for Class 0 to T Tauri stars \citep[e.g, ][]{Bachiller:1992kq, Mundt:1987fi} and $\dot{M}_{\rm acc}/\dot{M}_{\rm w}$ is between 2-10 from models of jet/wind formation \citep[e.g.,][]{Pelletier:1992fk,Shu:1994gr}. In this study, we chose the values of $v_{\rm w}$ and $\dot{M}_{\rm acc}/\dot{M}_{\rm w}$ as 150 km s$^{-1}$ and 10, respectively. Jet/wind models \citep[e.g., ][]{Najita:1994bs, Ferreira:1995ua} show that the values for $v_{\rm w}$ and $\dot{M}_{\rm acc}/\dot{M}_{\rm w}$ are closely related with each other and their multiplied value varies within a factor of 3 among those models \citep{Bontemps:1996vb}. Therefore our choice for $v_{\rm w}$ and $\dot{M}_{\rm acc}/\dot{M}_{\rm w}$ is expected to make the derived mass accretion rate to be uncertain by a factor of $\sim 3$. The $\alpha$ is assigned to be 0.5 by assuming that most of the material accretes through the disk \citep{Baraffe:2009jw}. The R can be assigned a value in the range 3-9 {\rsun}, but here we used the typical protostellar radius of 3 {\rsun} under the assumption that the protostellar radius should be much smaller than the disk radius because the accretion occurs from disk to the central object \citep[e.g., ][]{Dunham:2010fk}.

$M_{\rm acc}$ is the accreted mass which can be calculated by multiplying the mass accretion rate with the total dynamical timescale of the outflow. The total dynamical time $t_{\rm tot}$ can be inferred from the maximum of the dynamical time ($t_{\rm dyn_{\rm j}}$) corresponding to the distance of the outflow edge from the source, i.e., the size of outflow. Figure \ref{fig:Contourmap1} shows the outflow shapes from our targets, indicating that the majority of the VeLLOs (12 VeLLOs) were observed to cover most parts of the outflow features while the outflows of four other VeLLOs were partially observed. Therefore the total dynamical time scales for the former sources should be reasonably determined and found to be between $(0.5 - 16.5)\times10^{4}$ cos $i$/sin $i$ years, while the latter four sources whose outflows were not fully mapped may give a low limit to the total dynamical time.

Apart from the values of the outflow mass, all other quantities calculated by us for the outflow can be strongly affected by the inclination angle of the outflow direction. In fact, accurate determination of the inclination is always difficult. However, we found three sources (J041412, J162145, and J182920) for which the blue and red lobes of outflow are mostly overlapping, implying that their outflows are almost viewed face-on. We assume that the outflows of these sources have the inclination of $\sim$10{\arcdeg}. For three VeLLOs we adopted the values of the inclination angles that are already available in the literature from the previous studies. For example, inclination for the outflow of L328-IRS has been fairly well determined as $\sim66${\arcdeg} based on its ALMA observations \citep{Lee:2018iq} and inclinations for the outflows of IRAM 04191$+$1522 and L1521F-IRS are also known to be 50{\arcdeg} and 30{\arcdeg} from \citet{Andre:1999tr} and \citet{Takahashi:2013fm}, respectively. For the remaining nine sources, we adopted 57{\arcdeg}.3 as a mean inclination angle for random outflow orientations \citep{Bontemps:1996vb}. The inclination angles used in our calculations for all the sources are listed in the column 2 of Table \ref{tbl:outflow4}.

Taking these parameters into consideration, we estimated the mass accretion rate and the accretion luminosity. In Table \ref{tbl:outflow4} we list the important physical quantities of the outflows we estimated for a total of 16 VeLLOs. For three VeLLOs, namely, L1014-IRS, L1148-IRS and IC348-SMM2E, the values given in Table \ref{tbl:outflow4} are obtained from the literature. Further discussion on the outflow properties of the VeLLOs will be based on the quantities of these 19 VeLLOs. 

The mass accretion rates are found to span from $1.1\times10^{-8}$ {\msun} yr$^{-1}$ to $1.1\times10^{-5}$ {\msun} yr$^{-1}$ with a mode value of $8.8\times10^{-7}$ {\msun} yr$^{-1}$. The accretion luminosities are found to scatter between $6.4\times10^{-6}$ {\lsun} and 65 {\lsun} with a mode value of 0.6 {\lsun}. 

It should noted that for the sources where the entire area of the outflows are not fully covered in the mapping observations, the values of the outflow masses, the accreted masses, and the accretion luminosities may be lower limits. However, the forces and the mass accretion rates can still be reliable because they are the momentum-driven dynamical quantities \citep{Bontemps:1996vb} and are predominantly high towards the central region of outflow source \citep{Lee:2018iq}. 

The mass accretion rates of the sources obtained by us are prone to be uncertain mainly because of the uncertainties in the values of $f_{\rm ent}$, $\dot{M}_{\rm acc}/\dot{M}_{\rm w}$, $v_{\rm w}$, and the inclination angle of the outflows. For the reason that all these values are highly model-dependent and hence are not well constrained, the estimation of a realistic uncertainty in the mass accretion rate is not straightforward. Therefore we estimated a possible range for the uncertainty in the  mass accretion rate (similarly for the accretion luminosity) by considering a range in the inclination angles of 10-80 degree with respect to 57{\arcdeg}.3 \citep[e.g., ][]{Schwarz:2012hm,Dunham:2014gh}. This makes the mass accretion rate and the accretion luminosity uncertain by about a factor of 0.006-11.3 and 0.03-35 respectively. By taking these into consideration we proceed further to discuss our results.

\section{Discussions}\label{sec:dis}
\subsection{Identification of Candidate Proto-Brown Dwarfs}\label{sec:out1}

The formation mechanism of Brown Dwarfs (BDs) is still poorly understood mainly because of a very limited number of known precursors of BDs identified so far. As a result, finding additional number of the bona-fide proto-BDs would be extremely important to understand their formation mechanism. The basic idea to select a proto-BD candidate is to look for a source whose mass and the envelope mass are low enough so that at the end of its protostellar evolutionary period, the final mass of the central source remains substellar. 

The actual ``central mass'' of our sources cannot be directly obtained from this survey. Instead we estimated accreted mass of the sources and used them as an alternate estimate of the central mass of the sources. The envelope mass determines how much more material is available for the central object to accrete and buildup its mass in the future.

Based on the information of the accreted mass and the envelope mass of the VeLLOs and assuming a $\sim 20\%$ of the core-to-star efficiency (i.e. the fraction of the core mass that is converted into stars) \citep[e.g.,][]{Evans:2009bk}, we classified the VeLLOs studied here as proto-BD candidates if their mass satisfied the BD condition, $ M_{\rm acc}+0.2 \times M_{\rm env} < 0.08$ {\msun}. Figure \ref{fig:corr_Menv_Macc_dMacc} shows the distribution of the VeLLOs in an accreted mass versus envelope mass plot. The broken line satisfies the BD condition mentioned above. This figure indicates that a majority of the sources from our sample (in total 15 VeLLOs) are falling in the substellar mass regime even after taking into account the possible uncertainties in the envelope and the accreted masses shown using error bars in the plot. The remaining four VeLLOs are believed to be faint protostars which will eventually evolve to become normal low-mass stars. Out of four VeLLOs, two sources (IRAM 04191$+$1522 and L673-7-IRS) are known to be faint protostars from previous studies \citep{Andre:1999tr,Dunham:2010fk} and the other two VeLLOs (J033032 and J210221) are newly identified as likely protostars from this study. Of the 15 VeLLOs classified as proto-BD candidates, six VeLLOs, namely, L1014-IRS, L1148-IRS, L1521F-IRS, L328-IRS, IC348-SMM2E, and IRAS16253-2429 have been identified as possible precursors of BDs in previous studies \citep{Bourke:2005kg,Kauffmann:2011kz,Takahashi:2013fm,Lee:2013gx,Palau:2014kp,Hsieh:2016if}. The remaining nine VeLLOs are newly classified as the proto-BD candidates in this study. Remarks are given on the classification of the sources in the column 9 of Table \ref{tbl:outflow4}.

Note that the estimation of the accreted mass and the envelope mass can be affected by various uncertain parameters which are mostly unknown and hence our classification of the sources can be highly uncertain. The most influential parameter, but relatively easy to deal with, is the inclination of the outflow. In the Figure \ref{fig:corr_Menv_Macc_dMacc}, the error bars show the extent of the variation in the values of the accreted mass can occur if the inclination angles change from 10{\arcdeg} to 80{\arcdeg}. The values of the envelope mass is uncertain by a factor of $\sim 3$ mainly due to the uncertainty in the opacity properties \citep{Ossenkopf:1994tq}. Note that our classification could get affected by additional unknown factors also. For example, the accretion rate is believed to be highly variable, usually higher at the early stages of the formation of a protostar or proto-BD and lower towards the final stages by a few orders of magnitude within the protostellar evolutionary period \citep[e.g.,][]{Machida:2009gr}. In addition to these, ages of the sources which is one of the most difficult parameters to estimate could also affect the calculation of the final mass and their classification. 

A recent study conducted by us on the proto-BD candidate, L328-IRS, using the ALMA data is found to constrain the inclination angle of its outflow very well. Based on a simple analysis of its detailed image in symmetric pattern, the inclination-dependent physical quantities (including the mass accretion rate) were better constrained for this source \citep{Lee:2018iq}. Likewise, further studies of all the proto-BD candidates using the cutting-edge facilities such as ALMA are required to confirm their true nature.

\subsection{Physical Properties of Proto-BD candidates {\&} Their Implication}\label{sec:dis:out1}
\subsubsection{Mass Accretion Rate {\&} Accretion Luminosity}\label{sec:dis:pbd}
The mass accretion rates and the accretion luminosities of protostellar or substellar objects are important to our understanding of how the central object had accreted or is still accreting the material from the envelope. This section discusses how the above two quantities differ between proto-BD and protostellar objects. First of all, there exists a clear difference in the mass accretion rate between the protostars and the proto-BD candidates. In Table \ref{tbl:outflow4}, it is quite apparent that the proto-BD candidates mostly have mass accretion rates ($10^{-8}-10^{-7}$ {\msun} yr$^{-1}$) lower than the measured values ($>10^{-6}$ {\msun} yr$^{-1}$) in protostellar objects from this study and other literatures \citep{Bontemps:1996vb,Hatchell:2007cz,Curtis:2010ia}. 

The other important parameter, the accretion luminosity is also different between the proto-BD and the protostellar candidates. While the accretion luminosity inferred from the estimated mass accretion rate may give the information on the time-averaged accretion history over the dynamical time of the outflow in the past, the internal luminosity inferred from infrared flux may give us some information on the current accretion process. Therefore a comparison of the ratio of the accretion and internal luminosities in proto-BD and protostellar candidates can provide a useful insight into their past and present accretion activities. 

Figure \ref{fig:corr_L_rLL} displays the distribution of the ratio of accretion luminosity to the internal luminosity as a function of the internal luminosities for protostellar and proto-BD candidates. The values are found to be distributed over three orders of magnitude. For the uncertainty calculation for the ratios in the figure, we simply took the effect by the possible variation of the inclination angle between 10-80{\arcdeg} with respect to the assumed value (57{\arcdeg}.3) into account as discussed earlier in the section \ref{sec:res}. In those sources where the inclination angle has been determined by other studies, we used their values and uncertainties in the estimation of the ratio. Despite this uncertainty, the difference in the distribution of the ratio of accretion and internal luminosities in the protostellar and proto-BD candidates is quite evident. The values of $L_{\rm acc}/L_{\rm int}$ for the protostar candidates are usually over a few tens while those for the proto-BD candidates are mostly $\leq 1$. This may mean that the protostar candidates had higher (or episodic) accretion activities in the past and most of the proto-BD candidates have been accreting in a quiescent manner all along with a relatively low accretion rate of $10^{-8}-10^{-7}$ {\msun} yr$^{-1}$ or is presently experiencing higher (or episodic) accretion rate.

\subsubsection{Relation of Outflow Force against Luminosity {\&} Envelope Mass}\label{sec:dis:out2}

The previous outflow surveys of the embedded YSOs have found that the sources with fainter bolometric luminosities and/or smaller envelope masses tend to have weaker outflow forces \citep[e.g., ][]{Bontemps:1996vb, Hatchell:2007cz, Curtis:2010ia}. It is interesting to examine whether such a relation exists for the two groups of VeLLOs (proto-BD candidates and faint protostar candidates) studied here.

Figure \ref{fig:corr_F_Lbol} shows the outflow force versus the bolometric luminosity for the proto-BD candidates and the faint protostar candidates. The data for Class 0 and I protostars taken from \citet{Bontemps:1996vb}, \citet{Hatchell:2007cz}, and \citet{Curtis:2010ia} are also plotted in the same figure for comparison. It is apparent that the luminosity and the outflow forces for most of the members belonging to VeLLOs are smaller than those of protostars. The outflow forces of VeLLOs range between $10^{-8} - 10^{-5}$ {\msun} $\rm km~s^{-1}~yr^{-1}$, while those of the protostars are between $10^{-6}- 10^{-3}$ {\msun} $\rm km~s^{-1}~yr^{-1}$. The figure also shows that the VeLLOs with lower luminosity tend to have lower values of outflow forces, implying that the F-{\lbol} relationship found in protostars can also be applied to many of the VeLLOs in the lower {\lbol} regime. 

We note that VeLLOs are distributed in an inverse `L'-shaped pattern on the F-{\lbol} spaces. In Figure \ref{fig:corr_F_Lbol} the evolutionary tracks produced by the `toy' model are drawn for four initial envelope masses (0.08, 0.3, 0.6, and 1.2 {\msun}, from left to right) in colored curves \citep[See Figure 5, ][]{Bontemps:1996vb}. The figure indicates that the distribution of many of the VeLLOs seems to follow the evolutionary track of a protostar with the initial envelope mass, 0.08 {\msun}.

However, the distribution of the VeLLOs seen in Figure \ref{fig:corr_F_Lbol} can be interpreted somewhat differently if the distribution is considered for two separate groups, namely, proto-BD candidates and faint protostar candidates. We note that the faint protostar candidates tend to have weaker outflow forces similar to the trend seen in the case of more luminous protostars. This trend is found to be true for more than half of the proto-BD candidates which are having lower luminosity. Their outflow forces and internal luminosities are distributed in an inverse `L' shaped pattern and this distribution can be well fitted with a simple evolutionary track of a protostar with its initial envelope mass of $\sim$0.08 {\msun}. This may indicate that a significant number of the proto-BD candidates can be formed by the accretion and outflow processes as a downscaled version of low-mass star formation and the variation seen in the force (and also the mass accretion rate) in them could most likely be due to difference in their current evolutionary status. Such a rapid decrease in the mass accretion rates during the evolution periods of $\sim10^4$ yr in protostellar objects has been noticed from their MHD simulations by \citet{Machida:2009gr}. But it is interesting to see that a significant fraction (about 40\%) of the proto-BDs have much weaker outflow force than expected from the relations for protostars. This indicates that the formation mechanism for some of the BDs may be somehow different from that for the protostars.

This formation dichotomy of our proto-BD candidates can also be inferred from Figure \ref{fig:corr_F_Menv} where the outflow forces for the VeLLOs as a function of their envelope masses are compared with those of the protostars. The figure indicates that all the VeLLOs have envelope masses less than $\sim 1$ {\msun} and most of the VeLLOs have similar envelope masses comparable to those Class I protostars that are lying at the lower end of their distribution. The protostar candidates and more than half of the proto-BD candidates in our VeLLO samples are found to show a tendency in which the sources with less massive envelopes have less powerful outflows. However, we also notice that several ($\sim$6) proto-BD candidates do not follow this trend well. Their outflows are much weaker than those of the protostars having similar envelope masses. Therefore we believe that some of the proto-BDs candidates may be forming like normal protostars, while others may be forming differently from normal stars somehow in a manner that their outflows are much less powerful, compared with those of normal protostars. 

\subsubsection{On the Possible Formation Mechanisms of Brown Dwarfs}
How the group of proto-BD candidates with very low outflow force forms is a difficult question to address. There are several proposed mechanisms on how the BDs get lower mass compared with normal stars, 1) the interruption of accretion processes by the tidal shear and high velocity within the cluster, by ejection due to dynamical interaction in multiple systems, or by dissociation of envelope due to photoionizing radiation from OB stars, 2) the fragmentation by gravity in massive disk surrounding stars, or 3) the gravitational collapse in the smallest mass cores fragmented by turbulent compression \citep[e.g., ][]{Bonnell:2007vm, Whitworth:2007tq, Luhman:2012bg}. However, to observationally identify which mechanism is responsible for the formation of any specific target is not a simple task to do.

There are six sources in the group of proto-BD candidates with very low outflow forces. We made an attempt to investigate the regions surrounding the individual members of this group, using Simbad database, WISE, and Herschel data in order to examine whether there exist any special environment around them. We found that at least two targets, J182920 and IC348-SMM2E are located close to a cluster of stars (e.g., Serpens and IC348 cluster) where spectral type B or A stars exist. Thus it may be possible that the formation of these proto-BD candidates is affected by the presence of cluster environment. However, due to limitation in the currently available data, we could not identify the possible kind of mechanism (or environment) that may have caused the formation of other four sources (J041412, L1521F-IRS, J160115, and J162145). For example, we do not have any large mapping data required to investigate the possible association of the cluster, multiple systems, or OB stars with our sources which could influence the interruption of accretion processes in them. We also do not have either any high angular resolution mapping data to examine the mechanism of fragmentation in massive protostellar disk or any large mapping data in optically thick and thin molecular lines for investigation of gravitational collapse by turbulent compression. Therefore we believe that identification of the possible formation mechanism for the individual proto-BD candidates is beyond the scope of this work and leave it for future studies.

Why are the outflows in proto-BDs relatively weaker in comparison with those of the protostellar outflows? Low mass accretion rate and low envelope mass in the proto-BDs may be the main factors responsible for the weakening of outflows as previously shown. Magnetic field can also be an important factor which may affect the outflow force. However, the exact role of the magnetic field in powering of outflow is not simple to understand, mainly because of the very limited observations available for the VeLLOs or the proto-BD candidates. So far there are only five VeLLOs where magnetic field was observed \citep{Soam:2015ch,Soam:2015fe}. Out of them there are only three proto-BD candidates (L328-IRS, \citet{Soam:2015ch}; L1014-IRS, L1521F-IRS, \citet{Soam:2015fe}). Thus it is difficult to define how magnetic field acts on the proto-BD formation with this limited data. The only apparent thing known from the observations of magnetic field in the proto-BD candidates in comparison with normal protostars \citep[e.g., ][]{Wolf:2003jm} is that the plane-of-the-sky magnetic field strength (a few tens of $\rm \mu G$) of the proto-BD candidates (or VeLLOs) is smaller than that (a few hundred $\rm \mu G$) for protostars by one order of magnitude. Therefore we can postulate that a weak magnetic field in the proto-BDs might result in a less effective mass accretion processes and thus outflow activity to be less active than in the normal protostars.

In summary, we found that in more than half of the proto-BD candidates showing low luminosity and low envelope masses, the outflows become less powerful as expected for normal protostars. But there are other proto-BD candidates whose outflows are much less active than that predicted by the above behavior. Our results suggest that the formation process of the majority of the proto-BDs may be similar to that of normal stars, while a significant number of proto-BDs might be following a different formation process involving a lower accretion rate and less powerful outflows than normal protostars, although the nature of the mechanism under work is unclear at the moment due to the limited dataset that we have on them.

\section{Summary \& Conclusions}\label{sec:con}
We conducted the largest CO outflow survey toward the VeLLOs in 3-2 and/or 2-1 transitions of $^{12}$CO, $^{13}$CO, and/or C$^{18}$O lines using single dish telescopes. The main goal of the survey was to characterize the VeLLOs by investigating their outflow properties. Out of 68 VeLLOs observed in a single pointing mode, we found a total of 52 VeLLOs that show broad wing features in the CO spectra as a signature of outflow. Out of them, 16 VeLLOs are found to reveal clear outflow patterns in their mapping observations. Including three other previously studied VeLLOs, we analyzed the outflow properties for 19 VeLLOs. We draw the following conclusions from our analysis: 
\begin{itemize}
\item By considering a BD mass condition $\rm M_{acc}+0.2 \times M_{env} < 0.08$ {\msun} where a value of $\sim 20\%$ for the core-to-star efficiency is assumed, we were able to identify 15 VeLLOs as proto-BD candidates and four VeLLOs as likely faint protostar candidates. Thus our survey provides the largest sample of proto-BD candidates to date.

\item We found clear differences between protostar and proto-BD candidates. First, the proto-BD candidates have a mass accretion rate (mostly $10^{-8} -10^{-7}$ {\msun} yr$^{-1}$) much lower than that of the protostar candidates (mostly $\gtrsim10^{-6}$ {\msun} yr$^{-1}$). Second, the values of $L_{\rm acc}/L_{\rm int}$ for the proto-BD candidates are mostly $\leq 1$ while those for the protostar candidates are usually over a few tens. This may suggest that most proto-BD candidates either have been under small accretion activity in a quiescent manner all the way or are experiencing presently higher (or episodic) mass accretion than the past, while the protostar candidates have been under higher (or any episodic) accretion activities in the past than the present.

\item The two groups of sources we identified, proto-BD and faint protostar candidates, are in the regime of low luminosity and low envelope mass. We found that the faint protostar candidates well follow the previously known trends for the protostars where the protostars with less massive envelopes and/or smaller luminosities have weaker outflow forces. These trends were also found to be applicable to many of the proto-BD candidates in the lower luminosity or the light envelope mass regime. Interestingly, their outflow forces (or mass accretion rates) and internal luminosities are distributed in an inverse `L' shape pattern and this distribution is found to be well fitted with a simple evolutionary track of a protostar with its initial envelope mass of $\sim$0.08 {\msun}, implying the possibility that some of BDs are formed in a way similar to that of the normal stars formed through gravitational collapse in the dense core. But we also note that a significant fraction (about 40\%) of the proto-BDs have much weaker outflow force than expected by the relations for protostars, implying that the formation mechanism for some BDs may be somehow different from that for the protostars.
\end{itemize}

This survey provides the first statistical study conducted to understand the formation mechanism of the proto-BDs by using their outflow properties. However, because of the unknown mass of the central object and large uncertainties of their physical quantities related to the outflows, especially the mass accretion rate, the true nature of these sources are not very well constrained and thus require further study. High angular resolution observations in molecular lines with sensitive interferometers such as ALMA will be necessary to estimate the properties more precisely. Constraining the central mass and the mass accretion rate of the proto-BDs more accurately would help us to develop a better knowledge of the differences between the formation processes of stars and substellar objects.

\acknowledgments
This research has made use of the NASA$/$IPAC Infrared Science Archive, which is operated by the Jet Propulsion Laboratory, California Institute of Technology, under contract with the National Aeronautics and Space Administration. The SRAO telescope is operated by Seoul National University. The ASTE telescope is operated by National Astronomical Observatory of Japan (NAOJ). This material is based on work at the Caltech Submillimeter Observatory, which is operated by the California Institute of Technology. The James Clerk Maxwell Telescope is operated by the East Asian Observatory on behalf of the National Astronomical Observatory of Japan, Academia Sinica Institute of Astronomy and Astrophysics, the Korea Astronomy and Space Science Institute, the National Astronomical Observatories of China and the Chinese Academy of Sciences (Grant No. XDB09000000), with additional funding support from the Science and Technology Facilities Council of the United Kingdom and participating universities in the United Kingdom and Canada. This work was supported by the Basic Science Research Program through the National Research Foundation of Korea (NRF) funded by the Ministry of Education, Science, and Technology (NRF-2016R1A2B4012593). S. Kim is supported by the National Research Foundation of Korea (NRF) grant funded by the Korean government (MSIP)(no. 2017037333). M. Kim is supported by Basic Science Research Program through the National Research Foundation of Korea (NRF) funded by the Ministry of Education (2017R1A6A3A01075724).

{\it Facilities: SRAO, CSO, ASTE, \& JCMT}

\onecolumngrid 
\newpage  

\begin{deluxetable}{cCCcCCCc}
\tablecaption{List of Very Low Luminosity Objects \label{tbl:obj}}
\tablewidth{0pt}
\tablehead{
\colhead{Source} & \colhead{Ra.} & 
\colhead{Dec.} & \colhead{Region} & 
\colhead{Distance} & \colhead{$L_{\rm int}$} & 
 \colhead{$M_{\rm env}$}  & \colhead{Class}\\
\colhead{} & \colhead{($^h$ $^m$ $^s$)} & 
\colhead{({\arcdeg} {\arcmin} {\arcsec})} & \colhead{} & 
\colhead{(pc)} &  \colhead{(\lsun)} & 
\colhead{({\msun})} & \colhead{} 
}
\colnumbers 
\startdata
J032832 & 03~28~32.57 & +31~11~05.3 & Perseus & 250\pm 50 & 0.05\pm 0.01 & 0.10\pm 0.18 & I \\
J032839 & 03~28~39.10 & +31~06~01.8 & Perseus & 250\pm 50 & 0.02\pm 0.01 & 0.10\pm 0.22 & 0 \\
J033032 & 03~30~32.69 & +30~26~26.5 & Perseus & 250\pm 50 & 0.13\pm 0.03 & 0.30\pm 0.63 & 0 \\
IC348-SMM2E & 03~43~57.73 & +32~03~10.1 & Perseus & 250\pm 50 & 0.06\pm 0.01 & 0.03\pm 0.02 & 0 \\
J040134 & 04~01~34.36 & +41~11~43.1 & California & 450\pm 23 & 0.25\pm 0.05 & 0.02\pm 0.01 & Flat \\
J041412 & 04~14~12.30 & +28~08~37.2 & Taurus & 137\pm 10 & 0.06\pm 0.03 & 0.02\pm 0.03 & I \\
J041840 & 04~18~40.26 & +28~29~25.3 & Taurus & 137\pm 10 & \leq 0.01 & \leq 0.01 & I \\
IRAM 04191+1522 & 04~21~56.88 & +15~29~46.0 & IRAM04191+1522 & 140\pm 10 & 0.04\pm 0.01 & 0.05\pm 0.09 & 0 \\
J042513 & 04~25~13.24 & +26~31~45.0 & Taurus & 137\pm 10 & \leq 0.01 & \leq 0.01 & 0 \\
J042815 & 04~28~15.15 & +36~30~28.7 & California & 450\pm 23 & 0.05\pm 0.01 & 0.10\pm 0.01 & I \\
L1521F-IRS & 04~28~38.90 & +26~51~35.6 & Taurus & 140\pm 10 & 0.04\pm 0.01 & 0.40\pm 0.85 & 0 \\
J043014 & 04~30~14.96 & +36~00~08.5 & California & 450\pm 23 & 0.09\pm 0.03 & 0.30\pm 0.02 & I \\
J043055 & 04~30~55.99 & +34~56~47.9 & California & 450\pm 23 & 0.25\pm 0.06 & \leq 0.01 & I \\
J043411 & 04~34~11.50 & +24~03~41.4 & Taurus & 137\pm 10 & \leq 0.01 & \leq 0.01 & I \\
J043909 & 04~39~09.04 & +26~14~49.5 & Taurus & 137\pm 10 & \leq 0.01 & \leq 0.01 & 0 \\
J044022 & 04~40~22.47 & +25~58~32.9 & Taurus & 137\pm 10 & \leq 0.01 & \leq 0.01 & I \\
J080533 & 08~05~33.05 & -39~09~24.8 & BHR16 & 440\pm100 & 0.09\pm 0.05 & \leq 0.01 & I \\
J105959 & 10~59~59.71 & -77~11~18.4 & Chamaeleon I & 150\pm 15 & \leq 0.01 & \leq 0.01 & Flat \\
J110955 & 11~09~55.03 & -76~32~41.3 & Chamaeleon I & 150\pm 15 & 0.12\pm 0.03 & \leq 0.01 & Flat \\
J121406 & 12~14~06.51 & -80~26~25.3 & Chamaeleon III & 150\pm 15 & 0.03\pm 0.01 & \leq 0.01 & 0 \\
J125701 & 12~57~01.58 & -76~48~34.9 & Chamaeleon II & 178\pm 18 & 0.07\pm 0.01 & \leq 0.01 & I \\
J154051 & 15~40~51.62 & -34~21~04.7 & Lupus I & 150\pm 20 & \leq 0.01 & \leq 0.01 & I \\
J154216 & 15~42~16.99 & -52~48~02.2 & DC3272+18 & 250\pm 50 & 0.04\pm 0.02 & 0.04\pm 0.08 & 0 \\
J160115 & 16~01~15.55 & -41~52~35.4 & Lupus IV & 150\pm 20 & 0.10\pm 0.02 & 0.40\pm 0.81 & I \\
J162145 & 16~21~45.12 & -23~42~31.7 & Ophiuchus & 125\pm 25 & 0.08\pm 0.02 & \leq 0.01 & I \\
J162648 & 16~26~48.48 & -24~28~38.6 & Ophiuchus & 125\pm 25 & 0.04\pm 0.01 & 0.70\pm 0.16 & Flat \\
IRAS16253-2429 & 16~28~21.60 & -24~36~23.4 & Ophiuchus & 125\pm 25 & 0.10\pm 0.01 & 0.10\pm 0.12 & 0 \\
J180439 & 18~04~39.91 & -04~01~22.0 & Aquila & 260\pm 10 & \leq 0.01 & \leq 0.01 & I \\
CB130-3-IRS & 18~16~16.39 & -02~32~37.7 & CB130-3 & 270\pm 50 & 0.07\pm 0.02 & 0.20\pm 0.45 & 0 \\
L328-IRS & 18~16~59.47 & -18~02~30.5 & L328 & 217\pm 30 & 0.13\pm 0.01 & 0.09\pm 0.04 & 0 \\
J182855 & 18~28~55.84 & -01~37~34.9 & Aquila & 260\pm 10 & 0.10\pm 0.03 & \leq 0.01 & I \\
J182905 & 18~29~05.45 & -03~42~45.6 & Aquila & 260\pm 10 & 0.15\pm 0.03 & 0.10\pm 0.01 & I \\
J182912 & 18~29~12.11 & -01~48~45.4 & Aquila & 260\pm 10 & \leq 0.01 & \nodata & I \\
J182913 & 18~29~13.07 & -01~46~17.1 & Aquila & 260\pm 10 & 0.27\pm 0.03 & 0.30\pm 0.01 & I \\
J182920 & 18~29~20.97 & -01~37~14.2 & Aquila & 260\pm 10 & 0.10\pm 0.03 & 0.10\pm 0.01 & Flat \\
J182925 & 18~29~25.12 & -01~47~37.9 & Aquila & 260\pm 10 & 0.04\pm 0.01 & \leq 0.01 & Flat \\
J182933 & 18~29~33.69 & -01~45~10.3 & Aquila & 260\pm 10 & 0.03\pm 0.06 & 0.10\pm 0.01 & Flat \\
J182937 & 18~29~37.43 & -03~14~53.9 & Aquila & 260\pm 10 & 0.11\pm 0.03 & 0.10\pm 0.01 & I \\
J182943 & 18~29~43.96 & -02~12~55.3 & Aquila & 260\pm 10 & 0.10\pm 0.01 & 0.10\pm 0.01 & I \\
J182952 & 18~29~52.96 & -01~58~05.2 & Aquila & 260\pm 10 & 0.05\pm 0.01 & \leq 0.01 & 0 \\
J182958 & 18~29~58.35 & -01~57~40.2 & Aquila & 260\pm 10 & 0.04\pm 0.03 & \nodata & I \\
J183014 & 18~30~14.42 & -01~33~33.3 & Aquila & 260\pm 10 & 0.07\pm 0.03 & 0.60\pm 0.02 & 0 \\
J183015 & 18~30~15.63 & -02~07~19.6 & Aquila & 260\pm 10 & 0.22\pm 0.01 & 0.10\pm 0.01 & 0 \\
J183016 & 18~30~16.24 & -01~52~52.9 & Aquila & 260\pm 10 & 0.26\pm 0.01 & 0.04\pm 0.01 & 0 \\
J183017 & 18~30~17.47 & -02~09~58.5 & Aquila & 260\pm 10 & 0.15\pm 0.02 & 0.04\pm 0.01 & I \\
J183021 & 18~30~21.82 & -01~52~01.0 & Aquila & 260\pm 10 & 0.26\pm 0.05 & 0.40\pm 0.02 & I \\
J183027 & 18~30~27.58 & -01~54~39.3 & Aquila & 260\pm 10 & 0.25\pm 0.06 & \leq 0.01 & I \\
J183237 & 18~32~37.42 & -02~50~45.2 & Aquila & 260\pm 10 & 0.21\pm 0.05 & 0.10\pm 0.01 & I \\
J183242 & 18~32~42.48 & -02~47~56.5 & Aquila & 260\pm 10 & 0.17\pm 0.06 & 0.10\pm 0.01 & Flat \\
J183245 & 18~32~45.68 & -02~46~57.6 & Aquila & 260\pm 10 & 0.03\pm 0.06 & \leq 0.01 & I \\
J183329 & 18~33~29.45 & -02~45~58.3 & Aquila & 260\pm 10 & 0.09\pm 0.05 & 0.20\pm 0.01 & 0 \\
J183929 & 18~39~29.87 & +00~37~40.5 & Serpens & 429\pm  2 & 0.17\pm 0.04 & 0.40\pm 0.03 & I \\
L673-7-IRS & 19~21~34.82 & +11~21~23.4 & L673-7 & 300\pm100 & 0.04\pm 0.01 & 0.10\pm 0.15 & 0 \\
L1148-IRS & 20~40~56.66 & +67~23~04.9 & L1148 & 325\pm 25 & 0.12\pm 0.02 & 0.50\pm 1.09 & I \\
J210221 & 21~02~21.22 & +67~54~20.3 & Cepheus & 288\pm 50 & 0.23\pm 0.04 & \leq 0.01 & I \\
L1014-IRS & 21~24~07.58 & +49~59~08.9 & L1014 & 250\pm 50 & 0.09\pm 0.03 & 0.70\pm 1.39 & 0 \\
J214448 & 21~44~48.31 & +47~44~59.8 & IC5146 & 350\pm 35 & 0.14\pm 0.03 & 0.04\pm 0.01 & 0 \\
J214457 & 21~44~57.08 & +47~41~52.9 & IC5146 & 350\pm 35 & 0.02\pm 0.05 & 0.20\pm 0.03 & I \\
J214531 & 21~45~31.22 & +47~36~21.3 & IC5146 & 350\pm 35 & 0.21\pm 0.05 & \leq 0.01 & I \\
J214657 & 21~46~57.52 & +47~32~23.6 & IC5146 & 350\pm 35 & 0.07\pm 0.04 & \leq 0.01 & I \\
J214703 & 21~47~03.08 & +47~33~14.9 & IC5146 & 350\pm 35 & 0.10\pm 0.01 & \nodata & I \\
J214706 & 21~47~06.02 & +47~39~39.4 & IC5146 & 350\pm 35 & 0.06\pm 0.05 & 0.03\pm 0.01 & I \\
J214755 & 21~47~55.67 & +47~37~11.4 & IC5146 & 350\pm 35 & 0.19\pm 0.02 & 0.02\pm 0.01 & I \\
J214858 & 21~48~58.51 & +47~25~42.7 & IC5146 & 350\pm 35 & 0.20\pm 0.03 & \leq 0.01 & Flat \\
J215607 & 21~56~07.34 & +76~42~29.6 & Cepheus & 300\pm100 & 0.02\pm 0.02 & \leq 0.01 & Flat \\
J222933 & 22~29~33.35 & +75~13~16.0 & Cepheus & 300\pm100 & 0.03\pm 0.05 & \nodata & I \\
J222959 & 22~29~59.42 & +75~14~03.7 & Cepheus & 300\pm100 & 0.24\pm 0.05 & 0.02\pm 0.01 & I \\
L1251A-IRS4 & 22~31~05.58 & +75~13~37.2 & Cepheus & 300\pm100 & 0.23\pm 0.01 & 0.20\pm 0.02 & 0 \\
\enddata
\tablecomments{All information is from \citet{Kim:2016fh}. Column (1): Source name, Column (2), (3): coordinates in J2000, Column (4): Names for the cloud or core region, Column (5): Distance from us, Column (6): Internal luminosity measured from available 70 {\micron} flux, Column (7): Envelope mass measured from available 250 or 500 {\micron} flux, Column (8): Evolutionary stage classified by the bolometric temperature. }
\end{deluxetable}

\newpage  

\begin{deluxetable}{cccccccc}
\tablecaption{Summary of \textit{SRAO}, \textit{CSO}, \textit{JCMT}, {\&} \textit{ASTE} Observations \label{tbl:gen}}
\tabletypesize{\scriptsize}
\tablewidth{0pt}
\tablehead{
\colhead{Molecular lines} & \colhead{Frequency\tablenotemark{a}} & 
\colhead{Telescope} & 
\colhead{HPBW\tablenotemark{b}} & \colhead{$\eta_{\rm MB}$\tablenotemark{c}} &
Bandwidth & $\Delta \nu$ ($\Delta$v)\tablenotemark{d} &
\colhead{Observation date} \\
\colhead{} & \colhead{(MHz)} & 
\colhead{} & 
\colhead{(\arcsec)} & \colhead{(\%)} &
\colhead{(MHz)} & \colhead{(kHz)(km s$^{-1}$)} &
\colhead{} 
} 
\startdata
\multirow{4}{*}{\begin{minipage}{0.5in} $^{12}$CO2-1 \\ $^{13}$CO2-1 \\ C$^{18}$O2-1 \end{minipage}} & \multirow{4}{*}{\begin{minipage}{0.7in} 230538.0000 \\ 220398.6841 \\ 219560.3578 \end{minipage}} & SRAO 6 m & 48 & 61 & 100 & 48.8 (0.064) &  \begin{minipage}{1.3in}2008 Dec - 2010 Apr \end{minipage} \\
 & & \\
\cline{3-8}
& & CSO 10.4 m & 30 & 76 & 1,000 & 0.122 (0.16) &  \multirow{2}{*}{\begin{minipage}{1.3in} 2013 Oct \& Nov \\ 2014 Sep\end{minipage}} \\
& & \\
\hline
\multirow{5}{*}{\begin{minipage}{0.5in} $^{12}$CO3-2 \\ $^{13}$CO3-2 \\ C$^{18}$O3-2 \end{minipage}} & \multirow{5}{*}{\begin{minipage}{0.7in} 345795.9899 \\ 330587.9652 \\ 329330.5523 \end{minipage}} & ASTE 10 m & 22 & 60 & 128 & 0.125 (0.11) &  \multirow{3}{*}{\begin{minipage}{1.3in} 2013 Sep \\ 2014 Jun \& Sep \\ 2015 Nov \end{minipage}}\\
& & \\
& & \\
\cline{3-8}
& &
JCMT 15 m & 15 & 64 & 250 & 0.061 (0.05) &  \multirow{2}{*}{\begin{minipage}{1.3in} 2015 Sep, Nov, \& Dec \\ 2016 Mar, Apr, \& Jun \end{minipage}} \\
 & &\\
\enddata
\tablenotetext{a}{The frequencies of $^{12}$CO lines are from JPL catalog (\burl{http://spec.jpl.nasa.gov/home.html}). The frequencies of $^{13}$CO and C$^{18}$O lines are from \citet{Cazzoli:2004ff} and \citet{Cazzoli:2003bb}, respectively.}
\tablenotetext{b}{Angular resolution}
\tablenotetext{c}{Main beam efficiency}
\tablenotetext{d}{Frequency resolution (velocity resolution)}
\end{deluxetable}  

\begin{deluxetable}{ccccCCCCCCCccccCc}
\tablecaption{Results of Single-pointing and Mapping Observations for 68 VeLLOs \label{tbl:obs}}
\tabletypesize{\scriptsize}
\tablewidth{0pt}
\tablehead{
\colhead{}& \multicolumn{12}{c}{Single-pointing Observations} & \multicolumn{4}{c}{Mapping Observations} \\
\cmidrule(lr){2-13}\cmidrule(lr){14-17}
\colhead{Source} & \colhead{Tel.} & \colhead{J} & 
\multicolumn{2}{c}{$^{12}$CO} & \multicolumn{2}{c}{$^{13}$CO} & 
\multicolumn{3}{c}{C$^{18}$O} & \colhead{} & \multicolumn{2}{c}{CO profile} & \colhead{Tel.} & \colhead{J} & \colhead{Size} & \colhead{Out.}  \\
\cmidrule(lr){4-5}\cmidrule(lr){6-7}\cmidrule(lr){8-10}\cmidrule(lr){12-13}
\colhead{} & \colhead{} & \colhead{} & 
\colhead{$T_{\rm peak} $ } & \colhead{$\sigma_{\rm rms}$} & 
\colhead{$T_{\rm peak} $} & \colhead{$\sigma_{\rm rms}$} & 
\colhead{$T_{\rm peak}$} & \colhead{$\sigma_{\rm rms}$} &
\colhead{$\Delta$V} & \colhead{\vsys} & \colhead{Asym.} & \colhead{Wing}
& \colhead{} & \colhead{} & \colhead{} \\
\cmidrule(lr){4-9}\cmidrule(lr){10-11}
\colhead{} & \colhead{} & \colhead{} & 
\multicolumn{6}{c}{(K)} & 
\multicolumn{2}{c}{(km s$^{-1}$)} & \colhead{} & \colhead{}
& \colhead{} & \colhead{} & \colhead{($\arcsec \times \arcsec$)} & \colhead{}}
\colnumbers
\startdata
J032832 & SRAO & 2-1 & 14.7 & 0.09 & 5.1 & 0.19 & 0.9 & 0.07 & 2.0 & 7.4 & \nodata & Y & SRAO & 2-1 & 200\times200 & C \\
J032839 & SRAO & 2-1 & 14.3 & 0.08 & 5.2 & 0.15 & 1.7 & 0.08 & 1.4 & 6.8 & B & Y & SRAO & 2-1 & 200\times200 & C \\
J033032 & SRAO & 2-1 & 7.2 & 0.06 & 3.9 & 0.18 & 1.3 & 0.09 & 0.5 & 5.9 & B & Y & SRAO & 2-1 & 200\times200 & Y \\
IC348-SMM2E & JCMT & 3-2 & 22.8 & 0.14 & 9.8 & 0.25 & 5.3 & 0.29 & 0.8 & 8.8 & B & Y & JCMT & 3-2 & 90\times90 & C \\
J040134 & SRAO & 2-1 & 5.6 & 0.14 & 1.9 & 0.13 & 1.0 & 0.09 & 0.9 & -7.7 & \nodata & Y & SRAO & 2-1 & 200\times200 & C \\
J041412 & JCMT & 3-2 & 11.5 & 0.16 & 3.7 & 0.09 & 1.5 & 0.09 & 0.6 & 6.9 & R & Y & JCMT & 3-2 & 120\times120 & Y \\
J041840 & CSO & 2-1 & 7.3 & 0.08 & 4.8 & 0.07 & 1.7 & 0.08 & 0.5 & 7.3 & mc & \nodata & \nodata & \nodata & \nodata & mc \\
IRAM 04191+1522 & SRAO & 2-1 & 7.6 & 0.09 & 3.2 & 0.14 & 1.8 & 0.12 & 0.7 & 6.5 & B & Y & SRAO & 2-1 & 350\times400 & Y \\
J042513 & JCMT & 3-2 & 1.9 & 0.15 & \nodata & \nodata & \nodata & \nodata & \nodata & 6.5 & \nodata & Y & JCMT & 3-2 & 90\times90 & NY \\
J042815 & CSO & 2-1 & 4.9 & 0.10 & 3.4 & 0.10 & 0.7 & 0.09 & 0.6 & -0.5 & mc & \nodata & \nodata & \nodata & \nodata & mc \\
L1521F-IRS & CSO & 2-1 & 6.0 & 0.12 & 4.2 & 0.14 & 3.0 & 0.13 & 0.4 & 6.5 & B & Y & CSO & 2-1 & 60\times60 & Y \\
J043014 & CSO & 2-1 & 4.3 & 0.05 & 4.3 & 0.04 & 1.6 & 0.05 & 0.8 & -0.7 & B & Y & CSO & 2-1 & 120\times120 & C \\
J043055 & SRAO & 2-1 & 5.0 & 0.21 & 3.3 & 0.18 & 1.3 & 0.17 & 0.7 & -0.8 & B & Y & JCMT & 3-2 & 90\times90 & C \\
J043411 & JCMT & 3-2 & 4.4 & 0.13 & 1.4 & 0.29 & <3$\sigma$ & 0.32 & \nodata & 6.3 & \nodata & \nodata & JCMT & 3-2 & 90\times90 & N \\
J043909 & CSO & 2-1 & 5.6 & 0.12 & 2.1 & 0.12 & <3$\sigma$ & 0.13 & \nodata & 6.1 & R & Y & \nodata & \nodata & \nodata & \nodata \\
J044022 & JCMT & 3-2 & 2.6 & 0.16 & \nodata & \nodata & \nodata & \nodata & \nodata & 6.1 & \nodata & \nodata & JCMT & 3-2 & 90\times90 & N \\
J080533 & ASTE & 3-2 & 5.0 & 0.11 & \nodata & \nodata & \nodata & \nodata & \nodata & 9.2 & \nodata & \nodata & \nodata & \nodata & \nodata & \nodata \\
J105959 & ASTE & 3-2 & 4.6 & 0.08 & \nodata & \nodata & \nodata & \nodata & \nodata & 4.7 & \nodata & Y & ASTE & 3-2 & 60\times60 & NY \\
J110955 & ASTE & 3-2 & 6.1 & 0.07 & \nodata & \nodata & \nodata & \nodata & \nodata & 5.1 & \nodata & Y & ASTE & 3-2 & 60\times60 & C \\
J121406 & ASTE & 3-2 & 1.4 & 0.17 & \nodata & \nodata & \nodata & \nodata & \nodata & 2.0 & \nodata & \nodata & \nodata & \nodata & \nodata & \nodata \\
J125701 & ASTE & 3-2 & 3.3 & 0.09 & \nodata & \nodata & \nodata & \nodata & \nodata & 2.5 & \nodata & \nodata & \nodata & \nodata & \nodata & \nodata \\
J154051 & ASTE & 3-2 & 0.5 & 0.08 & \nodata & \nodata & \nodata & \nodata & \nodata & 4.8 & \nodata & \nodata & \nodata & \nodata & \nodata & \nodata \\
J154216 & ASTE & 3-2 & 4.4 & 0.05 & 3.3 & 0.07 & 0.9 & 0.08 & 0.6 & -0.1 & B & Y & ASTE & 3-2 & 120\times140 & Y \\
J160115 & ASTE & 3-2 & 9.0 & 0.06 & 4.4 & 0.06 & 0.9 & 0.09 & 0.9 & 4.1 & B & Y & ASTE & 3-2 & 40\times60 & Y \\
J162145 & JCMT & 3-2 & 22.0 & 0.11 & \nodata & \nodata & \nodata & \nodata & \nodata & 2.8 & \nodata & Y & JCMT & 3-2 & 90\times90 & Y \\
J162648 & SRAO & 2-1 & 34.3 & 0.12 & 12.2 & 0.24 & 5.3 & 0.14 & 1.4 & 3.2 & B & Y & SRAO & 2-1 & 100\times100 & C \\
  & ASTE & 3-2 & 21.9 & 0.07 & 15.6 & 0.10 & 3.7 & 0.16 & 1.4 & 3.3 & B & Y & \nodata & \nodata & \nodata & \nodata \\
IRAS16253-2429 & ASTE & 3-2 & 10.8 & 0.09 & 5.2 & 0.15 & 3.9 & 0.18 & 0.7 & 4.0 & B & Y & ASTE & 3-2 & 80\times160 & Y \\
 & SRAO & 2-1 & 16.9 & 0.12 & 5.9 & 0.29 & 4.3 & 0.12 & 0.7 & 3.8 & B & Y & \nodata & \nodata & \nodata & \nodata \\
J180439 & ASTE & 3-2 & 1.7 & 0.10 & \nodata & \nodata & \nodata & \nodata & \nodata & 7.0 & \nodata & \nodata & \nodata & \nodata & \nodata & \nodata \\
CB130-3-IRS & SRAO & 2-1 & 3.8 & 0.10 & 1.3 & 0.08 & 0.7 & 0.20 & 1.8 & 7.6 & \nodata & Y & SRAO & 2-1 & 200\times200 & Y \\
 & ASTE & 3-2 & 2.4 & 0.06 & 2.7 & 0.11 & 0.6 & 0.18 & 0.6 & 7.7 & \nodata & Y & \nodata & \nodata & \nodata & \nodata \\
L328-IRS & SRAO & 2-1 & 6.6 & 0.08 & 3.7 & 0.07 & 2.3 & 0.20 & 0.5 & 6.5 & B & Y & SRAO & 2-1 & 200\times300 & Y \\
J182855 & JCMT & 3-2 & 2.3 & 0.14 & 3.0 & 0.16 & 1.3 & 0.15 & 0.8 & 7.4 & \nodata & Y & JCMT & 3-2 & 90\times90 & C \\
J182905 & JCMT & 3-2 & 2.8 & 0.14 & 2.6 & 0.13 & 1.0 & 0.13 & 0.6 & 5.4 & \nodata & Y & JCMT & 3-2 & 90\times90 & C \\
J182912 & ASTE & 3-2 & 5.0 & 0.13 & \nodata & \nodata & \nodata & \nodata & \nodata & 7.0 & \nodata & \nodata & \nodata & \nodata & \nodata & \nodata \\
J182913 & JCMT & 3-2 & 4.7 & 0.13 & 3.2 & 0.11 & 0.9 & 0.13 & 1.2 & 7.0 & B & Y & JCMT & 3-2 & 90\times90 & C \\
J182920 & JCMT & 3-2 & 2.8 & 0.15 & 2.9 & 0.09 & 0.9 & 0.09 & 0.8 & 7.5 & B & Y & JCMT & 3-2 & 90\times90 & Y \\
J182925 & JCMT & 3-2 & 4.1 & 0.14 & 2.3 & 0.13 & 0.5 & 0.15 & 0.9 & 7.4 & B & Y & JCMT & 3-2 & 90\times90 & NY \\
J182933 & JCMT & 3-2 & 6.3 & 0.13 & 3.2 & 0.12 & 0.8 & 0.14 & 0.6 & 7.6 & B & Y & JCMT & 3-2 & 90\times90 & NY \\
J182937 & CSO & 2-1 & 2.2 & 0.11 & 1.6 & 0.12 & 0.9 & 0.10 & 0.4 & 6.1 & mc & \nodata & \nodata & \nodata & \nodata & mc \\
J182943 & JCMT & 3-2 & 2.7 & 0.17 & 1.5 & 0.11 & 1.3 & 0.11 & 0.6 & 7.6 & B & Y & JCMT & 3-2 & 180\times180 & Y \\
J182952 & CSO & 2-1 & 2.9 & 0.08 & 2.5 & 0.08 & 2.1 & 0.09 & 1.3 & 7.5 & \nodata & Y & CSO & 2-1 & 60\times120 & C \\
J182958 & ASTE & 3-2 & 3.5 & 0.05 & 1.3 & 0.16 & 1.3 & 0.27 & 1.2 & 7.7 & B & Y & ASTE & 3-2 & 80\times80 & C \\
J183014 & JCMT & 3-2 & 3.7 & 0.15 & 3.6 & 0.10 & 0.8 & 0.09 & 0.6 & 8.1 & mc & \nodata & JCMT & 3-2 & 90\times90 & mc \\
J183015 & JCMT & 3-2 & 3.6 & 0.13 & 3.0 & 0.13 & 2.9 & 0.13 & 0.8 & 6.9 & B & Y & JCMT & 3-2 & 90\times90 & C \\
J183016 & JCMT & 3-2 & 2.6 & 0.14 & 2.0 & 0.11 & 0.9 & 0.15 & 0.5 & 6.7 & B & Y & JCMT & 3-2 & 90\times90 & C \\
J183017 & JCMT & 3-2 & 4.9 & 0.14 & 3.7 & 0.12 & 2.1 & 0.15 & 1.0 & 6.7 & B & Y & JCMT & 3-2 & 90\times90 & C \\
J183021 & JCMT & 3-2 & 2.4 & 0.12 & 2.5 & 0.11 & 0.9 & 0.11 & 1.1 & 7.2 & \nodata & Y & JCMT & 3-2 & 90\times90 & C \\
J183027 & JCMT & 3-2 & 2.7 & 0.14 & 2.6 & 0.13 & 0.7 & 0.15 & 1.5 & 7.2 & B & Y & JCMT & 3-2 & 90\times90 & C \\
J183237 & JCMT & 3-2 & 4.2 & 0.13 & 2.5 & 0.14 & 1.5 & 0.14 & 0.5 & 6.3 & R & Y & JCMT & 3-2 & 90\times90 & Y \\
J183242 & ASTE & 3-2 & 3.6 & 0.05 & 3.0 & 0.11 & 1.1 & 0.15 & 0.5 & 6.4 & B & Y & ASTE & 3-2 & 40\times40 & C \\
J183245 & JCMT & 3-2 & 3.5 & 0.14 & 2.1 & 0.11 & 0.4 & 0.11 & 0.5 & 6.3 & mc & \nodata & JCMT & 3-2 & 90\times90 & mc \\
J183329 & JCMT & 3-2 & 2.6 & 0.15 & 2.3 & 0.10 & 1.4 & 0.10 & 0.4 & 7.4 & B & Y & JCMT & 3-2 & 90\times90 & C \\
J183929 & JCMT & 3-2 & 2.2 & 0.17 & 1.1 & 0.10 & 0.6 & 0.09 & 0.4 & 8.3 & R & Y & JCMT & 3-2 & 90\times90 & NY \\
L673-7-IRS & SRAO & 2-1 & 4.9 & 0.08 & 2.9 & 0.08 & 0.8 & 0.08 & 0.9 & 7.0 & R & Y & SRAO & 2-1 & 400\times400 & Y \\
L1148-IRS & SRAO & 2-1 & 4.4 & 0.07 & 1.5 & 0.09 & 0.6 & 0.14 & 0.1 & 2.6 & \nodata & Y & SRAO & 2-1 & 200\times200 & NY \\
J210221 & CSO & 2-1 & 6.0 & 0.10 & 4.2 & 0.09 & 1.5 & 0.11 & 0.6 & 2.8 & R & Y & CSO & 2-1 & 300\times300 & Y \\
L1014-IRS & SRAO & 2-1 & 2.9 & 0.09 & 1.9 & 0.13 & 1.0 & 0.07 & 0.5 & 4.1 & \nodata & Y & SRAO & 2-1 & 200\times200 & C \\
J214448 & JCMT & 3-2 & 3.5 & 0.17 & 2.3 & 0.10 & 1.0 & 0.11 & 0.8 & 4.5 & R & Y & JCMT & 3-2 & 90\times90 & C \\
J214457 & CSO & 2-1 & 5.2 & 0.12 & 3.2 & 0.12 & 1.2 & 0.12 & 0.7 & 3.2 & mc & \nodata & \nodata & \nodata & \nodata & mc \\
J214531 & CSO & 2-1 & 6.6 & 0.11 & 3.2 & 0.09 & 1.1 & 0.11 & 1.6 & 4.0 & R & Y & JCMT & 3-2 & 90\times90 & C \\
J214657 & JCMT & 3-2 & 4.1 & 0.14 & \nodata & \nodata & \nodata & \nodata & \nodata & 4.1 & \nodata & Y & JCMT & 3-2 & 90\times90 & C \\
J214703 & CSO & 2-1 & 4.7 & 0.11 & 3.1 & 0.10 & 1.8 & 0.11 & 0.7 & 3.9 & B & Y & JCMT & 3-2 & 90\times90 & C \\
J214706 & CSO & 2-1 & 5.7 & 0.09 & 4.0 & 0.09 & 1.4 & 0.11 & 0.7 & 4.0 & \nodata & Y & JCMT & 3-2 & 90\times90 & NY \\
J214755 & JCMT & 3-2 & 3.8 & 0.12 & \nodata & \nodata & \nodata & \nodata & \nodata & 3.2 & \nodata & Y & JCMT & 3-2 & 90\times90 & C \\
J214858 & JCMT & 3-2 & 4.9 & 0.16 & \nodata & \nodata & \nodata & \nodata & \nodata & 4.9 & \nodata & Y & JCMT & 3-2 & 90\times90 & C \\
J215607 & CSO & 2-1 & 3.6 & 0.13 & 2.1 & 0.13 & 0.6 & 0.12 & 1.1 & -5.9 & mc & \nodata & \nodata & \nodata & \nodata & mc \\
J222933 & CSO & 2-1 & 4.6 & 0.10 & 1.7 & 0.12 & 1.6 & 0.10 & 0.7 & -3.8 & B & \nodata & \nodata & \nodata & \nodata & \nodata \\
J222959 & JCMT & 3-2 & 2.9 & 0.14 & 1.7 & 0.14 & 1.0 & 0.17 & 1.0 & -3.9 & \nodata & Y & JCMT & 3-2 & 90\times90 & C \\
L1251A-IRS4 & SRAO & 2-1 & 7.2 & 0.07 & 1.9 & 0.17 & 0.9 & 0.07 & 0.5 & -4.2 & B & Y & SRAO & 2-1 & 200\times250 & Y \\
\enddata
\begin{minipage}{19.2cm}~\\\\\\
Note $-$ Column (1): Source name, Blank indicates the same target as the one in the row above, Column (2): Telescope used for single-pointing observation, Column (3): Rotational transition of molecular line used for single-pointing observation, Column (4): Peak temperature of $^{12}$CO line in $T_{\rm MB}$ scale, Column (5): Noise level of $^{12}$CO line in $T_{\rm MB}$ scale, Column (6): Peak temperature of $^{13}$CO line in $T_{\rm MB}$ scale, Column (7): Noise level of $^{13}$CO line in $T_{\rm MB}$ scale, Column (8): Peak temperature of C$^{18}$O line in $T_{\rm MB}$ scale, Column (9): Noise level of C$^{18}$O line in $T_{\rm MB}$ scale, Column (10): Line width derived by Gaussian fitting into C$^{18}$O line, Column (11): Systemic velocity obtained by Gaussian fitting to C$^{18}$O line, roughly to $^{12}$CO, or from nearby sources, Column (12): Description of line asymmetry: B: blue asymmetry, R: red asymmetry, mc: multiple velocity components, Column (13): Description of wing features in $^{12}$CO spectrum: Y: Existence of wing feature, Column (14): Telescope used for mapping observations, Column (15): Transition of molecular line used for mapping observation, Column (16): Mapping size, Column (17): Description of outflow features in $^{12}$CO mapping data: Y: Clear blue- or red-shifted outflow lobe feature, NY: Unclear outflow lobe feature and no YSOs within 100 arcsec radius based on SIMBAD astronomical database, C: Unclear outflow lobe feature and possible YSOs within 100 arcsec radius based on SIMBAD astronomical database, N: No outflow, Column (6)-(10), (14)-(17): `{\nodata}' indicates lack of available data, Column (12)-(13): `{\nodata}' indicates none of the features or lack of available data.
\end{minipage}%
\end{deluxetable}  
\clearpage  
\begin{deluxetable}{ccCcCCCCC}
\tablecaption{Parameters for Outflow Identification in 16 VeLLOs \label{tbl:outflow1}}
\tabletypesize{\scriptsize}
\tablewidth{0pt}
\tablehead{
\colhead{Source} & 
\multicolumn{2}{c}{Blue} & \multicolumn{2}{c}{Red} & 
\multicolumn{2}{c}{Blue} & \multicolumn{2}{c}{Red} \\ 
\cmidrule(lr){2-3} \cmidrule(lr){4-5} 
\cmidrule(lr){6-7} \cmidrule(lr){8-9} 
\colhead{} &
\colhead{$v_{max}$} & \colhead{$v_{min}$} & 
\colhead{$v_{min}$} & \colhead{$v_{max}$} &  
\colhead{$C_{\rm min}$} & \colhead{$C_{\rm step}$} & 
\colhead{$C_{\rm min}$} & \colhead{$C_{\rm step}$} \\
\cmidrule(lr){2-5}\cmidrule(lr){6-9}
\colhead{} & 
\multicolumn{4}{c}{(km s$^{-1}$)} & \multicolumn{4}{c}{(K km s$^{-1}$)} }
\colnumbers
\startdata
J033032 & -0.7 & 4.3 & 8.5 & 9.7 & 2.31 & 0.64 & 0.34 & 0.06 \\
J041412 & 2.7 & 5.8 & 8.1 & 10.2 & 2.07 & 0.61 & 1.34 & 0.52 \\
IRAM 04191+1522 & -2.0 & 5.6 & 7.2 & 13.3 & 3.47 & 3.02 & 2.24 & 2.46 \\
L1521F-IRS & 3.1 & 4.8 & 8.5 & 10.1 & 0.55 & 0.04 & 0.29 & 0.06 \\
J154216 & -6.7 & -1.6 & 1.1 & 10.0 & 1.08 & 0.99 & 0.97 & 0.38 \\
J160115 & 1.3 & 3.1 & 5.2 & 6.4 & 0.77 & 0.07 & 0.31 & 0.04 \\
J162145 & -6.9 & 1.6 & 3.7 & 11.1 & 1.25 & 2.34 & 3.21 & 2.61 \\
IRAS16253-2429 & -0.9 & 0.7 & 5.3 & 9.2 & 0.58 & 0.17 & 5.43 & 1.67 \\
CB130-3-IRS & 2.3 & 5.1 & 11.6 & 14.3 & 1.27 & 0.18 & 0.71 & 0.09 \\
L328-IRS & 4.9 & 5.7 & 7.4 & 8.3 & 0.42 & 0.10 & 0.47 & 0.14 \\
J182920 & 2.5 & 5.7 & 10.0 & 13.5 & 2.58 & 0.24 & 1.50 & 0.50 \\
J182943 & \nodata & \nodata & 9.2 & 14.1 & \nodata & \nodata & 1.65 & 0.32 \\
J183237 & 2.9 & 4.5 & 10.1 & 11.2 & 0.50 & 0.07 & 0.49 & 0.09 \\
L673-7-IRS & 2.0 & 5.0 & 9.5 & 13.6 & 0.46 & 0.35 & 1.25 & 0.76 \\
J210221 & -6.7 & 2.0 & 3.6 & 10.1 & 8.43 & 4.36 & 7.17 & 3.60 \\
L1251A-IRS4 & -8.8 & -6.4 & -2.6 & -1.3 & 0.52 & 0.10 & 1.37 & 0.13 \\
\enddata
\tablecomments{Column (1): Source name, Column (2): Maximum velocity of $^{12}$CO line for blue-shifted outflow lobe, Column (3): Minimum velocity of $^{12}$CO line for blue-shifted outflow lobe, Column (4): Minimum velocity of $^{12}$CO line for red-shifted outflow lobe, Column (5): Maximum velocity of $^{12}$CO line for red-shifted outflow lobe, Column (6): Minimum contour level of blue-shifted outflow lobe in $^{12}$CO contour map, Column (7): Interval contour level of blue-shifted outflow lobe in $^{12}$CO contour map, Column (8): Minimum contour level of red-shifted outflow lobe in $^{12}$CO contour map, Column (9): Interval contour level of red-shifted outflow lobe in $^{12}$CO contour map.}
\end{deluxetable}
\clearpage  
\begin{deluxetable}{cCCCCCCCc}
\tablecaption{Outflow Properties and Classification of 19 VeLLOs \label{tbl:outflow4}}
\tabletypesize{\scriptsize}
\tablewidth{0pt}
\tablehead{
\colhead{Source} & \colhead{Inc.} & \colhead{$t_{\rm tot}$} & \colhead{F} &
\colhead{$\dot{M}_{\rm acc}$} 
& \colhead{$M_{\rm acc}$} & \colhead{$L_{\rm acc}$} & \colhead{$\frac{L_{\rm acc}}{L_{\rm int}}$} & \colhead{Class.}\\
\colhead{} & \colhead{({\arcdeg})} & \colhead{($10^{4}$ yr)} & \colhead{($10^{-6}$ {\msun} km s$^{-1}$ yr$^{-1}$)} & 
\colhead{($10^{-6}$ {\msun} yr$^{-1}$)} & \colhead{({\msun})} & \colhead{({\lsun})}
& \colhead{} & \colhead{}  
}
\colnumbers
\startdata
J033032 & 57.3 & 3 & 8\pm1 & 2\pm1 & \geq0.07\pm0.01 & \geq0.9\pm0.1 & \geq6 & Protostar \\
J041412 & 10.0 & 7 & 0.2\pm0.1 & 0.05\pm0.01 & 0.004\pm0.001 & 0.001\pm0.001 & 0.02 & Proto-BD \\
IRAM 04191+1522 & 50.0 & 7 & 9\pm1 & 2\pm1 & 0.2\pm0.1 & 2\pm1 & 69 & Protostar \\
L1521F-IRS & 30.0 & 1 & 0.04\pm0.01 & 0.01\pm0.01 & 0.0001\pm0.0001 & 0.000006\pm0.000001 & 0.0002 & Proto-BD \\
J154216 & 57.3 & 2 & 3\pm1 & 0.9\pm0.1 & 0.02\pm0.01 & 0.09\pm0.01 & 2 & Proto-BD \\
J160115 & 57.3 & 0.3 & 0.2\pm0.1 & 0.06\pm0.01 & 0.0002\pm0.0001 & 0.00006\pm0.00001 & 0.0006 & Proto-BD \\
J162145 & 10.0 & 2 & 0.4\pm0.1 & 0.1\pm0.1 & 0.003\pm0.001 & 0.002\pm0.001 & 0.02 & Proto-BD \\
IRAS16253-2429 & 57.3 & 1 & 5\pm1 & 1\pm1 & \geq0.02\pm0.01 & \geq0.1\pm0.1 & \geq1 & Proto-BD \\
CB130-3-IRS & 57.3 & 1 & 3\pm1 & 0.9\pm0.1 & 0.01\pm0.01 & 0.06\pm0.01 & 0.9 & Proto-BD \\
L328-IRS & 66.0 & 7 & 3\pm1 & 0.8\pm0.1 & 0.06\pm0.01 & 0.2\pm0.1 & 1 & Proto-BD \\
J182920 & 10.0 & 5 & 0.3\pm0.1 & 0.09\pm0.02 & 0.005\pm0.001 & 0.002\pm0.001 & 0.02 & Proto-BD \\
J182943 & 57.3 & 1 & 3\pm1 & 0.9\pm0.1 & 0.010\pm0.001 & 0.05\pm0.01 & 0.5 & Proto-BD \\
J183237 & 57.3 & 0.5 & 1\pm1 & 0.3\pm0.1 & \geq0.002\pm0.001 & \geq0.003\pm0.001 & \geq0.01 & Proto-BD \\
L673-7-IRS & 57.3 & 7 & 6\pm1 & 1\pm1 & 0.1\pm0.1 & 1\pm1 & 30 & Protostar \\
J210221 & 57.3 & 10 & 40\pm7 & 10\pm1 & \geq1\pm1 & \geq65\pm12 & \geq283 & Protostar \\
L1251A-IRS4 & 57.3 & 4 & 2\pm1 & 0.6\pm0.1 & 0.03\pm0.01 & 0.08\pm0.01 & 0.4 & Proto-BD \\
IC348-SMM2E & 57.3 & 0.06 & 0.3\pm0.1 & 0.08\pm0.01 & 0.00005\pm0.00001 & 0.00002\pm0.00001 & 0.0004 & Proto-BD \\
L1148-IRS & 57.3 & 0.9 & 3\pm1 & 0.8\pm0.1 & 0.007\pm0.001 & 0.03\pm0.01 & 0.2 & Proto-BD \\
L1014-IRS & 57.3 & 0.07 & 0.9\pm0.1 & 0.2\pm0.1 & 0.0002\pm0.0001 & 0.0002\pm0.0001 & 0.002 & Proto-BD \\
\enddata
\begin{minipage}{19.0cm}~\\\\\\
Note $-$ 
Column (1): Source name, Column (2): inclination of outflow as the separation angle between outflow axis and the line-of-sight, Column (3): Total dynamical timescale of the entire outflow, Column (4): Outflow force, Column (5): Mass accretion rate of a central object inferred from outflow, Column (6): Accreted mass into a central object over a dynamical time inferred from outflow, Column (7): Accretion luminosity of a central object over a dynamical time inferred from outflow, Column (8): Ratio of accretion luminosity to internal luminosity, Column (9): Classification of VeLLOs. All the quantities derived for the outflows are from this study except for those for L1014-IRS, L1148-IRS, and IC348-SMM2E which are from other studies \citep{Bourke:2005kg, Kauffmann:2011kz, Palau:2014kp}.
Inclinations for the outflows of L328-IRS, IRAM 04191$+$1522, and L1521F-IRS, are from the following references; \citet{Lee:2018iq}, \citet{Andre:1999tr}, and \citet{Takahashi:2013fm}, respectively. Inclinations for the outflows of J041412, J162145, and J182920 for which the blue and red lobes of outflow are almost overlapped are assumed to be 10{\arcdeg}. Inclinations for the remaining nine sources are assumed to 57{\arcdeg}.3 as a mean inclination angle for random outflow orientations. Note that physical quantities for the outflow of J182943 showing only the red component of the outflow were estimated by doubling those of the red component.
\end{minipage}
\end{deluxetable}

\begin{figure}
\includegraphics[angle=0,scale=1.0]{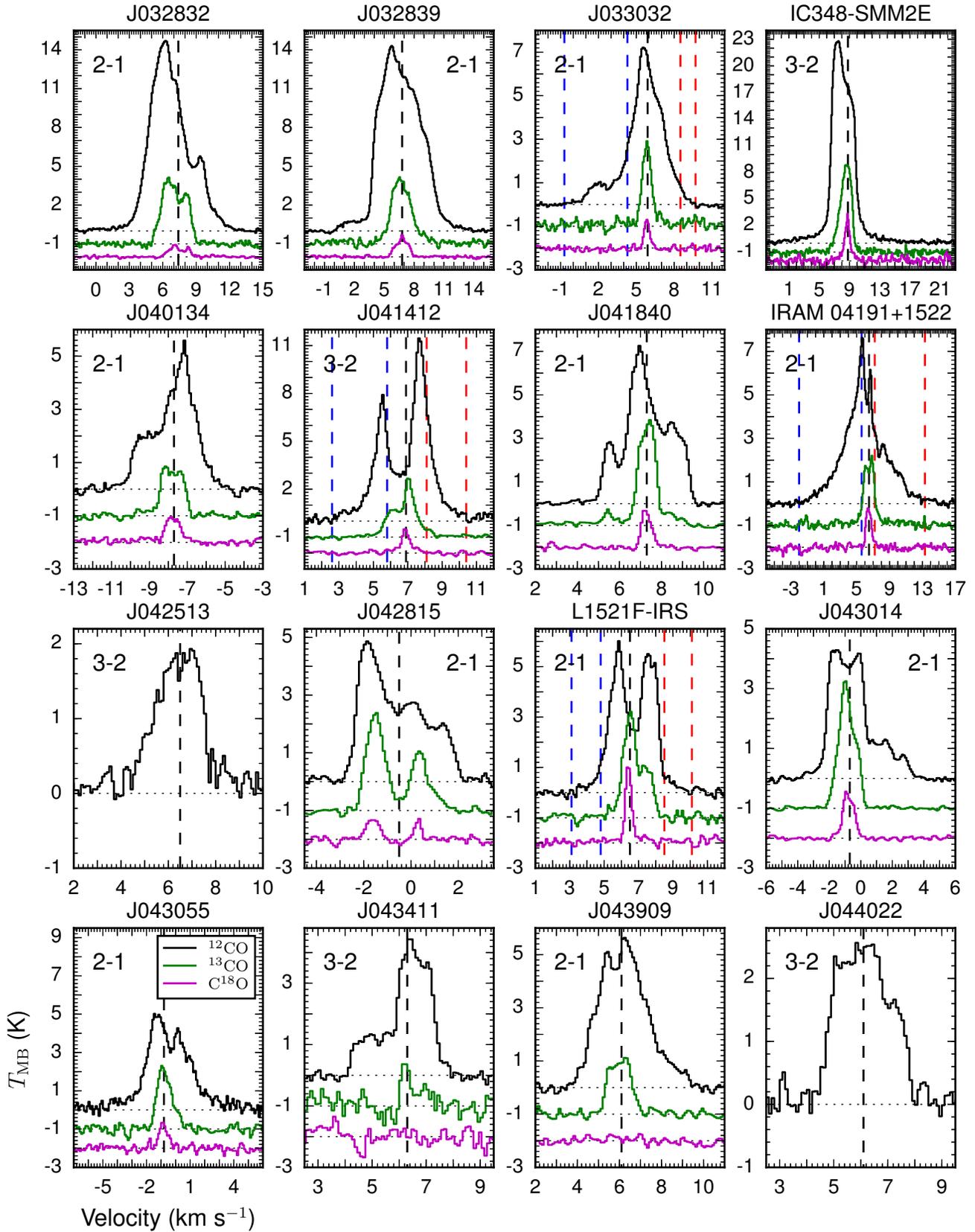} 
\caption{Spectra of 68 VeLLOs in the rotational transitions 2-1 or 3-2 of $^{12}$CO, $^{13}$CO, and C$^{18}$O. Each spectrum is from given coordinate of a VeLLO. A horizontal dotted line indicates the main beam temperature of 0 K. Black vertical line is to indicate the systemic velocity. Blue and red vertical lines indicate the velocity ranges of blue- and red-shifted outflow lobes found in mapping observation, respectively. X and Y axes correspond to a velocity and a main beam temperature, respectively. \label{fig:Lineprofile1}}
\end{figure}  
\setcounter{figure}{0}  
\begin{figure}
\includegraphics[angle=0,scale=1.0]{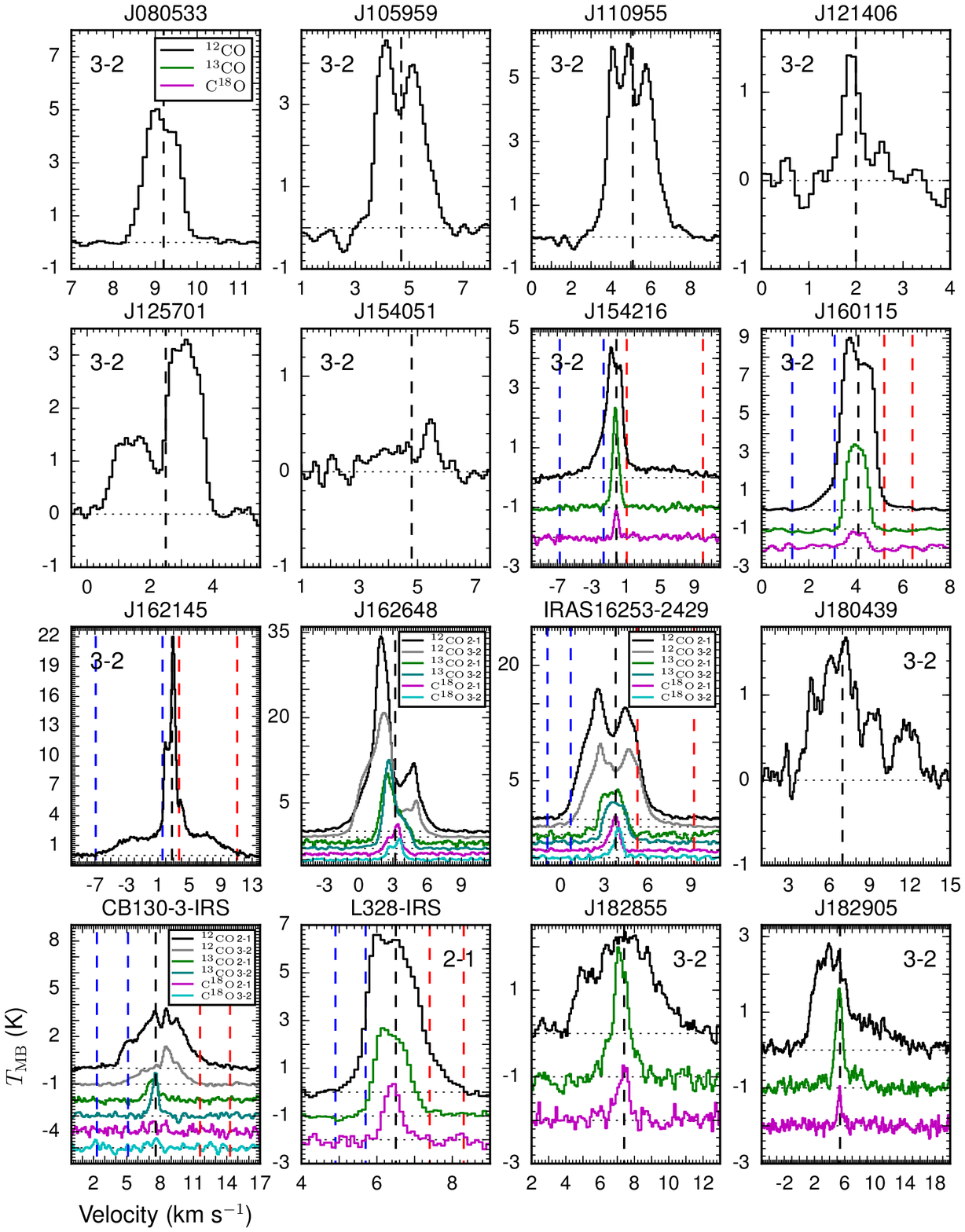} 
\caption{(Continued.) \label{fig:Lineprofile2}}
\end{figure}  
\setcounter{figure}{0}  
\begin{figure}
\includegraphics[angle=0,scale=1.0]{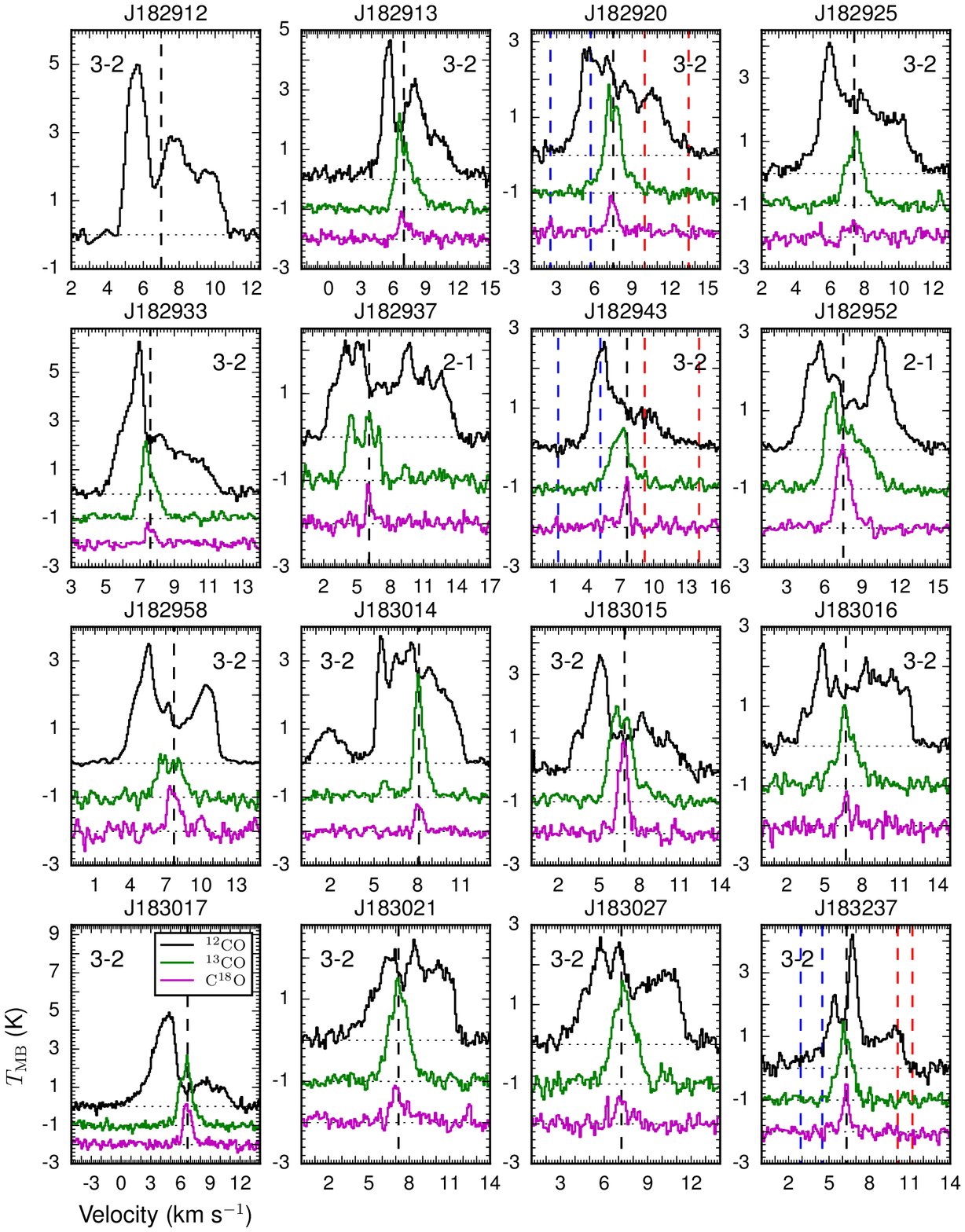} 
\caption{(Continued.) \label{fig:Lineprofile3}}
\end{figure}  
\setcounter{figure}{0}  
\begin{figure}
\includegraphics[angle=0,scale=1.0]{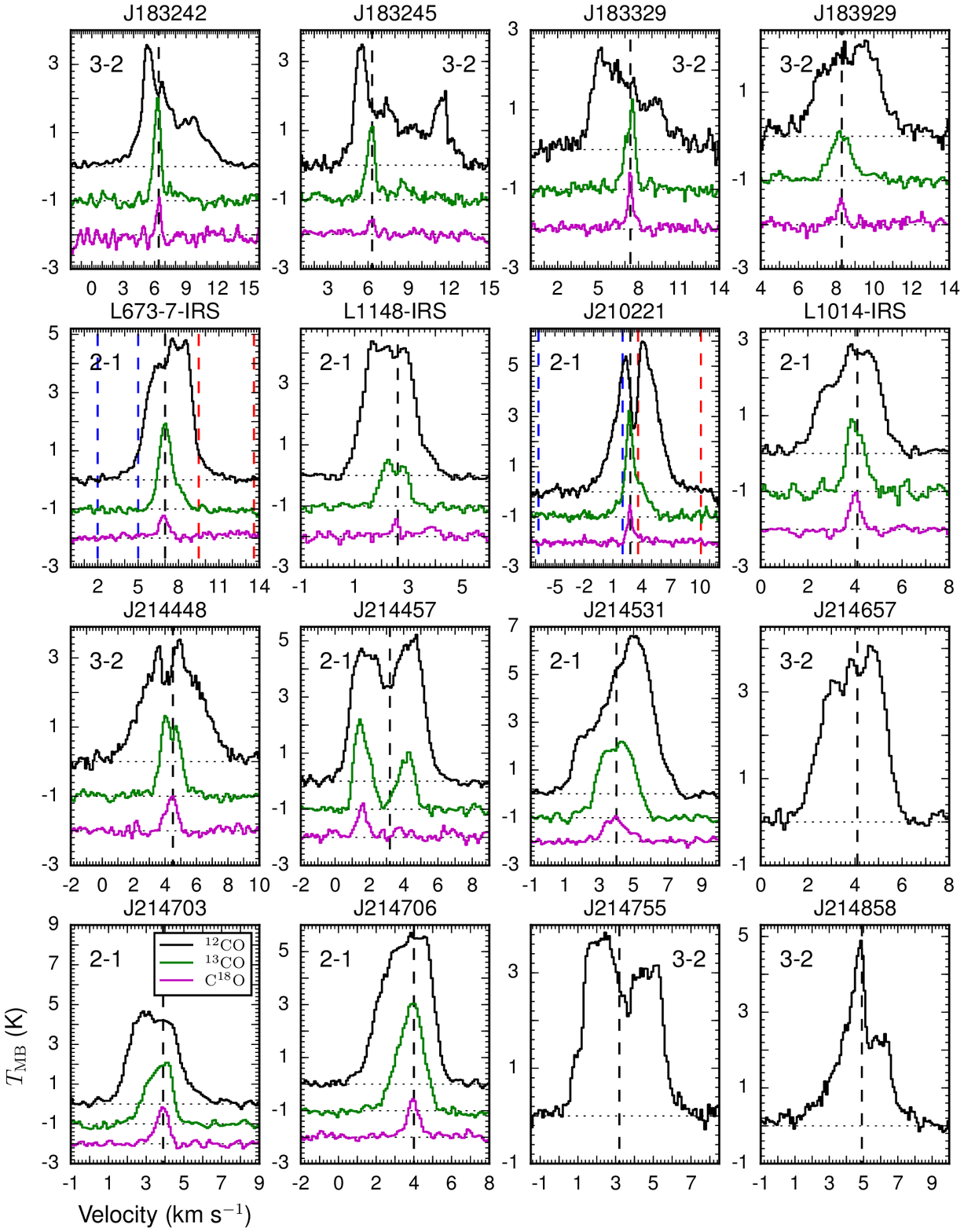} 
\caption{(Continued.) \label{fig:Lineprofile4}}
\end{figure}  
\setcounter{figure}{0}  
\begin{figure}
\includegraphics[angle=0,scale=1.0]{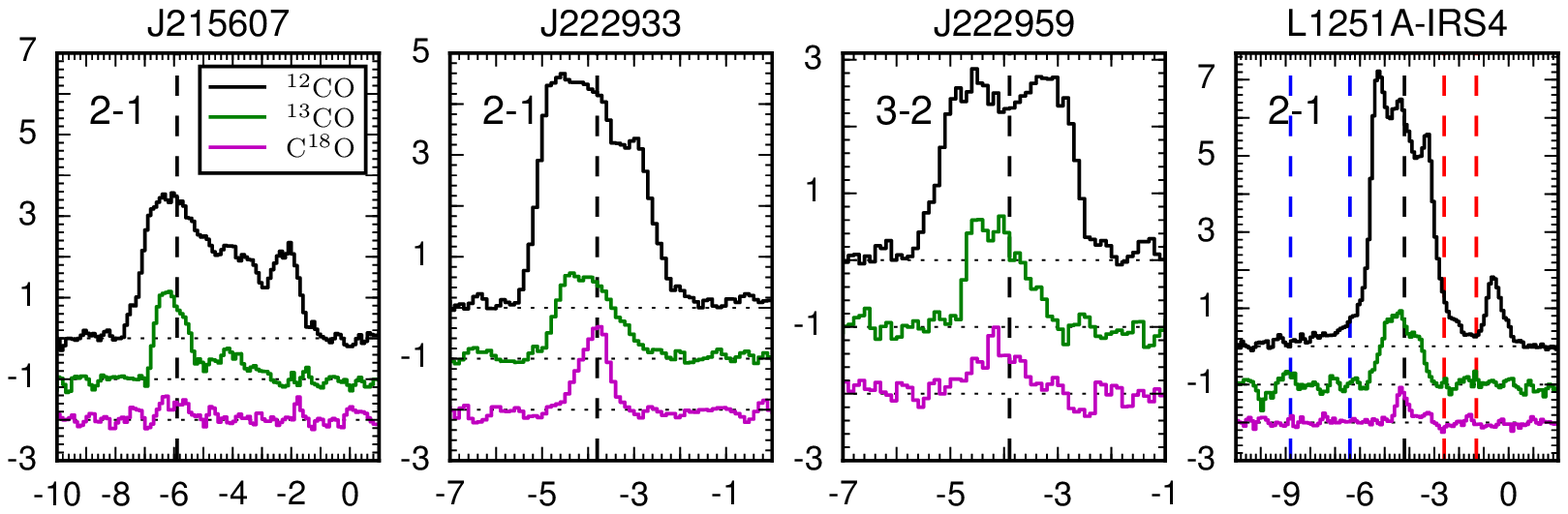} 
\caption{(Continued.) \label{fig:Lineprofile5}}
\end{figure}  
\setcounter{figure}{1}  
\begin{figure}
\includegraphics[angle=0,scale=0.2]{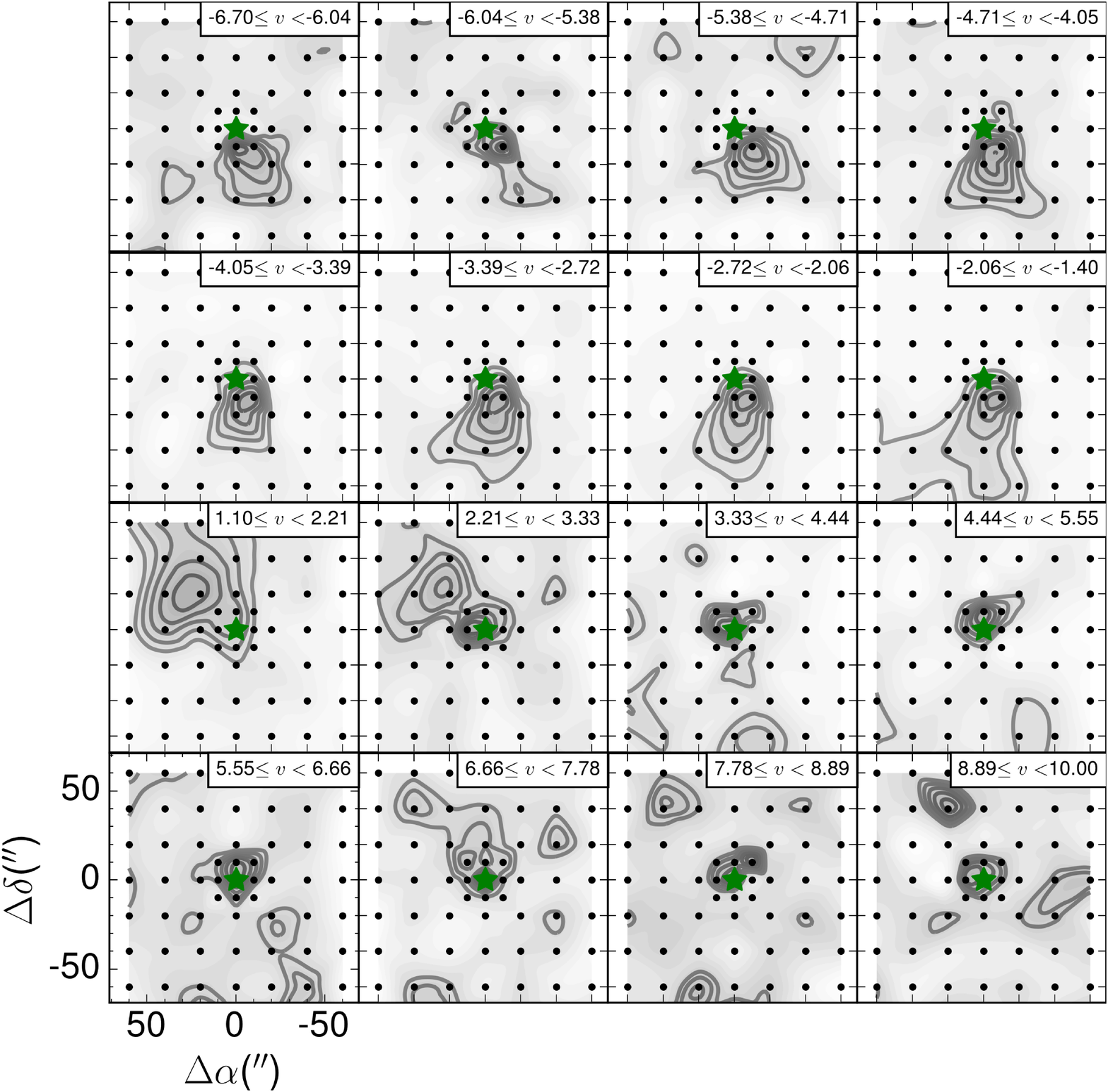} 
\caption{Channel map of J154216 as an example of an identification of CO outflow in a VeLLO. Each panel shows an integrated intensity map in contour and grey scale from velocity interval listed at the top of each panel. In this Figure, the velocity intervals of 0.66 and 1.11 km s$^{-1}$ were used for publication, but in real visual inspection, the velocity interval was assigned to 0.1 km s$^{-1}$. Black dots indicate positions for mapping observations. Star symbol indicates a position of VeLLO. \label{fig:chmap}}
\end{figure}  
\begin{figure}
$\begin{array}{cccc}
\includegraphics[angle=0,scale=0.065]{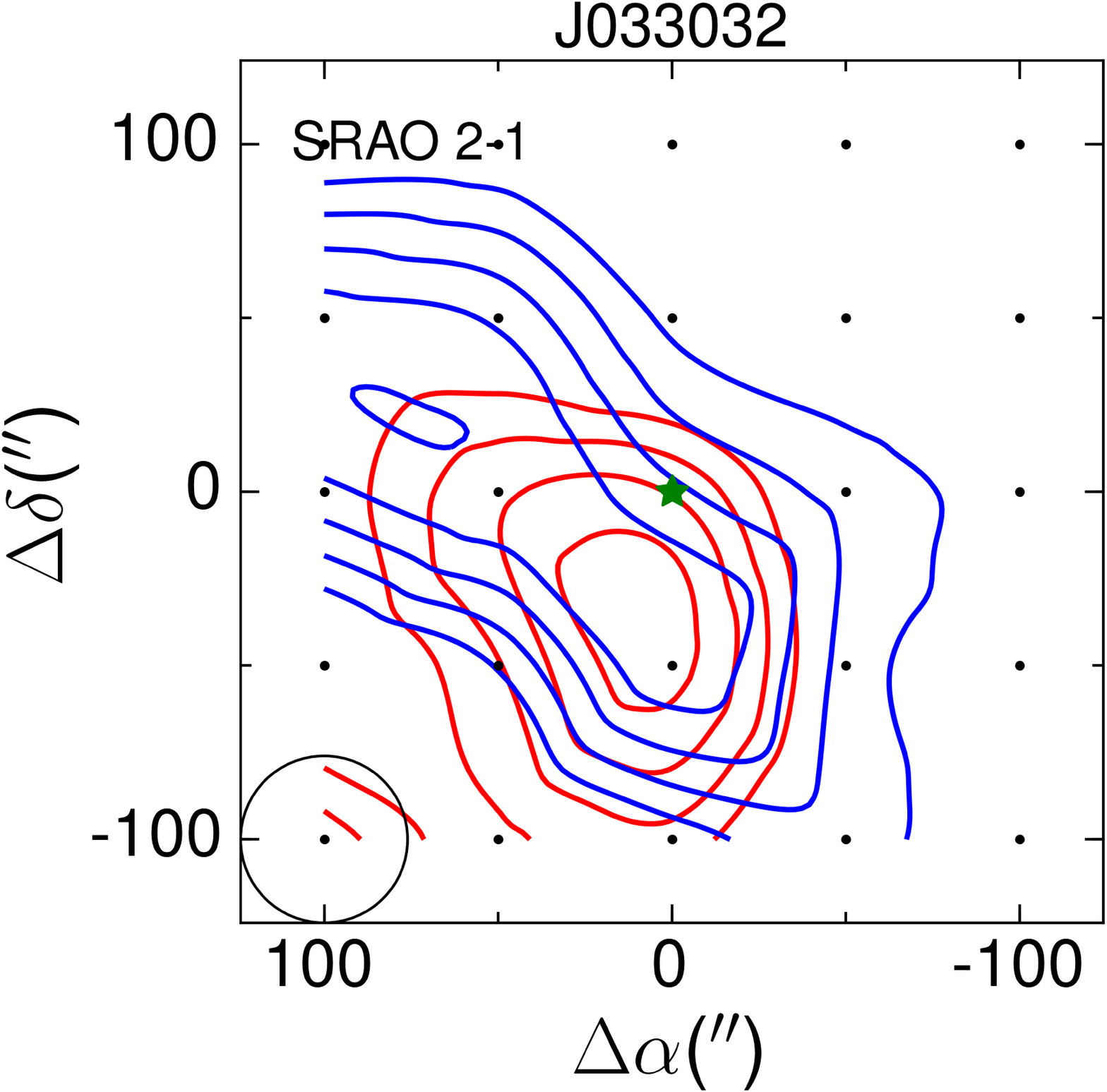} &
\includegraphics[angle=0,scale=0.065]{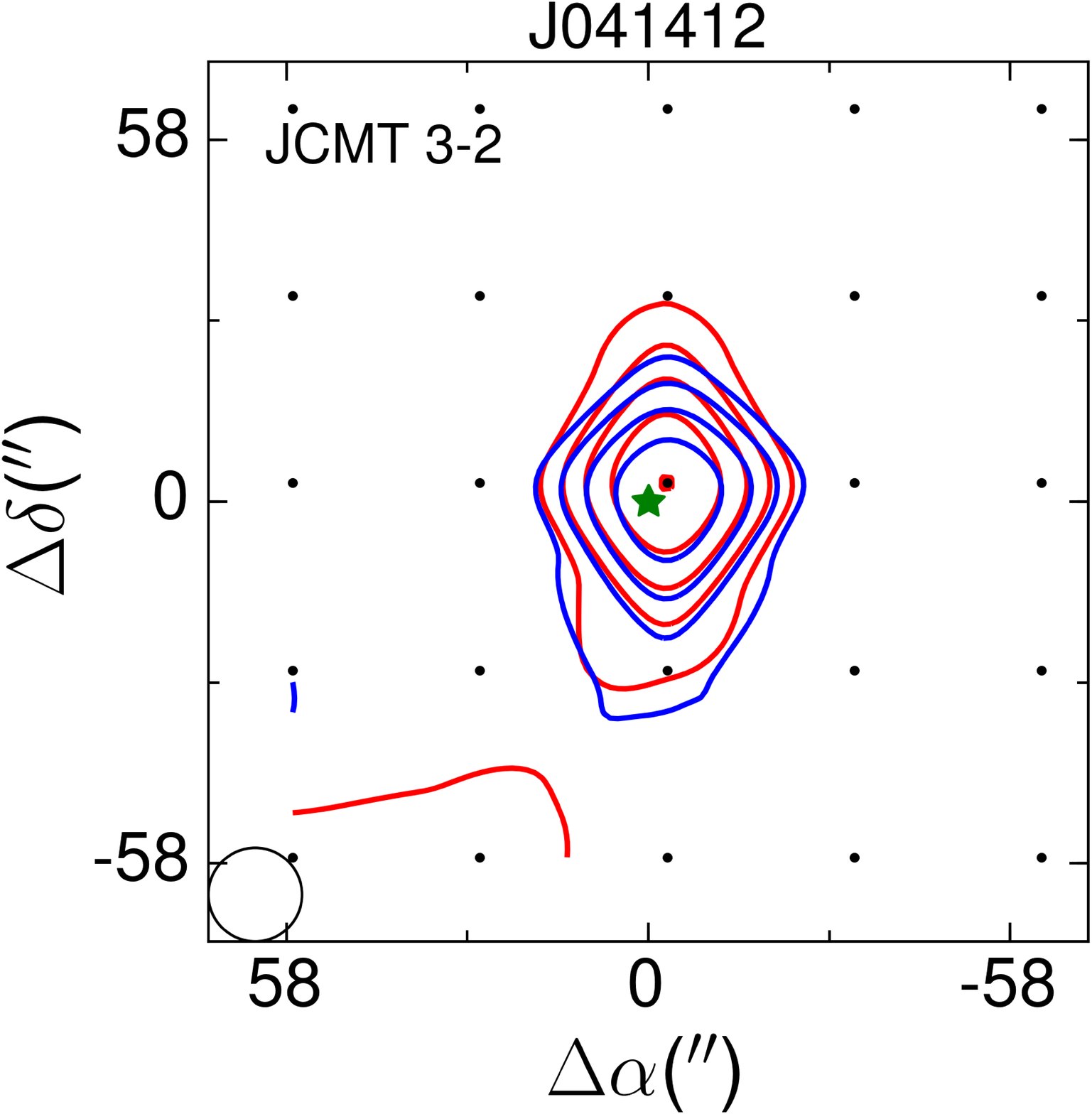} &
\includegraphics[angle=0,scale=0.065]{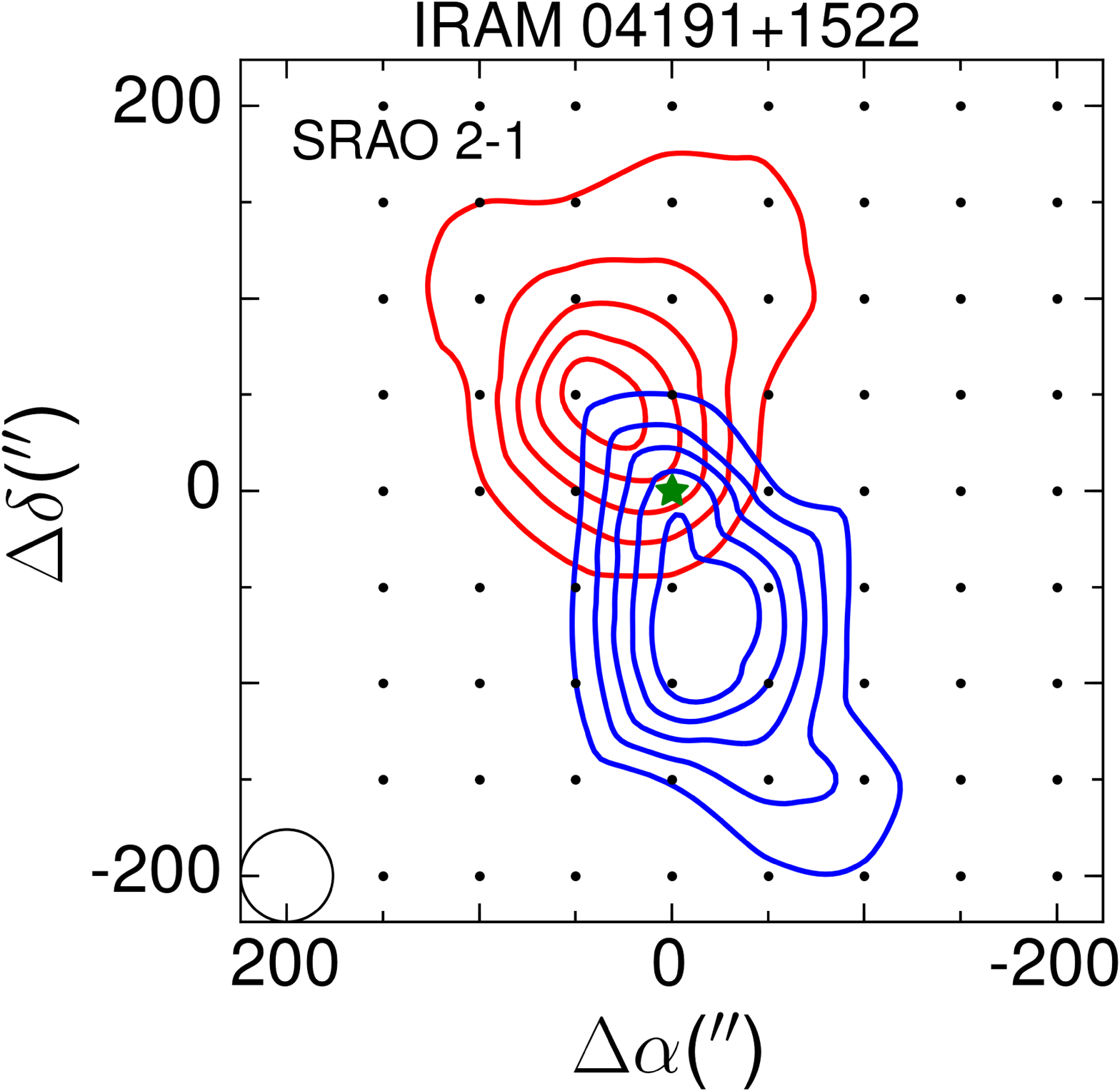} &
\includegraphics[angle=0,scale=0.065]{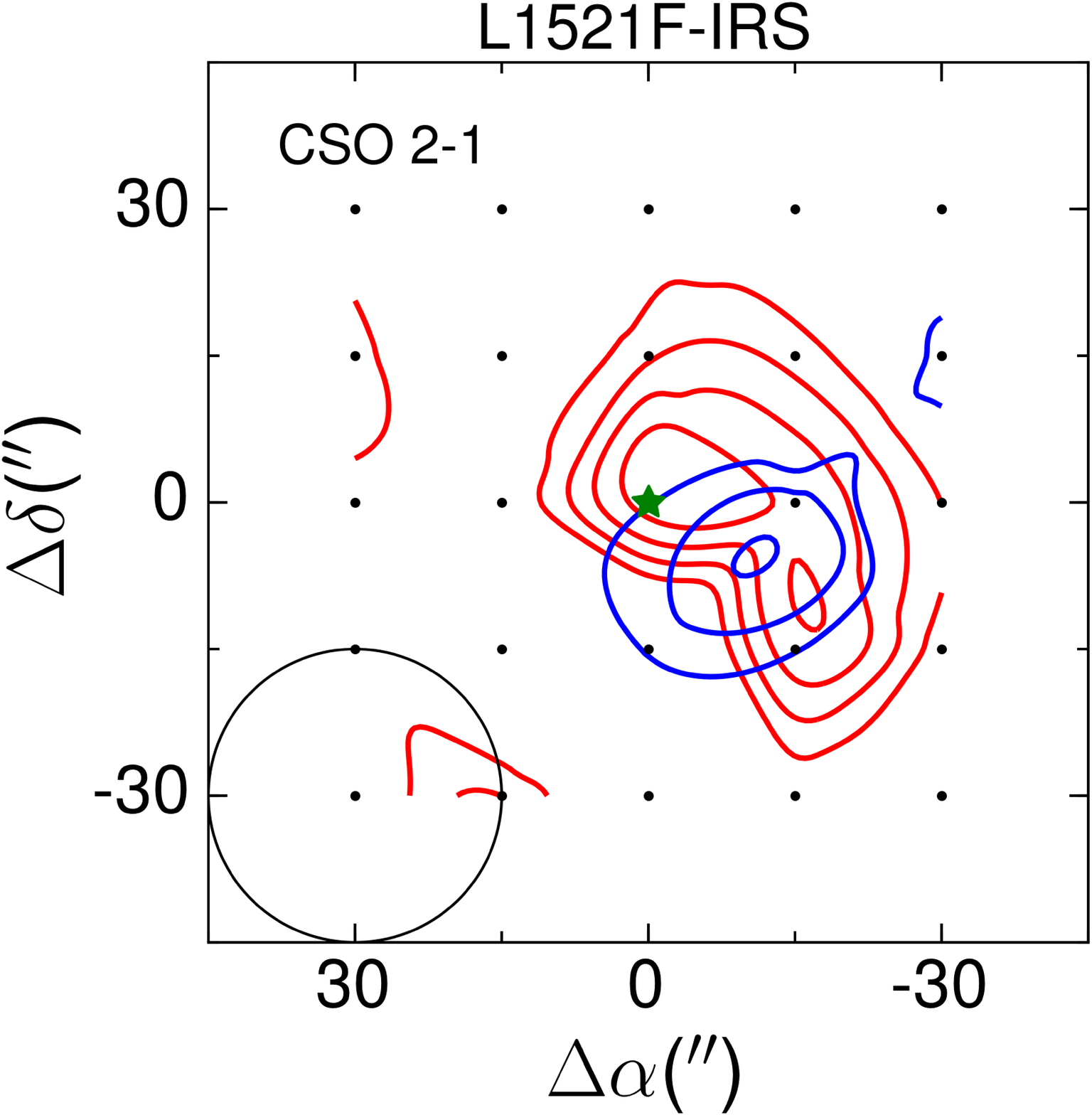} \\
\includegraphics[angle=0,scale=0.065]{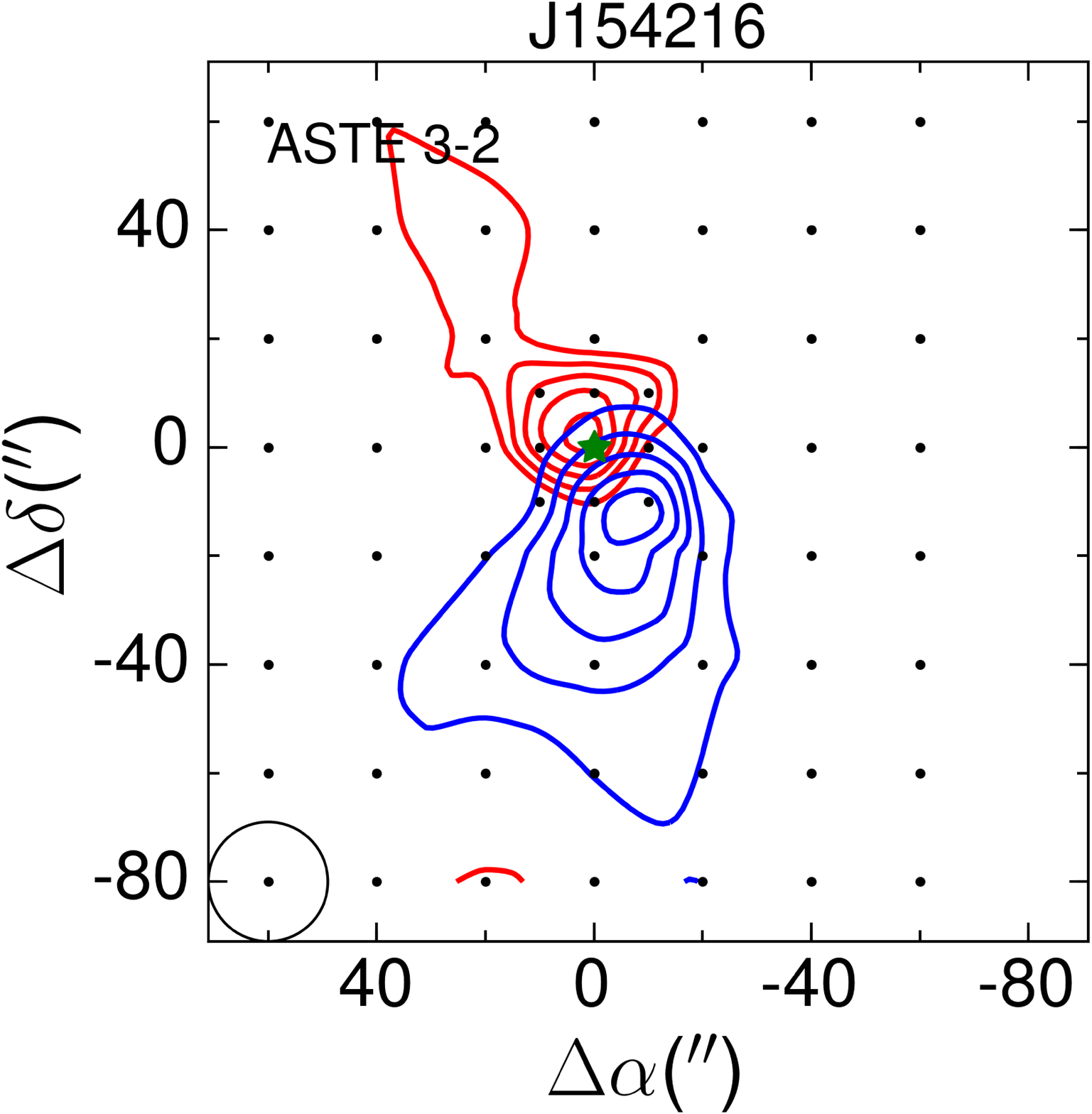} &
\includegraphics[angle=0,scale=0.065]{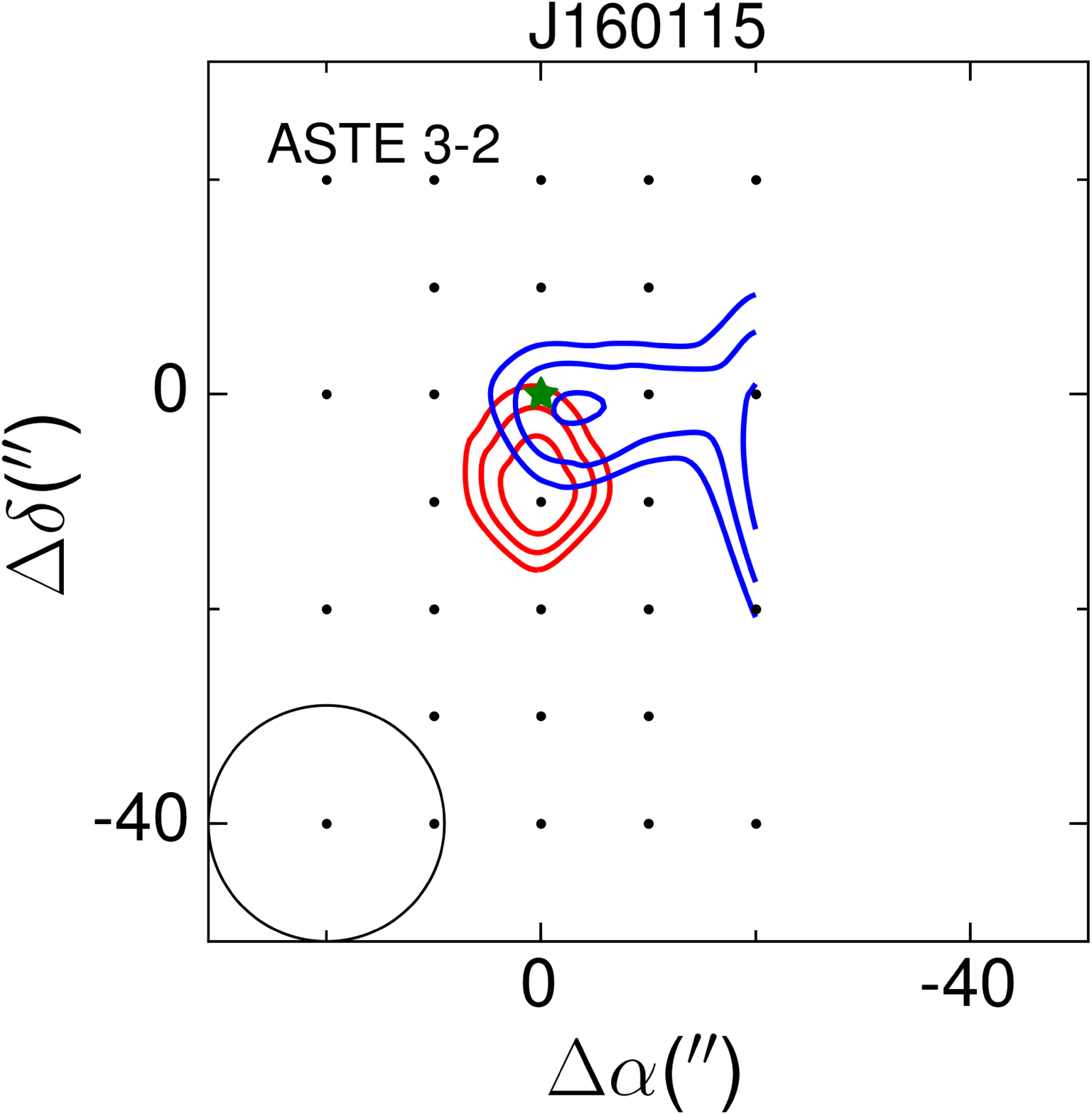} &
\includegraphics[angle=0,scale=0.065]{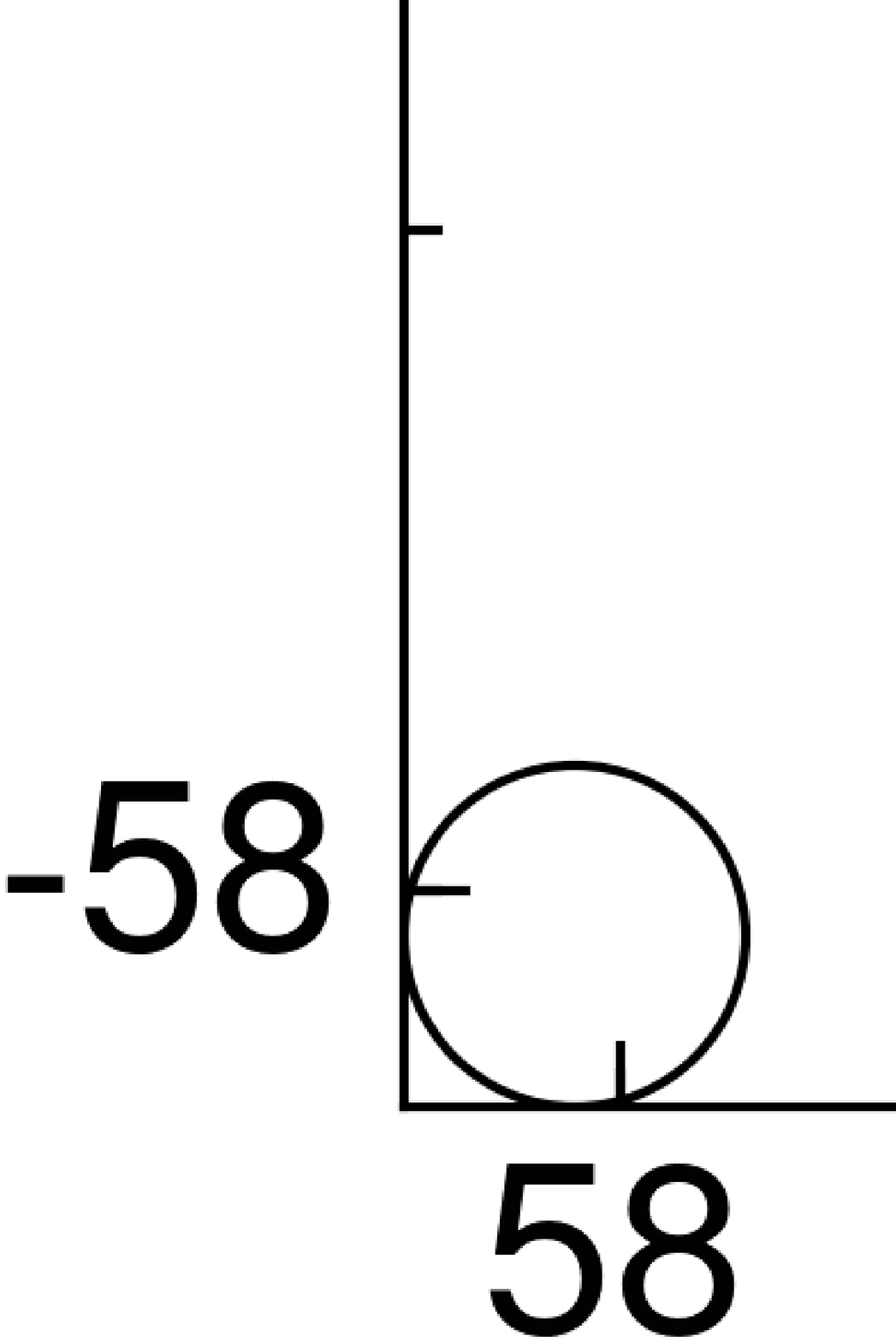} &
\includegraphics[angle=0,scale=0.065]{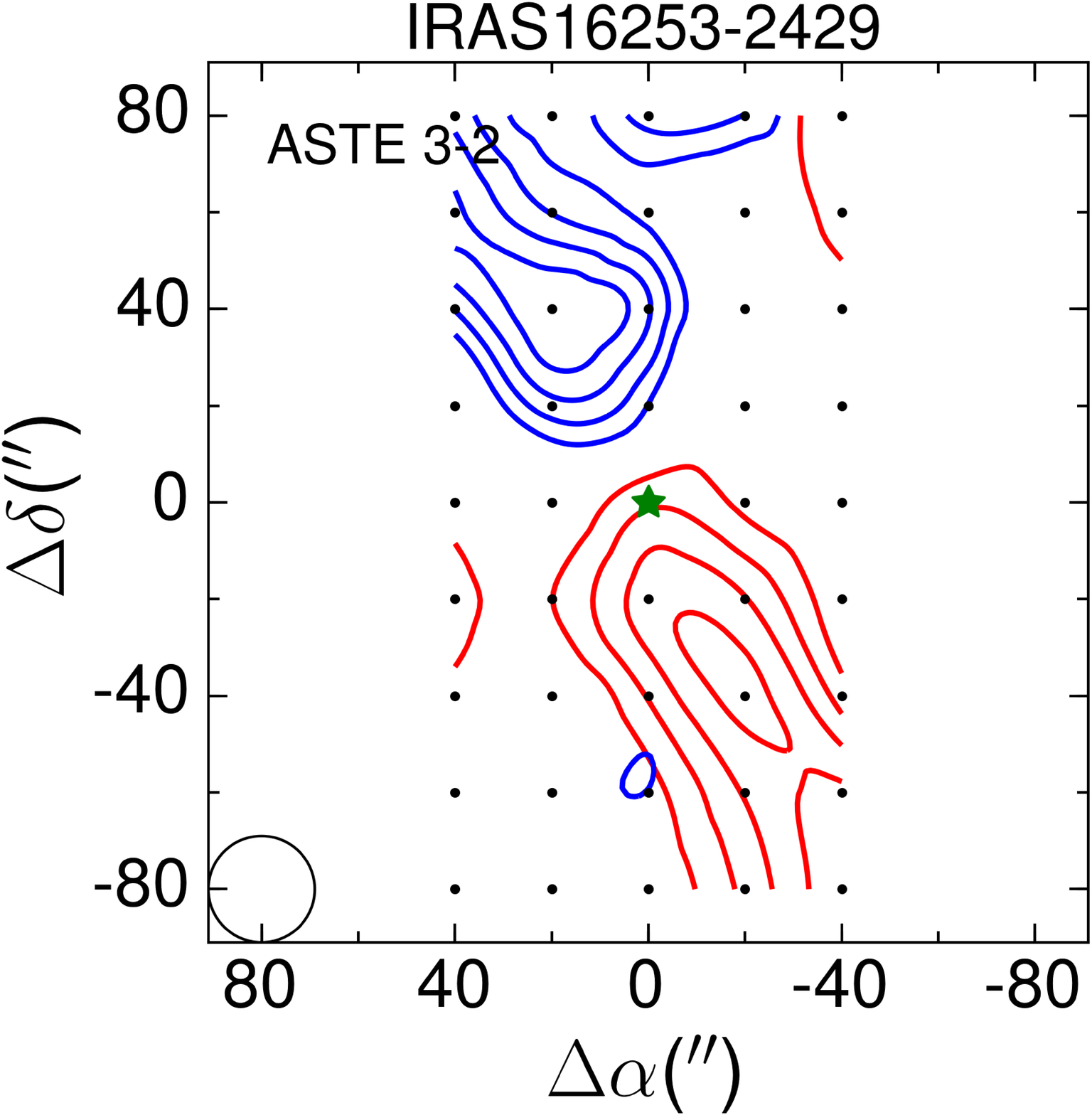} \\
\includegraphics[angle=0,scale=0.065]{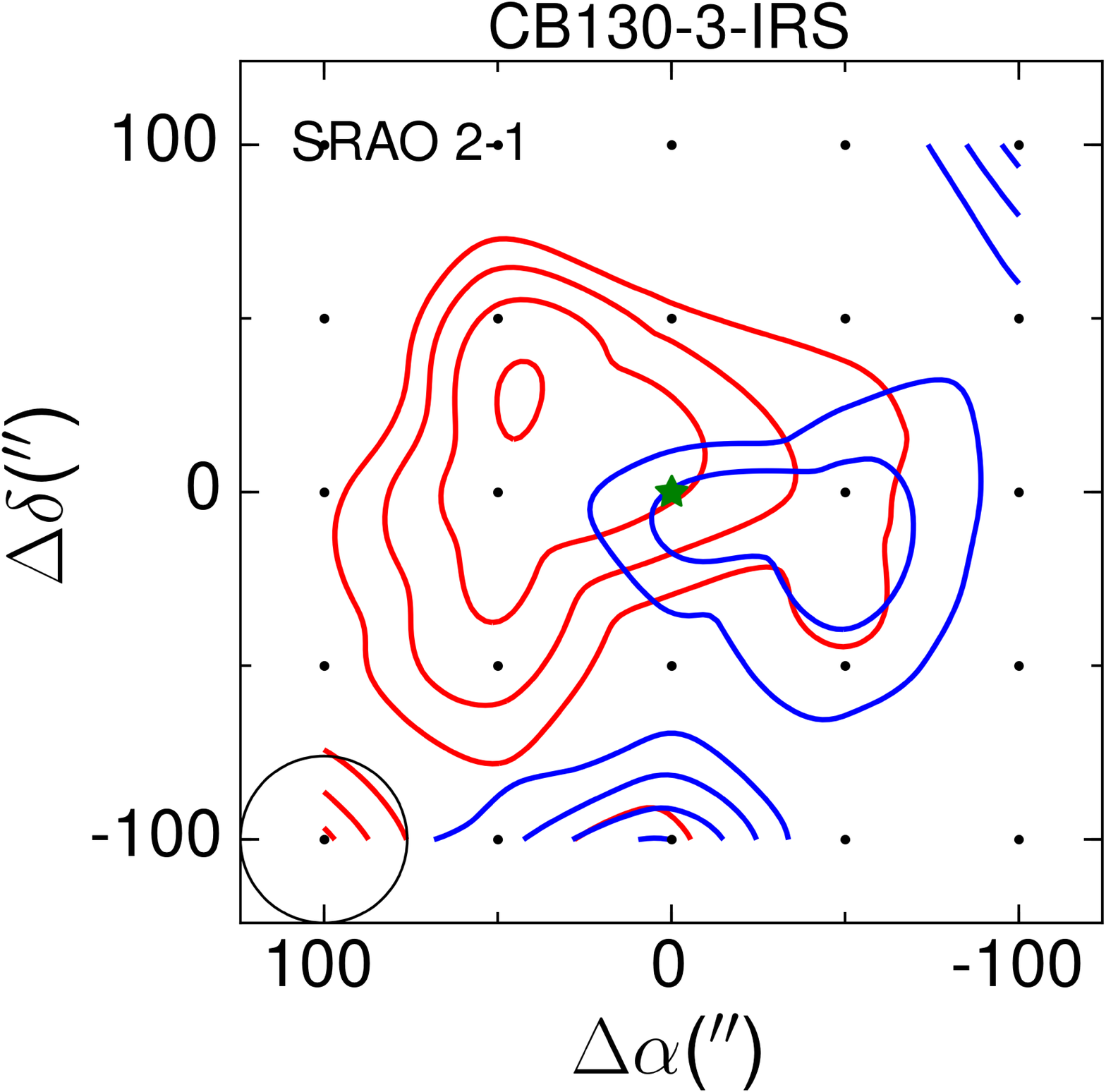} &
\includegraphics[angle=0,scale=0.065]{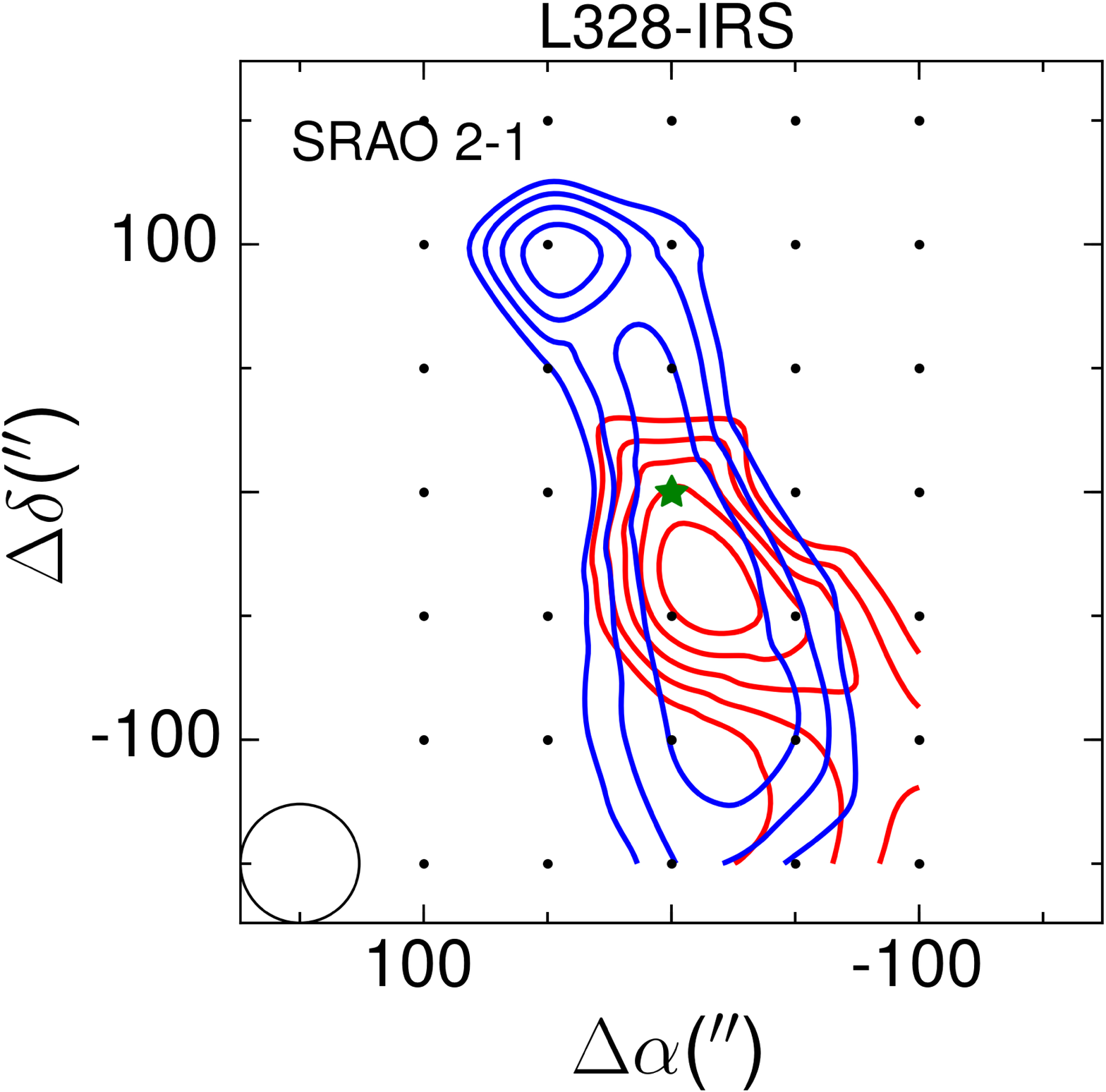} &
\includegraphics[angle=0,scale=0.065]{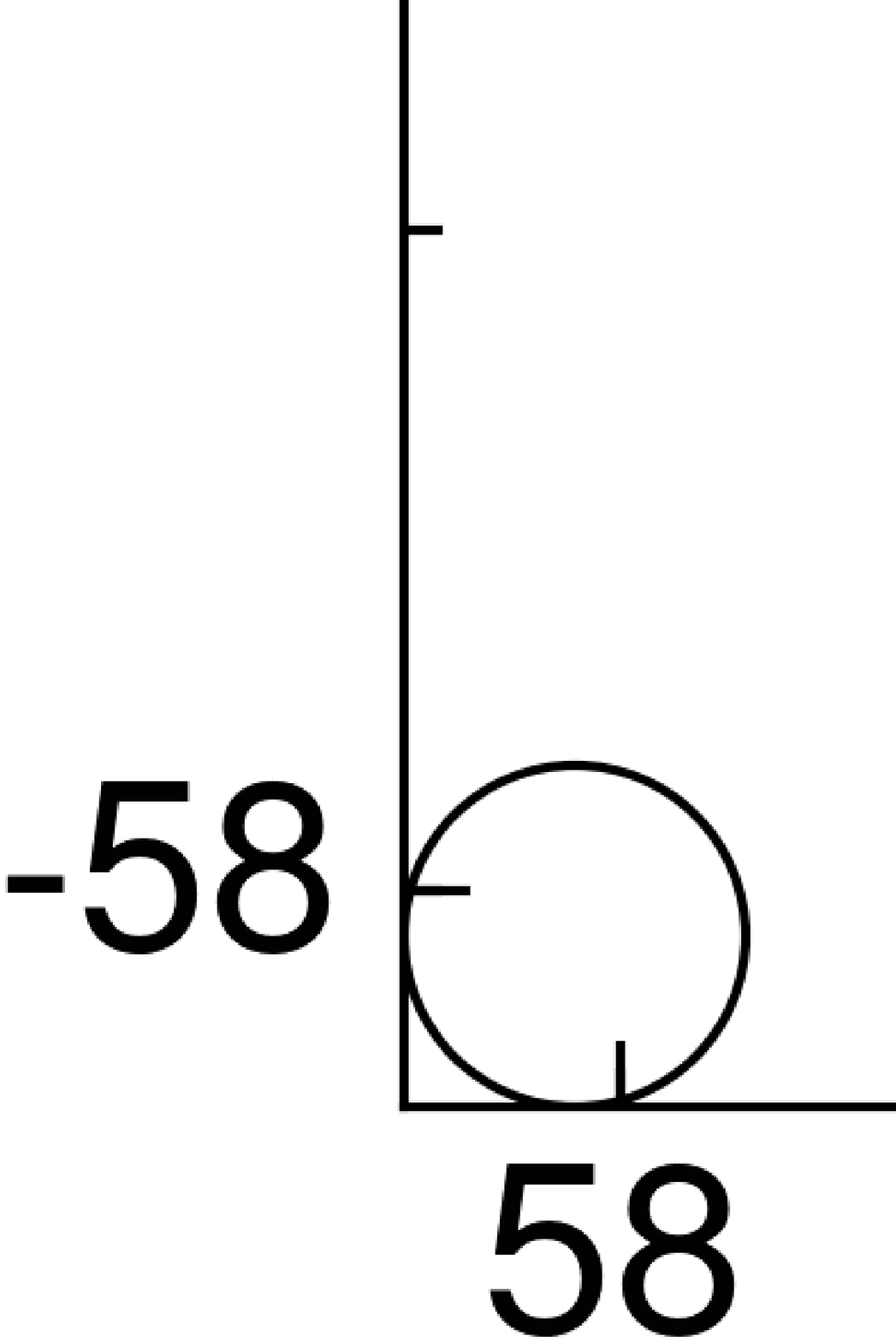} &
\includegraphics[angle=0,scale=0.065]{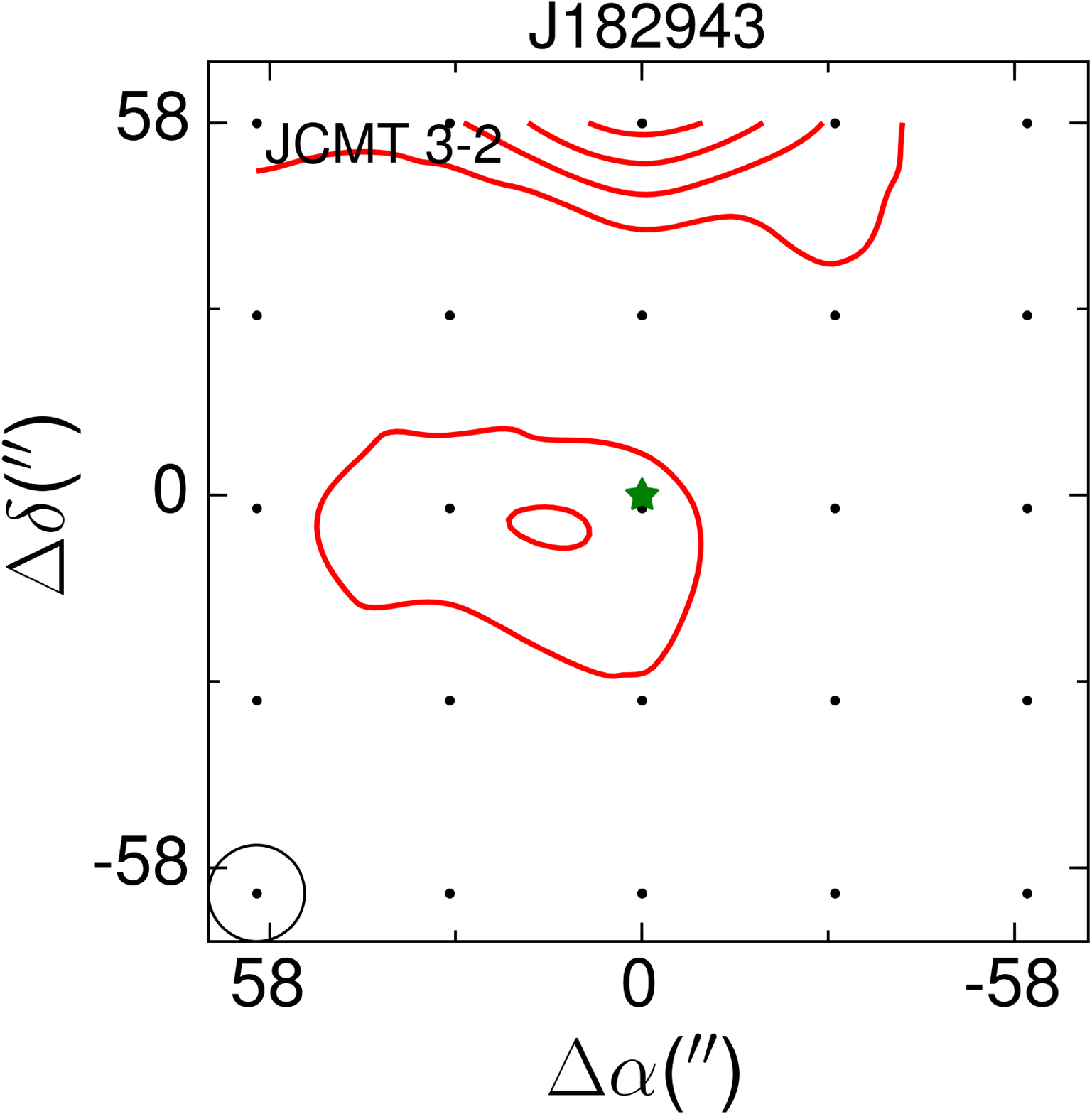} \\
\includegraphics[angle=0,scale=0.065]{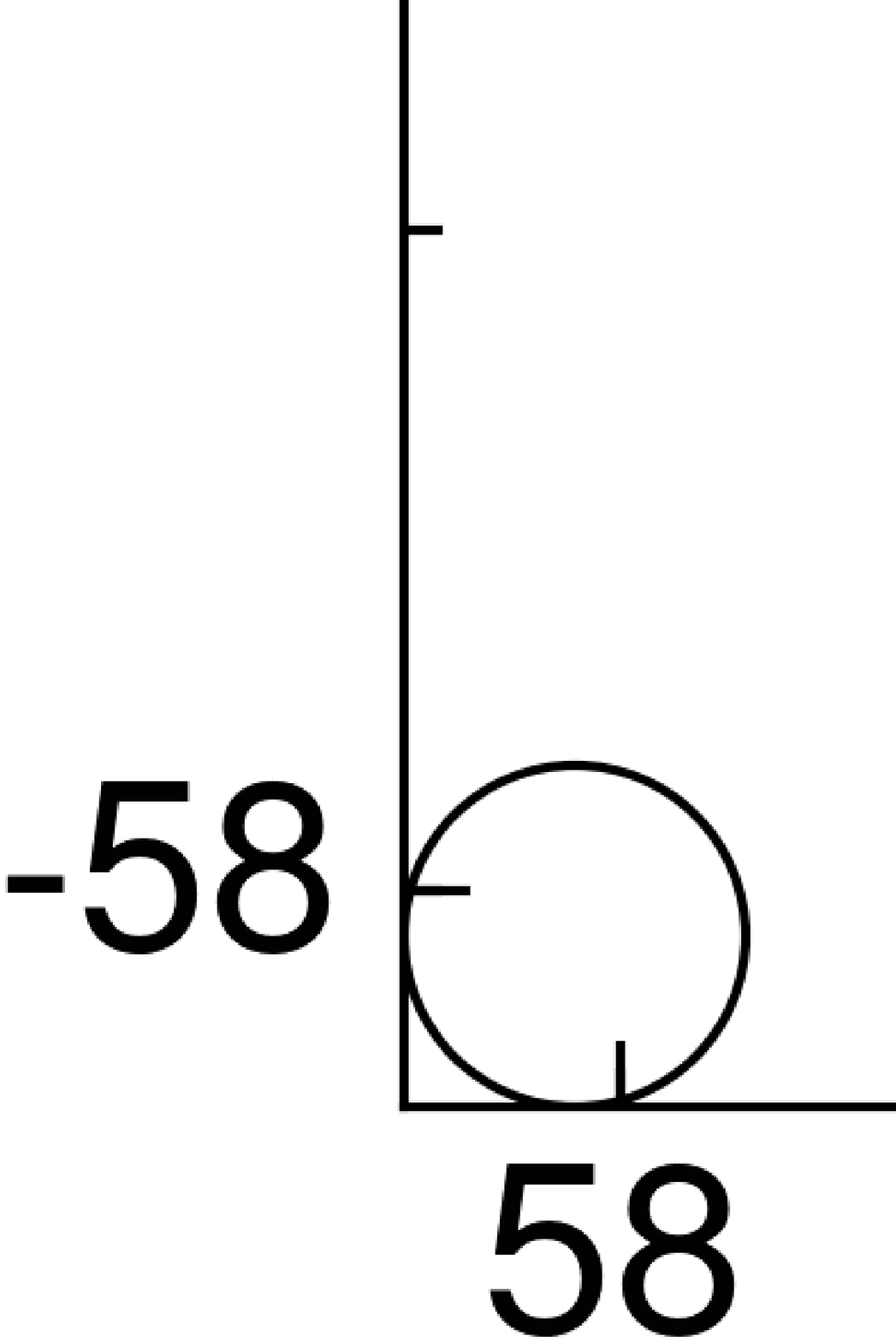} &
\includegraphics[angle=0,scale=0.065]{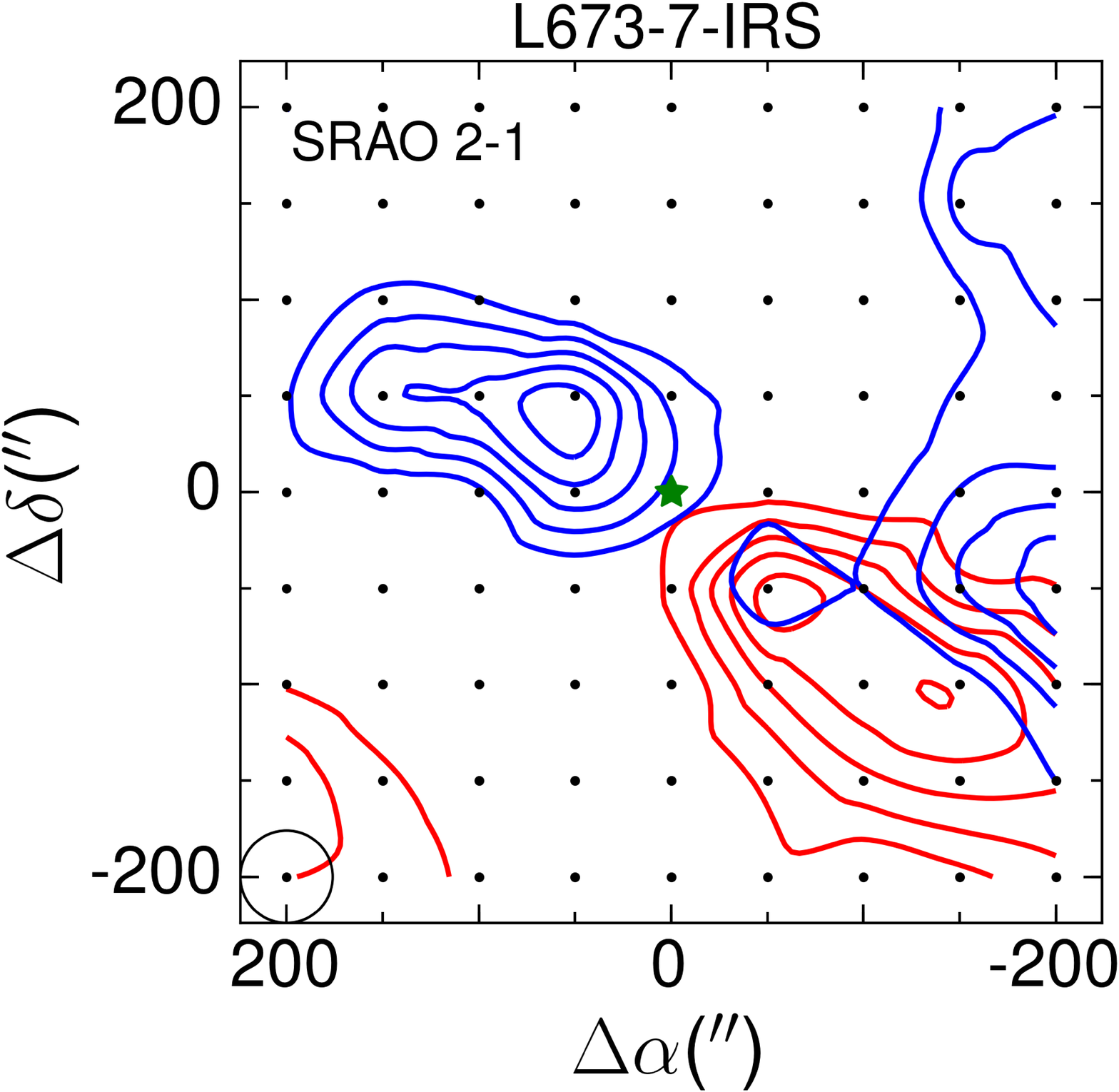} &
\includegraphics[angle=0,scale=0.065]{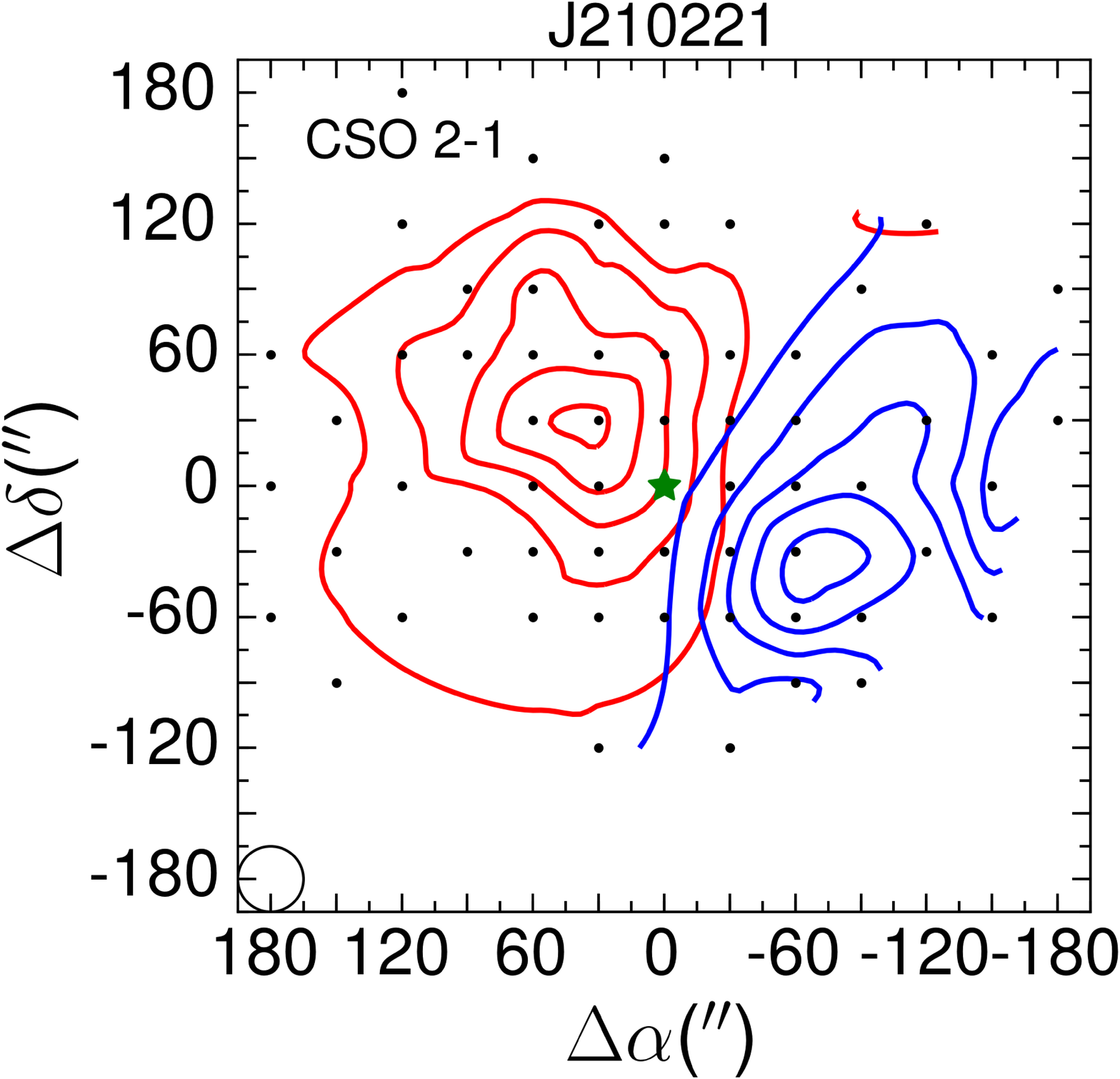} &
\includegraphics[angle=0,scale=0.065]{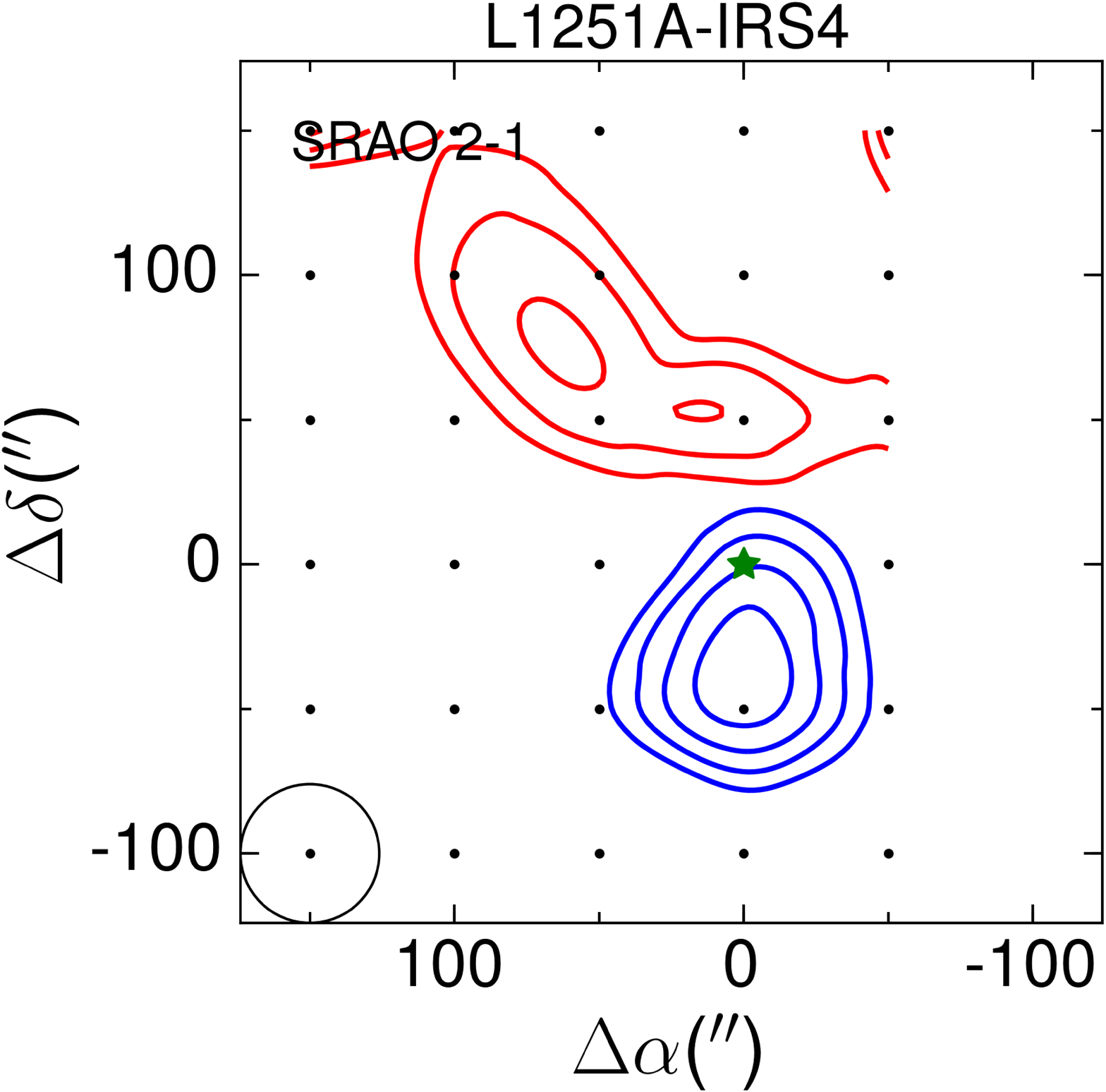} 
\end{array}$
\caption{Contour maps for clear outflow pattern of 16 VeLLOs. Each contour map displays the blue or red component of outflows in contours, respectively. Most VeLLOs show bipolar outflow patterns, but one VeLLO, J182943, presents only red component of the outflow. In three VeLLOs (J041412, J162145, and J182920), the blue and red components of their outflow appear overlapping. In J041412, L1521F-IRS, J154216, IRAS16253-2429, CB130-3-IRS, J182943, J183237, L673-7, J210221, and L1251A-IRS4, some emission around the boundaries of their contour maps could originate in nearby YSOs. Of 16 VeLLOs, the full extent of outflow are almost presented in the contour map of most VeLLOs, but partially shown in the four VeLLOs (J033032, IRAS16253-2499, J183237, and J210221). Black points represent the positions of mapping observation. Star symbol indicates a position of VeLLO. Circle symbol indicates beam size of mapping observation. The velocity ranges, minimum contour levels, and contour level steps used for the blue and the red lobes of the outflows are listed in Table \ref{tbl:outflow1}, respectively.\label{fig:Contourmap1}}
\end{figure}  
\begin{figure}
\center
\includegraphics[angle=0,scale=0.6]{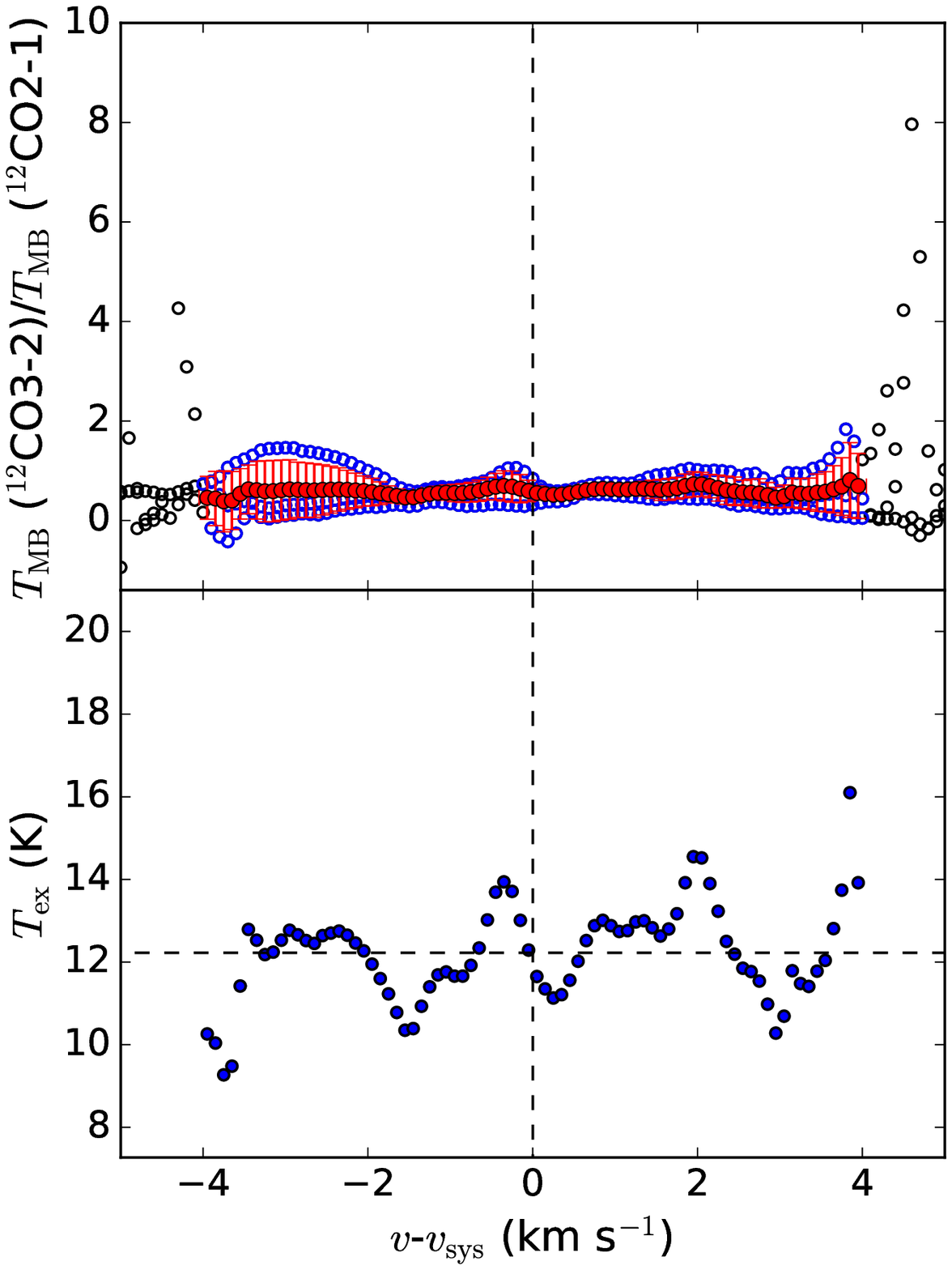}
\caption{Diagrams for deriving the excitation temperature as a function of the spectral line velocity for $^{12}$CO lines. Upper panel indicates the ratios of main beam temperatures as a function of velocity in two $^{12}$CO spectral lines obtained from three VeLLOs (J162648, IRAS16253-2429, and CB130-3). The black circles are all the data and blue circles represent the data selected out of them for the calculation of excitation temperature. Red circles with error bar indicate the average values of data in velocity interval of 0.1 km s$^{-1}$ and their dispersion. Bottom panel represents the resultant excitation temperatures in velocity interval of 0.1 km s$^{-1}$. A vertical line indicates a systemic velocity normalized to 0 km s$^{-1}$ and a horizontal line represents an excitation temperature of 12.2 K which is an average of the excitation temperatures over the velocity ranges of $|v-v_{\rm sys}| \leq 4$ km s$^{-1}$. 
\label{fig:tex}}
\end{figure}  
\begin{figure}
$\begin{array}{cc}
\includegraphics[angle=0,scale=0.6]{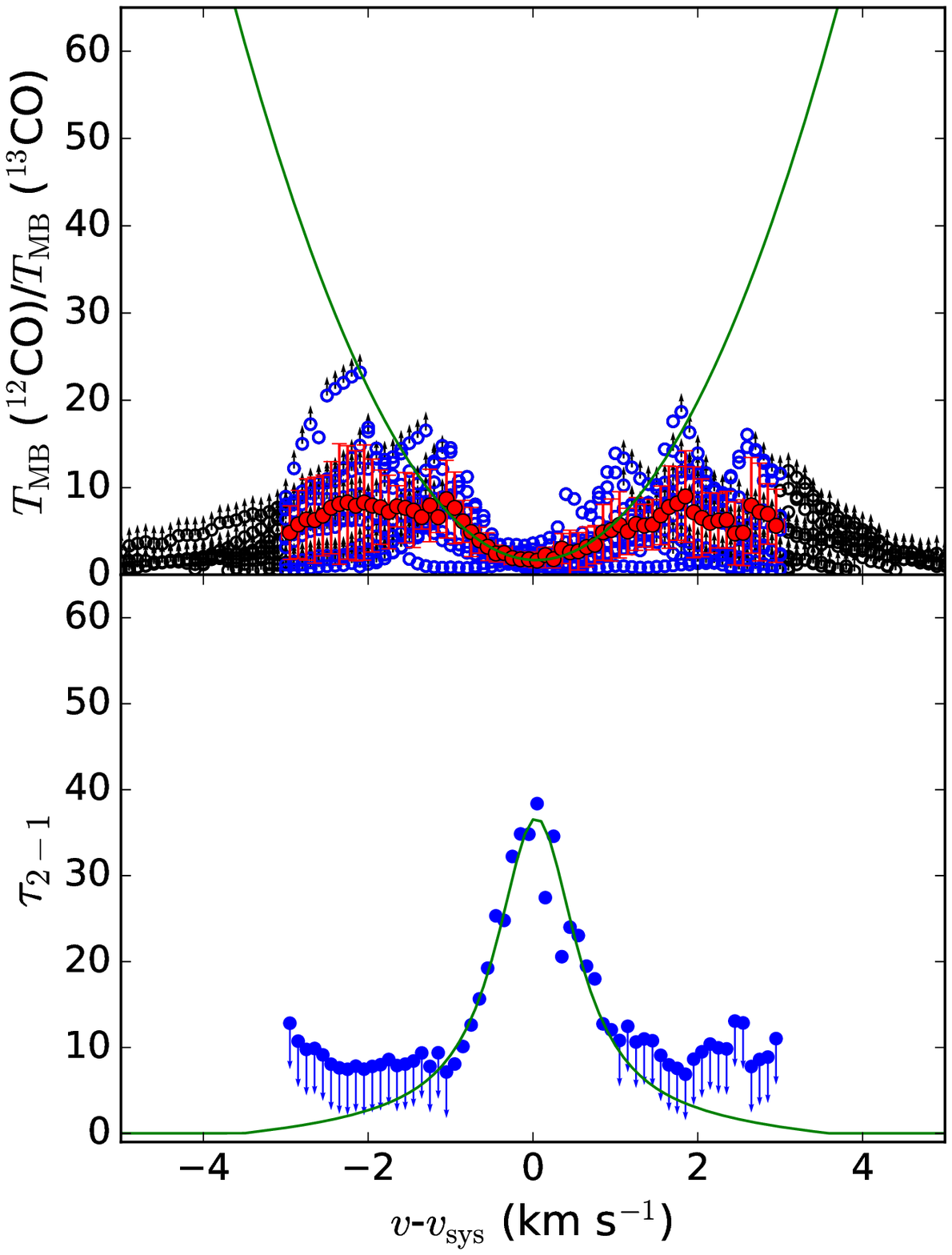} &
\includegraphics[angle=0,scale=0.6]{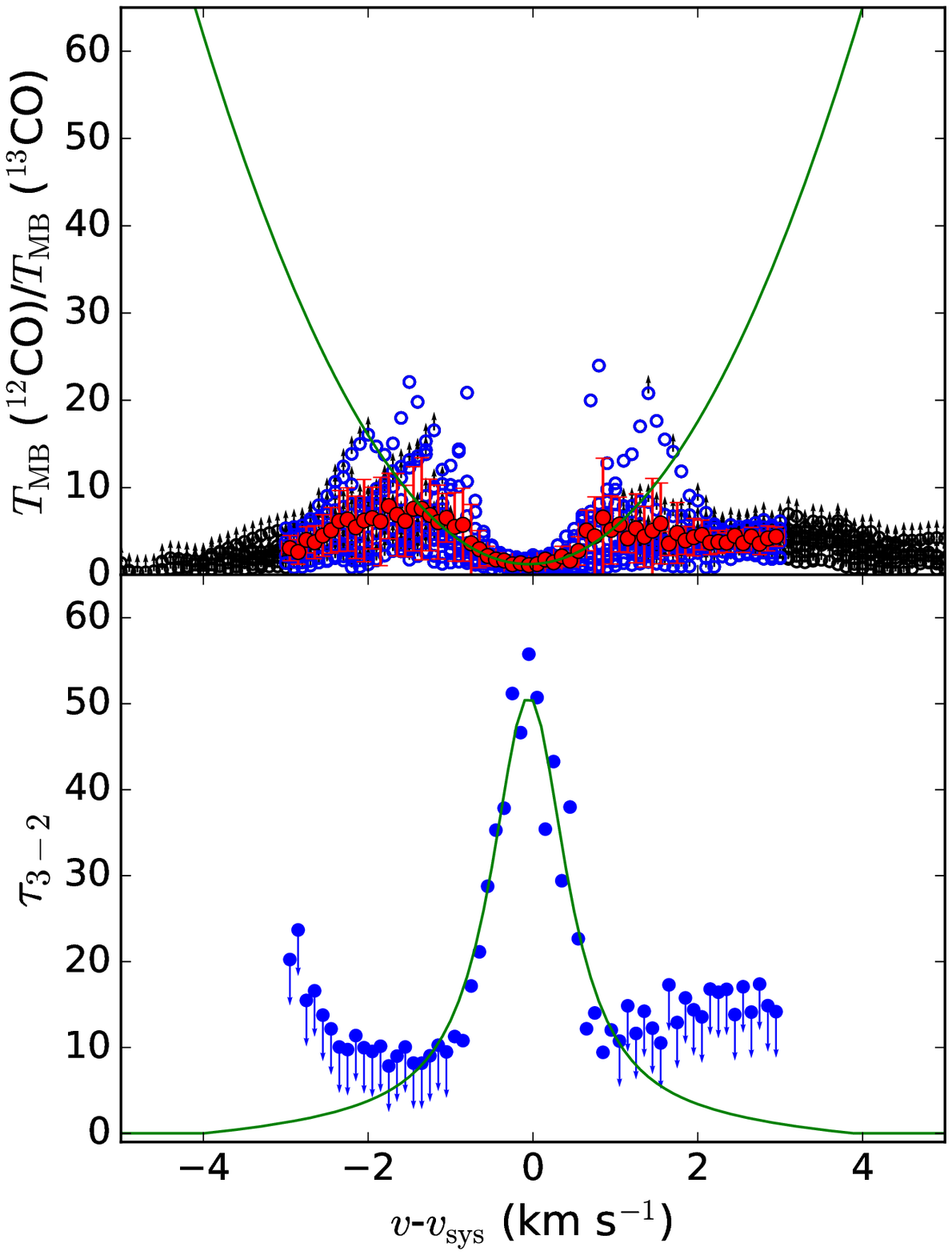} 
\end{array}$
\caption{Diagrams for deriving the optical depths as a function of the spectral line velocity for $^{12}$CO and $^{13}$CO lines. Top panels indicate the ratios of main beam temperatures in $^{12}$CO and $^{13}$CO average spectra obtained at (0,0) positions of observed VeLLOs. The black circles are all the data and blue circles represent the data selected for the calculation of optical depth over the velocity ranges of $|v-v_{\rm sys}| \leq 3$ km s$^{-1}$. Red circles with error bar indicate the average values of data in velocity interval of 0.1 km s$^{-1}$ and their dispersion. The up arrows in this panel indicate the lower limit for the temperature ratio due to non detection in $^{13}$CO line emission in velocity channel. In these calculations we assumed 3$\sigma$ intensity of $^{13}$CO line emission. The green lines represent the best-fit second-order polynomial for red circled data over $|v-v_{\rm sys}| \leq 1$ km s$^{-1}$ for 14 VeLLOs in J=2-1 (left) and over $|v-v_{\rm sys}| \leq 0.8$ km s$^{-1}$ for 12 VeLLOs in J=3-2 (right) so that the temperature ratio for the calculation of optical depth at high velocity wing where $^{13}$CO line is not detected can be inferred from this polynomial function. Bottom panels indicate the resultant optical depths in the velocity interval of 0.1 km s$^{-1}$. The down arrows in this panel indicate the upper limit for the optical depth due to non detection in $^{13}$CO line emission in velocity channel. In these calculations we assumed 3$\sigma$ intensity of $^{13}$CO line emission. The green lines represent optical depths as a function of velocity derived from the best-fit second-order polynomial for the temperature ratios. These optical depth values will be used to correct the optical depth effect of observing tracers ($^{12}$CO lines) in the calculation of column density obtained from the optically thin assumption of the tracers.
\label{fig:tau}}
\end{figure}  
\begin{figure}
\center
\includegraphics[angle=0,scale=0.8]{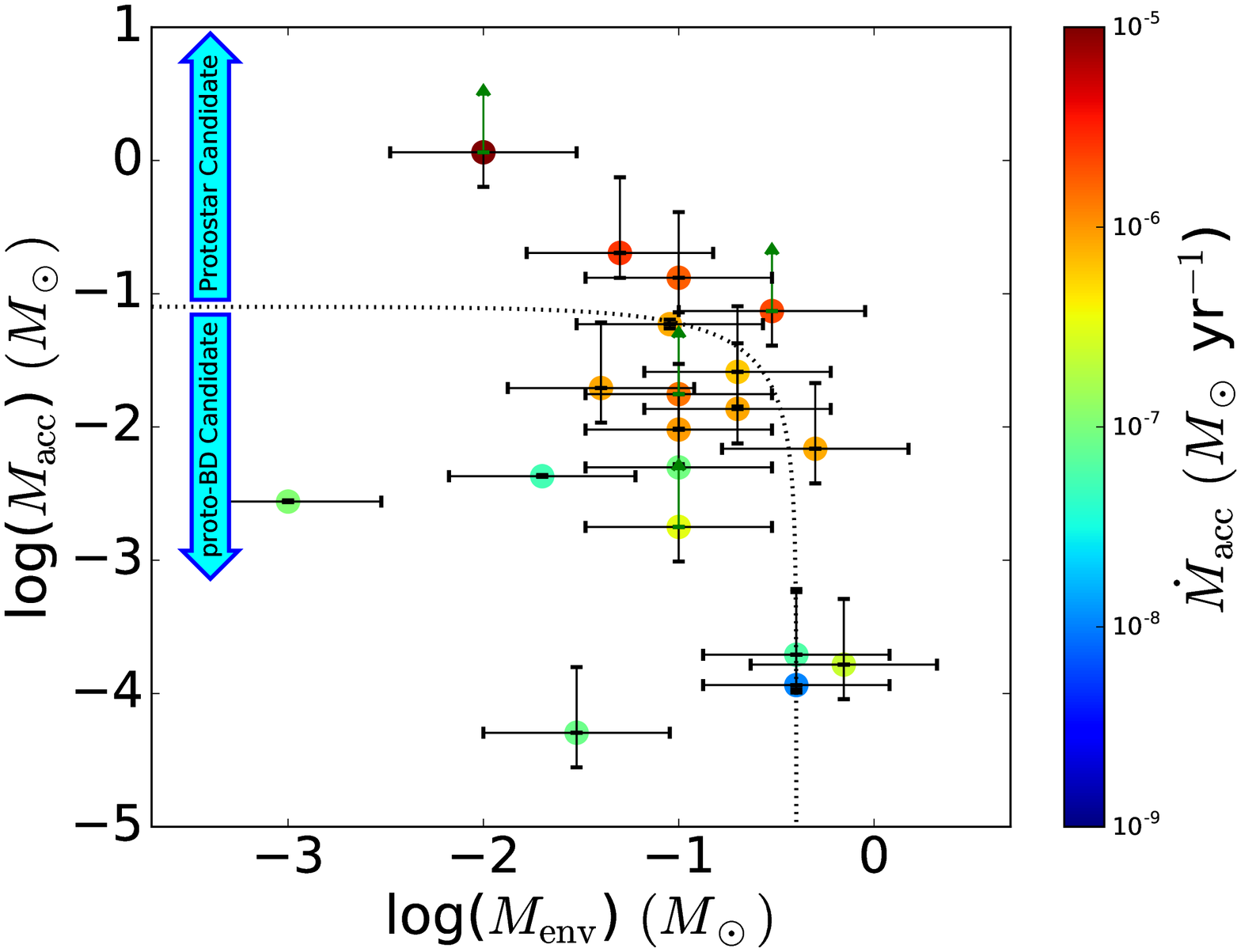}
\caption{Distribution of 19 VeLLOs in the domain of their accreted masses ($M_{\rm acc}$), mass accretion rates ($\dot{M}_{\rm acc}$), and envelope masses ($M_{\rm env}$). The colors of symbols and color bar are to indicate the mass accretion rates. The arrows mean to show the lower limit of the accreted mass because of partial coverage of the outflows in mapping observation. The error bars of x axis indicate the uncertainties of envelope masses that an opacity error of a factor of 3 are taken. The error bars of y axis represent the errors of accreted masses taking inclination uncertainties into account. Dotted line is to show a BD condition $M_{\rm acc}+0.2 \times M_{\rm env} = 0.08$ {\msun} for dividing the stellar and substellar regime. Here we adopted 0.2 by assuming a core-to-star efficiency of 20{\%}. 
\label{fig:corr_Menv_Macc_dMacc}}
\end{figure}  
\begin{figure}
\center
\includegraphics[angle=0,scale=0.8]{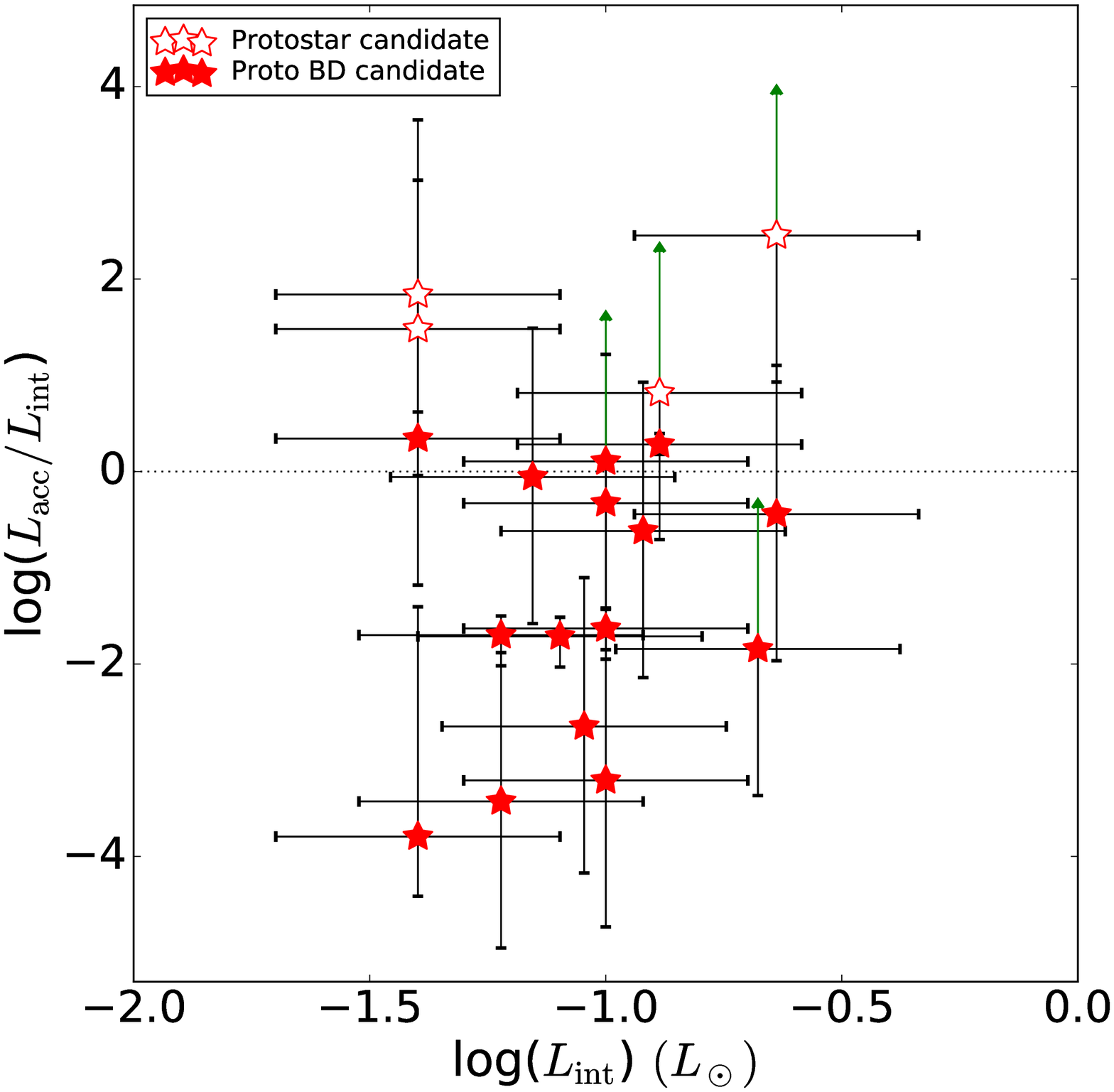}
\caption{Fractional accretion luminosities of the VeLLOs with respect to their internal luminosities. Open star and filled star indicate protostar candidate and proto brown dwarf candidate among 19 VeLLOs, respectively. The error bars of x axis indicate the uncertainties of internal luminosity with a factor of 2. The error bars of y axis represent the errors of {\lacc}/{\lint} taking inclination uncertainties into account. The arrows are to the lower limits of the fractional accretion luminosities because of partial coverage of the outflows in mapping observation. A horizontal line corresponds to 1 in the fractional values of {\lacc}/{\lint}. \label{fig:corr_L_rLL}}
\end{figure}  
\begin{figure}
\center
\includegraphics[angle=0,scale=0.8]{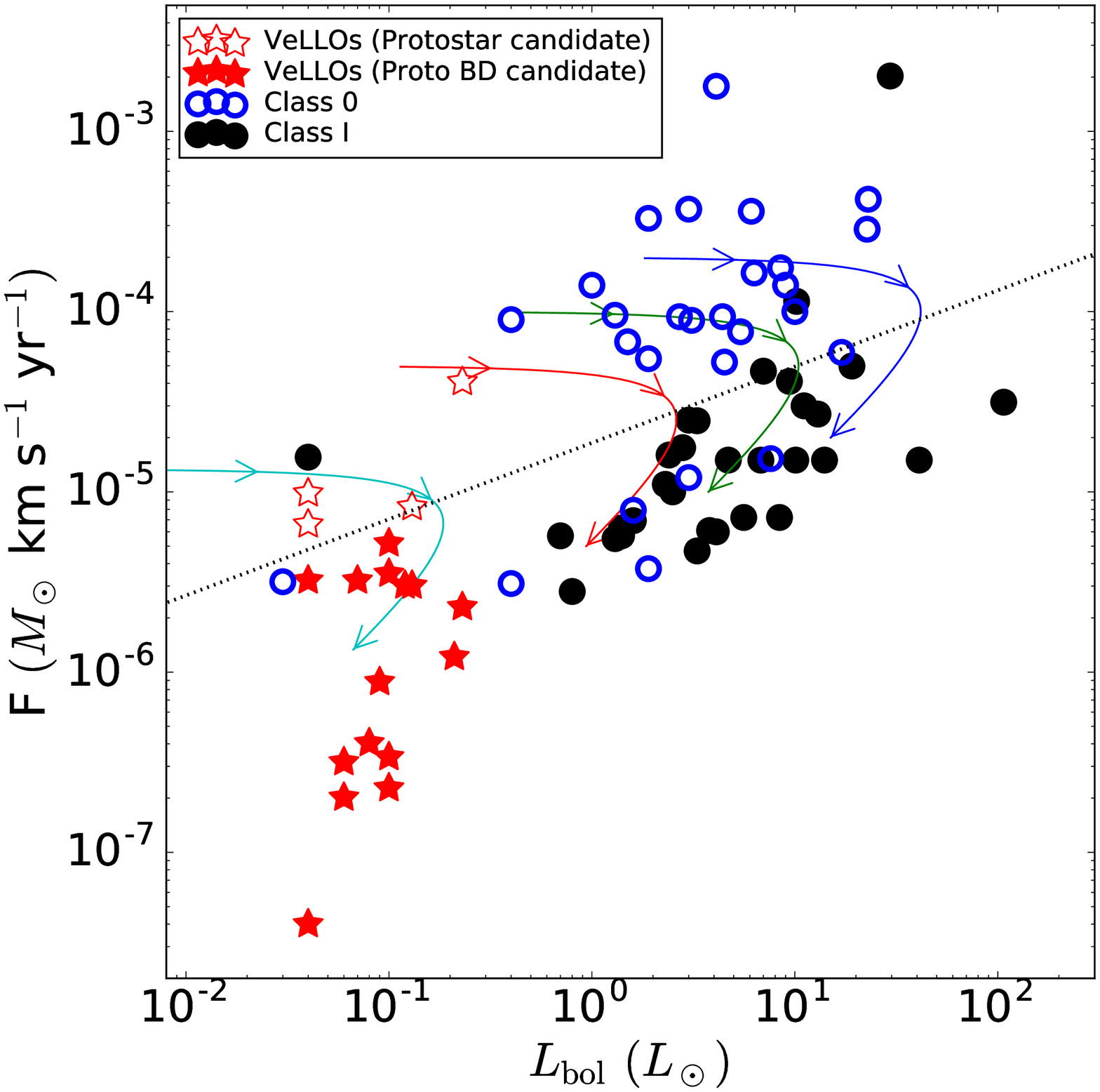}
\caption{Outflow forces of the VeLLOs against their luminosities in comparison with those of Class 0 and I protostars. Open star and filled star indicate protostar candidate and proto brown dwarf candidate among 19 VeLLOs from this study, respectively. Blue circle and black solid circle indicate Class 0 and I protostars from \citet{Bontemps:1996vb}, \citet{Hatchell:2007cz}, and \citet{Curtis:2010ia}, respectively. The outflow forces of most sources are plotted as the corrected values of opacity and inclination, and those of source obtained by interferometers are additionally plotted with correction for missing flux. The luminosities of VeLLOs are given with their internal luminosities, while those of Class 0 and I protostars are shown as their bolometric luminosities. The colored curves represent evolutionary tracks of protostars described by the `toy' model for four initial envelope masses \citep[0.08 {\msun}: cyan, 0.3 {\msun}: red, 0.6 {\msun}: green, and 1.2 {\msun}: blue, from 10$^3$ year to $2\times10^5$ year, See Table 5,][]{Bontemps:1996vb}. A line represents a `best fit' correlation found for protostars by $\rm log(F)=0.4 log(M_{\rm env})-4.7$. \label{fig:corr_F_Lbol}}
\end{figure}  
\begin{figure}
\center
\includegraphics[angle=0,scale=0.8]{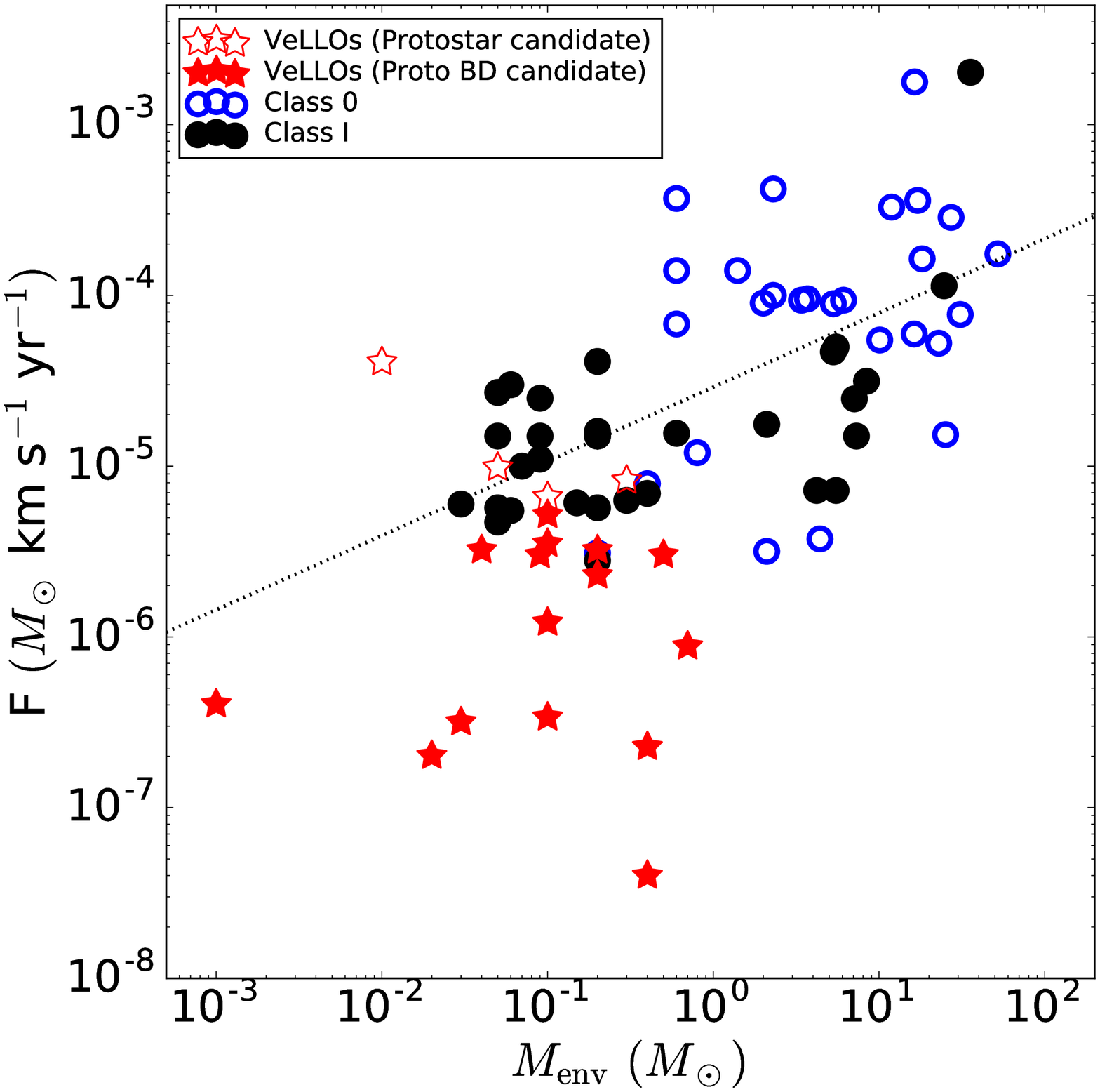}
\caption{Outflow forces of the VeLLOs against their envelope mass in comparison with those of Class 0 and I protostars. Open star and filled star indicate protostar candidate and proto brown dwarf candidate among 19 VeLLOs from this study, respectively. Blue circle and black solid circle indicate Class 0 and I protostar from \citet{Bontemps:1996vb}, \citet{Hatchell:2007cz}, and \citet{Curtis:2010ia}, respectively. Dotted line represents a `best fit' correlation found for protostars by $\rm log(F)=0.4 log(M_{\rm env}) -4.5$.
\label{fig:corr_F_Menv}} 
\end{figure}  
\end{document}